\documentclass[10pt]{article}

\usepackage[english]{babel}
\usepackage{gensymb}
\usepackage[letterpaper,top=2cm,bottom=2cm,left=3cm,right=3cm,marginparwidth=1.75cm]{geometry}

\usepackage{amsmath}
\usepackage{graphicx}
\usepackage[colorlinks=true, allcolors=blue]{hyperref}
\usepackage{float}	%force figure in place
\usepackage{placeins}
\usepackage{subcaption}
\usepackage{natbib}
\usepackage{authblk}

\title{Flow Characterization of the Delft Multiphase Flow Tunnel}

\author[1]{Lina Nikolaidou\thanks{Corresponding author: m.nikolaidou@tudelft.nl}}
\author[1]{Angeliki Laskari}
\author[1,2]{Tom van Terwisga}
\author[1]{Christian Poelma}
\affil[1]{Delft University of Technology}
\affil[2]{Maritime Research Institute of the Netherlands (MARIN)}
\date{}

\begin{document}
\maketitle

\begin{abstract}
At the end of 2020, a new cavitation tunnel was commissioned at the Ship Hydrodynamics laboratory of TU Delft, replacing its 1960s predecessor. Since this was a new facility, a flow characterization campaign was performed to investigate the flow quality in the test section. To that end, velocity measurements were performed in the test section using Laser Doppler Anemometry. Velocities in the range of $2.13~\mathrm{m/s}$ to $9~\mathrm{m/s}$ were measured and a linear relation was found between the freestream velocity and the rotational frequency of the thruster. Long term measurements at the center of the test section, did not reveal any large scale fluctuations of the mean velocity. The freestream turbulence intensity was found to lie between 0.5\% - 0.6\% throughout the test section, after removing the measurement noise. Local measurements in various planes in the test section confirmed that the flow is uniform ($u_{local}< U_{\infty} \times 1\%$), with few outliers near the side walls, due to the turbulent boundary layer. Finally, preliminary measurements of the turbulent boundary layer (TBL) indicated that the TBL originates upstream of the test section and its growth is not strictly canonical. Smaller TBL thickness was found in the side wall compared to the top wall.
\end{abstract}

\section{The Multiphase Flow Tunnel}

\begin{figure}[htp!]
\centering
\includegraphics[width= 0.95\textwidth]{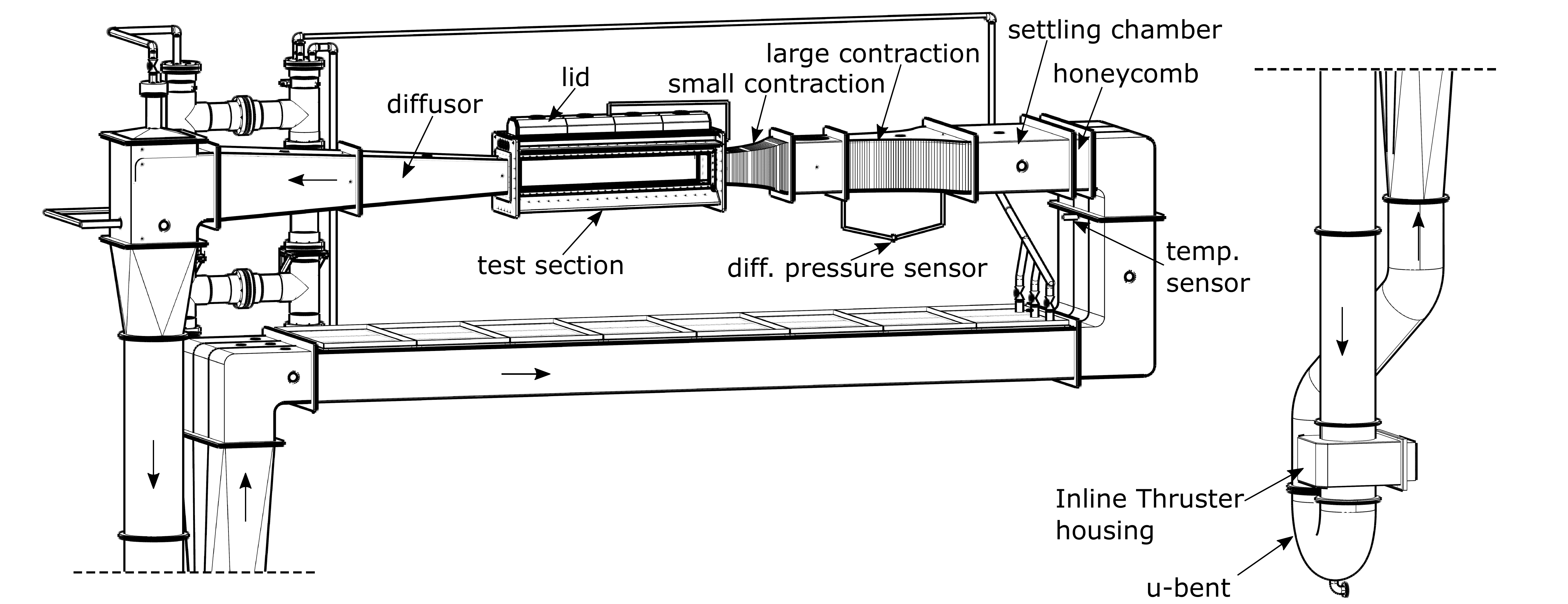}
\caption{Sketch of the Multiphase Flow Tunnel. Flow in the test section is from right to left. The sketch is adapted by the original one by P. Poot.}
\label{mpft_archit}
\end{figure}

The total volume of the tunnel is approximately $V_{tunnel}=17~\mathrm{m}^{3}$ and the flow is driven by a Voith Inline Thruster (VIT; Voith, Heidenheim, Germany) with a total power of 110 KW. Prior to reaching the test section and with the aim to attenuate the turbulence, the flow goes through a honeycomb, a settling chamber and two contractions (see \autoref{mpft_archit}). The honeycomb section is $200~\mathrm{mm}$ long, and is comprised of hexagon shaped cells. The large (first) contraction is asymmetric, with a cross section varying from $A_0=1500~\times500~\mathrm{mm}^2$ upstream to $A_1=500~\times500~\mathrm{mm}^2$ downstream. The small (second) contraction is symmetric, and it produces a further reduction of the cross-section to $A_2=300\times300~\mathrm{mm}^2$. The cross section dimensions of the test section change from $w=300~\mathrm{mm}$ $\times$ $h=300~\mathrm{mm}$ at the inlet to $w=300~\mathrm{mm}$ $\times$ $h_1=320~\mathrm{mm}$ at the outlet, due to a sloping bottom wall. The length of the test section in $2.14~\mathrm{m}$. The test section is firmly closed from the top with a lid featuring four viewing windows. For the purpose of the current experiments, a smooth polycarbonate flat plate of length $L = 1.95~\mathrm{m}$ was positioned flush with the top wall of the second contraction. Downstream of the test section, the flow enters a two-part diffuser. By the end of the diffuser the cross section is $668~\mathrm{mm}\times668~\mathrm{mm}$. The total length of the diffuser is approximately $3~\mathrm{m}$. Next, the flow is directed to the inline thruster located two floors beneath the test section level, to avoid cavitation.  Within the interests of air lubrication and cavitation research, the Multiphase Flow Tunnel (MPFT) features a degassing system.

\section{Experimental setup \& Methods}
\label{expsetupLDA}

Laser Doppler Anemometry (LDA) is a widely used technique to measure velocity in liquid flows. Using this technique, the liquid flow velocity is calculated by measuring the velocity of small particles (tracers). The measurement principle behind LDA is that it detects the Doppler shift in light scattered by these moving particles, with the shift being directly proportional to their velocity. Among the advantages of LDA is that it is a non-intrusive technique with the possibility to measure each velocity component separately and also in reverse flow conditions. Moreover, the system response is linear and the calibration straightforward. The measurement volume can be small (typically $0.1\times0.1\times1~\mathrm{mm}^3$) resulting in a high spatial resolution. In addition, the high temporal resolution, high sampling rate and small sample volume  makes it suitable for turbulence measurements. From a practical point of view, LDA was also suitable for the MPFT flow characterization campaign because it provides the flexibility to perform measurements in various 2D planes without the need to perform multiple calibrations (like eg. in PIV). Furthermore, the test section was closed from the top so that the limited accessibility made LDA an attractive choice.

Over the years several optical modes for the LDA are developed (reference beam system, cross-beam system). In our measurements, we used the cross-beam system and analyzed it with the ``fringe model''. These are briefly described next.

\begin{figure}[htp!]
\centering
\includegraphics[width= 0.8\linewidth]{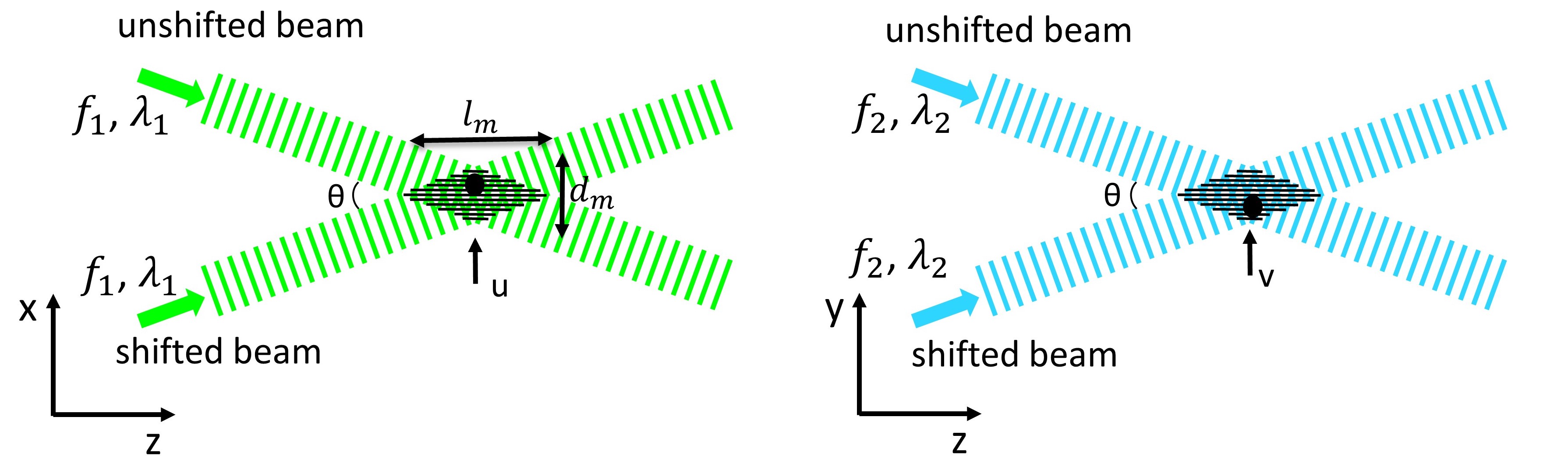}
\caption{Schematic representation of the ``fringe model'' in two planes.}
\label{laserbeamLDA}
\end{figure}

When two laser beams with wavelength $\lambda$ of equal intensity cross, there is a formation of interference fringes in the crossing region. As the frequency of the light $f$ is the same in the two beams, these fringes are stationary. If a solid particle is passing through the intersection of the beams, it will cross successively bright and dark fringes. Light will be scattered only when a bright fringe is crossed and an oscillatory signal of a certain amplitude and frequency will be produced, depending on how fast or slow the particle velocity is and how thick the fringes are. The optical setup may be regarded then as an interferometer. The fringe spacing, $d_f$ is an important parameter of the LDA system. It depends only on the wavelength of light, $\lambda$, and on the angle of the two laser beams $\theta$:

\begin{equation}
    d_f=\frac{\lambda}{2 \times \sin{\theta /2}}
\end{equation}

Provided that the frequency $f_{D}$ of the oscillatory signal is acquired by the LDA system, the velocity of a moving particle is then calculated with:

\begin{equation}
    u=d_f \times f_D
    \label{fd}
\end{equation}

On the above equation the streamwise velocity $u$ is used, but the vertical velocity $v$ is acquired in the same way, when a particle traveling in the vertical direction crosses the fringes (\autoref{laserbeamLDA}). 

In practice the Doppler frequency follows from the analysis of a photodetector output signal that oscillates with frequency $f_D$. As a result, it is not possible to determine the sign of the velocity component $u$ in \autoref{fd}. To correct for this "directional ambiguity", one of the incident laser beams is frequency shifted by a known value, $f_s$. This causes the fringe pattern to move at speed $f_s \times d_f$. Particles crossing the measurement volume will now have a frequency either above or below $f_s$, depending on their direction. The direction of motion is then known.

\begin{figure}[htp!]
\centering
\includegraphics[width= \linewidth]{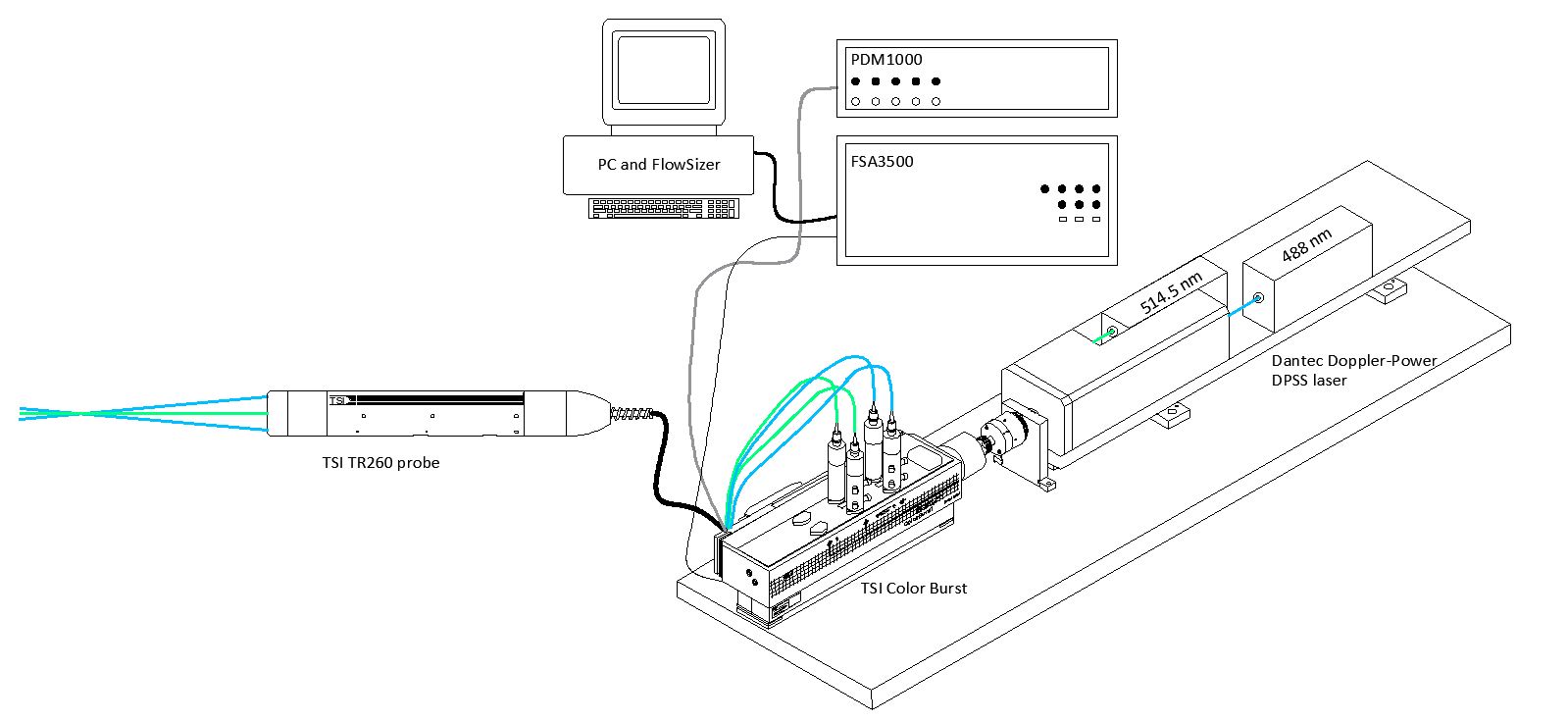}
\caption{Components of our LDA system. Adapted from \cite{TSI_manual}.}
\label{LDA_setup}
\end{figure}

Having introduced the basic working principles of the LDA system used, the experimental setup is subsequently explained. The velocity measurements were performed with a two-component LDA system operating in a backward-scatter configuration (\autoref{LDA_setup}). Two Dantec Doppler-Power DPSS argon-ion lasers were used as laser sources. These lasers produce laser beams of $\lambda=514.5$ nm (green) and $\lambda=488$ nm (blue). The mixed color laser beam is then directed to the TSI Color Burst which uses an acousto-optic Bragg cell to split the incoming laser beam into its green and blue components while simultaneously producing shifted (and un-shifted) versions of these beams. Thus, after passing through the Bragg cell there are two blue beams (one shifted and one un-shifted) and two green beams (one shifted and one un-shifted). Each beam is coupled into an optical fiber with the manipulators on top of the TSI Color burst. The Bragg cell is driven by a typically 40 MHz signal coming from a driver that is installed in the FSA3500 processor. The four beams are then directed to the laser probe (TSI TR260), where light transmitting optics focus the four laser beams into the measurement volume, while separate optics collect the (back-) scattered light. The fiber that goes into the light bar contains the back-scattered mixed color signal collected by the fibre probe. Inside the light bar the back-scattered light is split into its components (green light at 514.5 nm and blue light at 488.0 nm) and then imaged on the surface of two photodetectors (one for each color). The photodetectors (PDM100) convert the optical signal into an electronic signal that is subsequently handled by the processor's electronics (FSA3500).  The FLOWSIZER$^{TM}$ software
was used for data acquisition and processing of the signal. An oscilloscope was also used to assess the signal quality.

\begin{figure}[htp!]
\centering
\includegraphics[width= 0.65\linewidth]{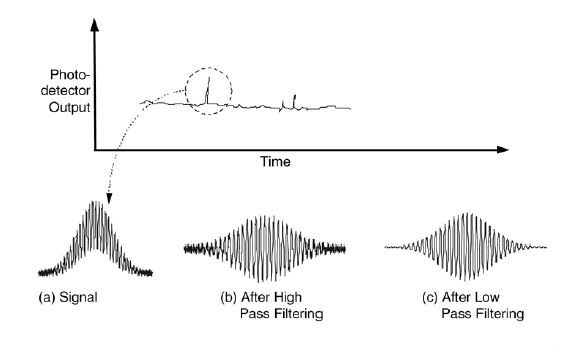}
\caption{Schematic representation of the different signal components. Source: \cite{TSI_manual}.}
\label{LDA_signals}
\end{figure}

As explained earlier, the light scattered by a particle passing through the fringes in the measurement volume is optically collected and focused onto the photomultiplier tube (PMT). The PMT produces an electrical current proportional to this light flux. Variations in this electrical current caused by the particle passing the fringes are subsequently analyzed to determine the velocity of the particle. It must be noted that the signal is discontinuous and is produced only when a tracer particle crosses the measurement volume. The signal output from the PMT can typically be decomposed to the following components (\autoref{LDA_signals}): 

\begin{enumerate}
    \item a low frequency ``pedestal'' caused by the particle passing through the focused Gaussian-intensity laser beams.
    \item a Doppler frequency, $f_d$, signal that is superimposed on the pedestal and has a regular sinusoidal pattern related to how fast the particle crosses the fringes in the measurement volume. 
    \item wide bandwidth electronic noise generated by stray light, the PMT, and associated electronics. 
\end{enumerate} 

Signal (a) in \autoref{LDA_signals} is the sum of the pedestal component, the Doppler component and the noise from the photodiode. Since the pedestal is undesirable for good signal processing, it is removed by simply filtering the signal through a high-pass filter. The resulting signal, with pedestal-removed, is shown in (b) in \autoref{LDA_signals}. High frequency noise can be reduced by low-pass filtering the noise at a frequency which must be greater than $f_D$ in order to avoid filtering the Doppler signal.

\begin{table*}[!ht]
\small
\centering

\begin{tabular}{c c c c}
& & \\ % put some space after the caption
\hline
parameter & definition & \multicolumn{2}{c}{laser beam }\\
 &  & green & blue\\
\hline
f (mm) & focal length & \multicolumn{2}{c}{250}\\
$\theta$ ($\degree$) & beams angle (air) & \multicolumn{2}{c}{11.422}\\ 
D (mm) & initial beam diameter & \multicolumn{2}{c}{2.65}\\
d (mm) & beams separation distance & \multicolumn{2}{c}{50}\\
$\lambda$ (nm) & wavelength & 514.5 & 488\\
$d_f$ ($\mu$m) & fringe spacing & 2.5853 & 2.4522\\
$d_{e^{-2}}$ ($\mu$m) & $e^{-2}$ beam diameter & 61.8 & 58.62\\
$d_m$ ($\mu$m) & measurement volume diameter & 62 & 58.91\\
$l_m$ ($\mu$m) & measurement volume length & 621 & 589\\
$N_{FR}$ & number of fringes & 24 & 24\\
V ($\mathrm{mm}^3$) & measurement volume & 0.0012 & 0.0011\\
\hline
\end{tabular}
\caption{Summary of the characteristics of the measurement volume.}
\label{LDA_parameters}
\end{table*}

The summary of the geometrical characteristics of the measurement volume are shown in \autoref{LDA_parameters}. The formulas and further details can be found in \cite{TSI_manual}. The measurement volume indicating the spatial resolution of the LDA is then 0.058  $\times$ 0.062 $\times$ 0.621 $\mathrm{mm}^3$.

To optimize the raw signal, the following parameters were modified: photomultiplier voltage, downmixing frequency, signal to noise ratio and burst threshold such that the burst efficiency remained always $>70\%$. Software coincidence mode was used. As a result, the mean sampling rate varied between 50 Hz and 150 Hz depending on the measurement position.

The calibration for this backward-scatter LDA system is quite straightforward. At all cases the axis was perpendicular to the tunnel wall (optical windows). Since the transmitting-receiving probe is self-aligned, the four beams were focused at one point. Each day before measurements the laser beams were projected on a A4 sheet in low intensity, to make sure that the fiber optics were aligned with the laser beams. Modifications were done if needed using the 4-knobs of Bragg cell. 

Hollow glass spheres of mean diameter $10$ ${\mu} m$ were used as seeding particles. The amount of seeding particles needed was evaluated with a help of an oscilloscope. Too few seeding particles resulted in very low data rate, while too many introduced measurement noise and decreased the water transparency.

To ensure precise positioning of the measurement volume, the laser probe was mounted on a two-way traverse system, allowing accurate adjustments in the $y-z$ plane (0.5 mm accuracy). The positioning of the laser probe with respect to the desired measurement volume location in water, was determined by taking into consideration the refractive indices of the tunnel wall ($n_r\approx1.495$) and water ($n_r\approx1.33$).  In the streamwise direction, the probe was positioned manually (0.5 cm accuracy).  

Along with the LDA measurements, simultaneous measurements of the water  temperature, the differential pressure over the contraction and the absolute pressure in three positions were taken (see \autoref{mpft_archit}). These were controlled with a LabVIEW™ program. More specifically, the differential sensor (PD-33X manufactured by KELLER®) was used to measure the pressure drop $\Delta p$ across the large contraction. Subsequently, with the application of Bernoulli and continuity equation across the large and small contractions, the velocity $U_{\Delta p}$ at the test section inlet can be determined using:

\begin{equation}
\label{udp}
    U_{\Delta p}=\sqrt{\frac{2{\Delta p}}{\rho \left(1-\left(\frac{A_1}{A_2}\right)^2\right)\left(\frac{A_0}{A_1}\right)^2}}
\end{equation}

where $A_0$ is the cross-sectional area at the inlet of the large contraction (= $0.735~\mathrm{m}^2$), $A_1$ is the cross-sectional area at the inlet of the small contraction (= $0.2507~\mathrm{m}^2$), $A_2$ is the cross-sectional area at the inlet of the test section (= $0.09~\mathrm{m}^2$) and $\rho$ is the water density. The temperature was used to calculate the density $\rho$ and the viscosity according to \cite{ittc2006}.

\section{Measurement Plan}
\label{measurement plan}

\begin{figure}[htp!]
\centering
\includegraphics[width= 0.8\linewidth]{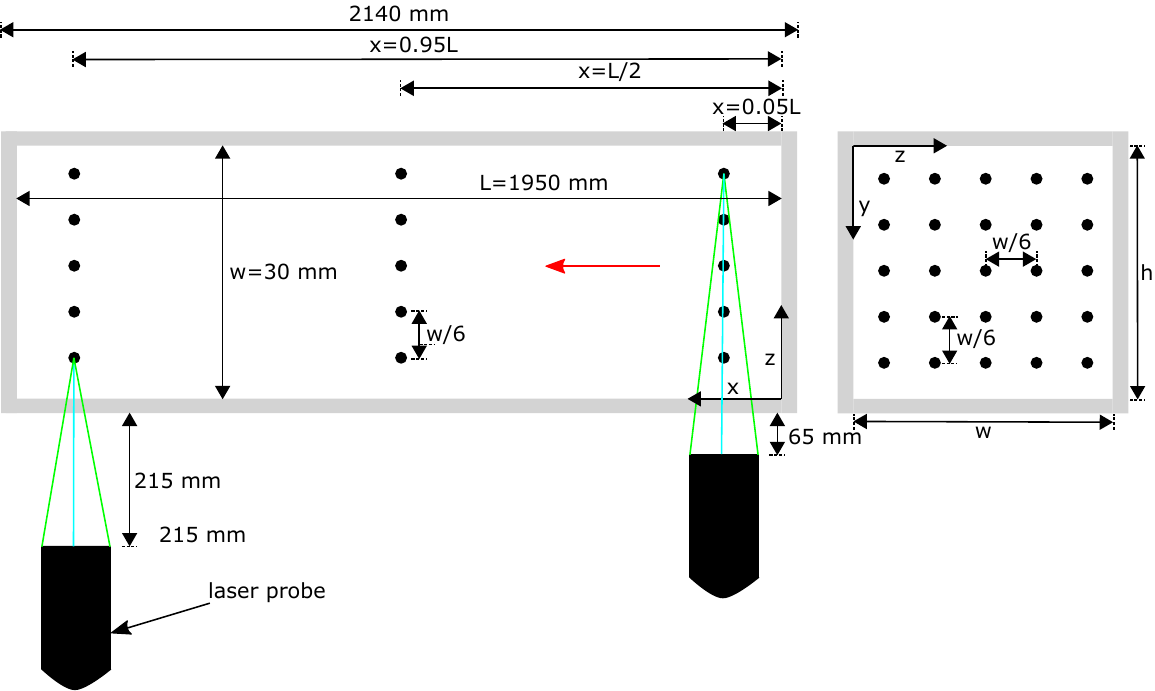}
\caption{Schematic representation of the flow uniformity measurement positions (not to scale). Flow is from right to left.}
\label{ts}
\end{figure}

To assess the flow uniformity in the test section and characterize the flow quality, four types of measurements were conducted. Firstly, local velocity measurements were taken on a $5 \times 5$ grid  in the spanwise-wall-normal plane ($y-z$) to assess the flow uniformity. These were performed in three streamwise locations (see \autoref{ts}): just downstream of the inlet ($x = 100$ mm $\approx 0.05L$), in the mid-span of the tunnel ($x=975$ mm $\approx L/2$) and just upstream of the outlet ($x=1850$ mm $\approx 0.95L$). These measurements were performed for three rotational frequencies (RPM) of the VIT: 200 RPM ($U_{\infty}=3.5$ m/s), 285 RPM ($U_{\infty}=5$ m/s) and 400 RPM ($U_{\infty}=7$ m/s), resulting in a total of 225 measurement points. Each grid point measurement lasted about 3 minutes depending on the location. Secondly, boundary layer measurements at the top and side walls were performed at the same locations and RPM with the local velocity measurements. It must be noted that accurate boundary layer measurements with LDA require special treatment of the incoming laser beam and also different raw signal optimization from measurements away from the wall. Nevertheless, preliminary measurements were taken to estimate the boundary layer growth and thickness. The wall was found by manually positioning the measurement volume at the wall, and then traversing the probe away from it in 0.5 mm steps, until the data rate sharply increased, indicating reflections by the transparent wall. Next, long measurements (60 minutes per point) at $x=L/2$ and at the midpoint of the $y-z$ plane ($y=h/2$ and $z=w/2$) were performed to assess the long term dynamics of the flow for three characteristic frequencies (200, 285 and 550 RPM). Lastly, shorter measurements (30 minutes per point) were performed for a wider range of frequencies (125-500 RPM) at the inflow of the test section ($x=0.05L$ and at the midpoint of the $y-z$ plane), to quantify the freestream turbulent intensity, establish the RPM - $U_{\infty}$ correlation and calibrate the differential pressure sensor. For all cases, bias corrected statistics were calculated \citep{buchhave1978bias}.

\section{Results}
\subsection{Boundary layer measurements}
\label{BLmeasurements}

 \begin{figure*}[!ht]
\begin{center}
\begin{subfigure}{0.4\linewidth}
  \centering
  \includegraphics[width=\linewidth]{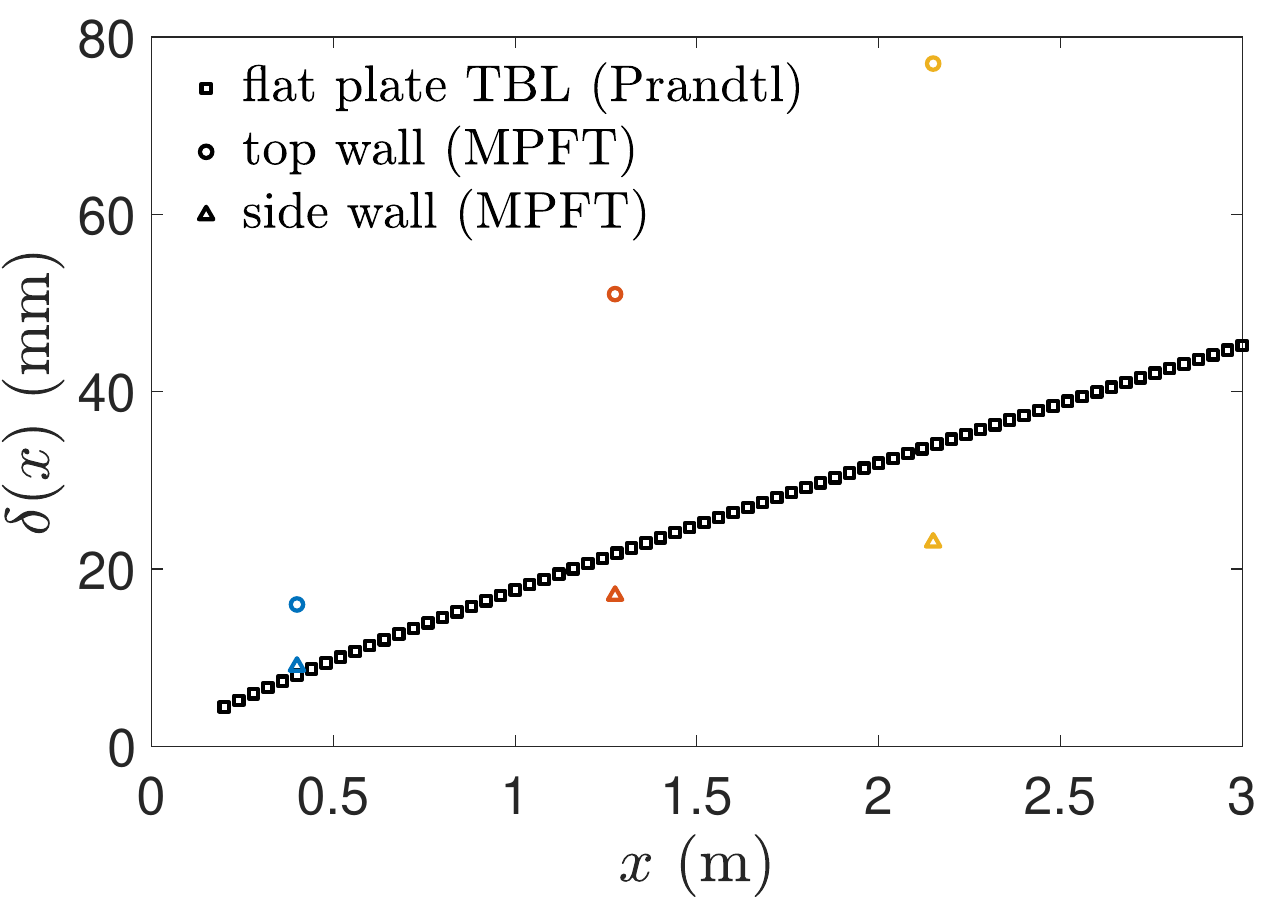}
  \caption{}
  \label{tbl_growth}
  \end{subfigure}
  \begin{subfigure}{0.4\textwidth}
  \centering
    \includegraphics[width=\linewidth]{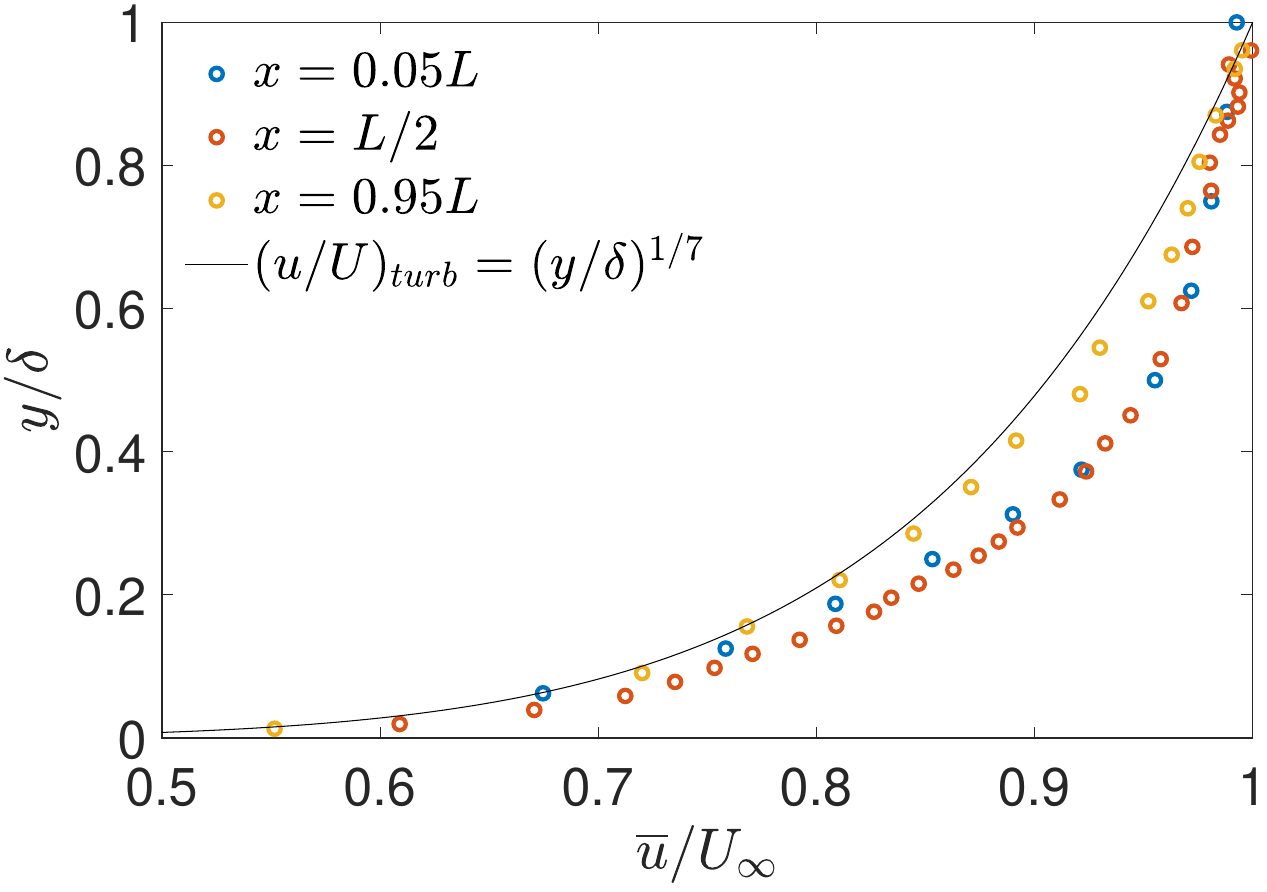}
      \caption{}
      \label{tbl_profile}
      \end{subfigure}
      \caption{TBL measurements for $U_{\infty}=5$ m/s: (a) top and side wall TBL thickness measurements in three streamwise positions and (b) top wall normalized mean streamwise velocity profiles in the same positions.}
      \label{tbl plots}
      \end{center}
  \end{figure*}

Boundary layer measurements of the top wall were performed for $U_{\infty}$ of 3.5, 5 and 7 m/s (or 200, 285 and 400 RPM). The boundary layer thickness was defined as the wall normal distance
where the velocity is $99\%$ of the freestream velocity. $U_{\infty}$ and $\delta$ were
then determined iteratively with $U_{\infty}$ defined as the mean of all data points with $y>\delta$. The friction velocity was determined using the Clauser chart method  \citep{clauser1956turbulent}. In \autoref{tbl_growth}, measurements are shown for $U_{\infty}=5$ m/s along with a power-law approximation for the turbulent boundary layer (TBL) growth \citep{white2006viscous}. The measurements points do not follow a $\delta \sim x^{6/7}$ growth. Fitting an exponent $n$ to the equation $\delta/x=0.16/Re_x^{1/n}$, resulted in $\delta \sim x^{10/11}$ with a virtual origin 30 mm prior to the inlet, indicating a faster TBL growth than the canonical one. TBL profiles of the same velocity (\autoref{tbl_profile}), also deviate from the one-seventh power law except for the most downstream measurement position. 

Several effects can lead to the TBL growth deviating from the canonical state. Firstly, since the flat plate is flush with the contraction top wall, the TBL does not originate from the leading edge of the plate; instead, it starts developing already at the tunnel walls upstream of the test section. Secondly, due to the sloping bottom across the test section, a mild adverse pressure gradient is present in the flow (see also \autoref{flow uniformity measurements}), affecting the TBL.

Side wall measurements were performed for $U_{\infty}=5$ m/s to investigate asymmetry effects (\autoref{tbl_growth}). Considerably lower $\delta$ were found in the side walls than the top ones along with a slower TBL growth rate ($\delta \sim x^{5/6}$). Geometrical asymmetries in the tunnel prior to the inlet possibly contribute to this discrepancy (see large contraction in \autoref{mpft_archit}). A summary of the top and side TBL measurements can be found in \autoref{summary_tbl}.

\begin{table*}[!ht]
\centering
\begin{tabular}{c c c c c}
& & \\
 \hline
 \hline
 \multicolumn{5}{c}{$U_{\infty}=3.5$ m/s \textit{top} TBL} \\
 \hline
 \hline
x  & $Re_{x}\times10^6$ & $\delta_{99}$ (mm) & $u_{\tau}$ (m/s) & $\nu/ u_{\tau}$ ($\mu$m )\\
\hline
0.05L & 1.7 & - & - & - \\
L/2 & 5.4 & 53 & 0.1236 & 7.33\\
0.95L & 9.1 & -  & - \\
\hline
\hline
\multicolumn{5}{c}{$U_{\infty}=5$ m/s \textit{top} TBL} \\
 \hline
 \hline
x  & $Re_{x}\times10^6$ & $\delta_{99}$ (mm)   & $u_{\tau}$ (m/s) & $\nu/ u_{\tau}$ ($\mu$m )\\\hline
0.05L & 2.4 & 16 & 0.1912 & 4.38 \\
L/2 & 7.6 & 51 & 0.1755 & 4.77\\
0.95L & 12.8 & 77 & 0.1588 & 5.19\\
\hline
\hline
\multicolumn{5}{c}{$U_{\infty}=7$ m/s \textit{top} TBL} \\
 \hline
 \hline
x  & $Re_{x}\times10^6$ & $\delta_{99}$ (mm)  & $u_{\tau}$ (m/s) & $\nu/ u_{\tau}$ ($\mu$m )\\\hline
0.05L & 3.4 & 12  & - & -\\
L/2 & 10.8 & 52  & 0.2456 & 3.14\\
0.95L & 18.1 & - & - & -\\
\hline
\hline
\multicolumn{5}{c}{$U_{\infty}=5$ m/s \textit{side} TBL} \\
 \hline
 \hline
x  & $Re_{x}\times10^6$ & $\delta_{99}$ (mm)  & $u_{\tau}$ (m/s) & $\nu/ u_{\tau}$ ($\mu$m )\\\hline
0.05L & 2.4 & 9  & - & -\\
L/2 & 7.6 & 17  & - & -\\
0.95L & 12.8 & 23  & - & -\\
\hline
\hline
\end{tabular}
\caption{Summary of the top and side TBL measurements.}
\label{summary_tbl}
\end{table*}

\FloatBarrier

\subsection{Flow uniformity measurements}
\label{flow uniformity measurements}

\begin{figure}[!t]
\centering
\resizebox{0.75\textwidth}{!}{%
\begin{minipage}{\textwidth}
 %\textbf{$x=0.05L$ \hspace{2cm} $x=L/2$ \hspace{2cm} $x=0.95L$}\par\medskip
\noindent
\begin{tabular}{p{0.36\linewidth} p{0.16\linewidth} p{0.32\linewidth}}
\centering \textbf{$x=0.05L$} &
\centering \textbf{$x=L/2$} &
\centering \textbf{$x=0.95L$}
\end{tabular}
\par\medskip
\begin{subfigure}{0.34\linewidth}
  \centering
  \includegraphics[width=\linewidth]{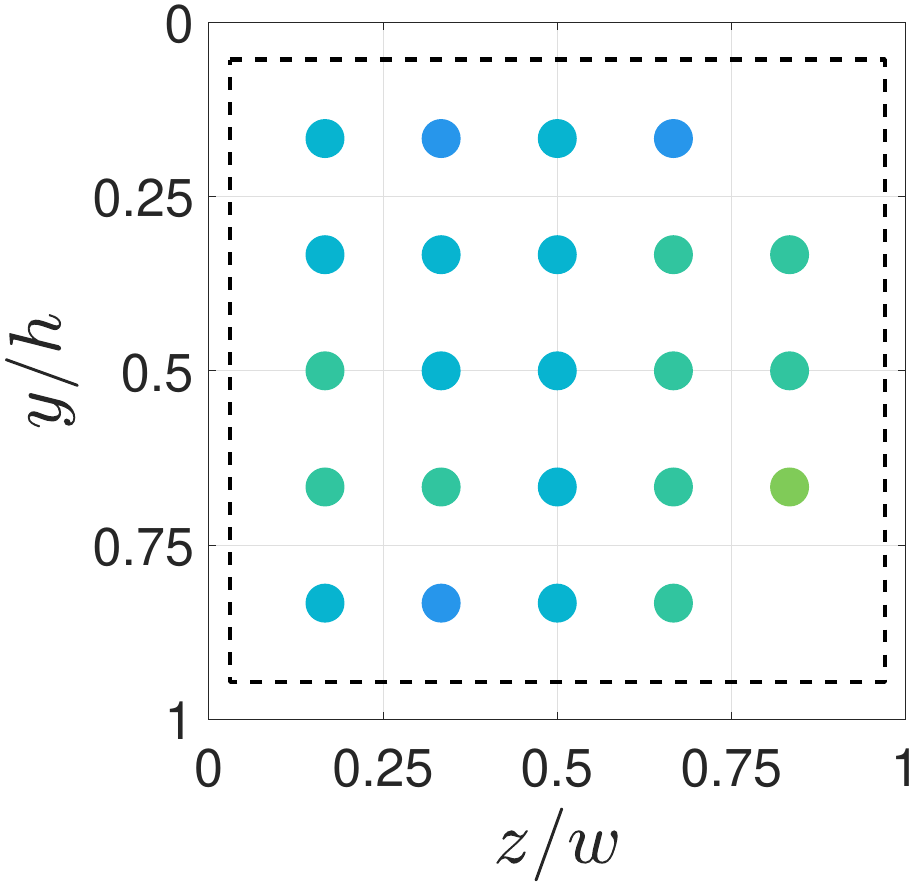}
  \caption{}
  \label{UMEAN_up}
  \end{subfigure}
  %\hspace{0.01\textwidth}
  \begin{subfigure}{0.27\linewidth}
  \centering
    \includegraphics[width=\linewidth]{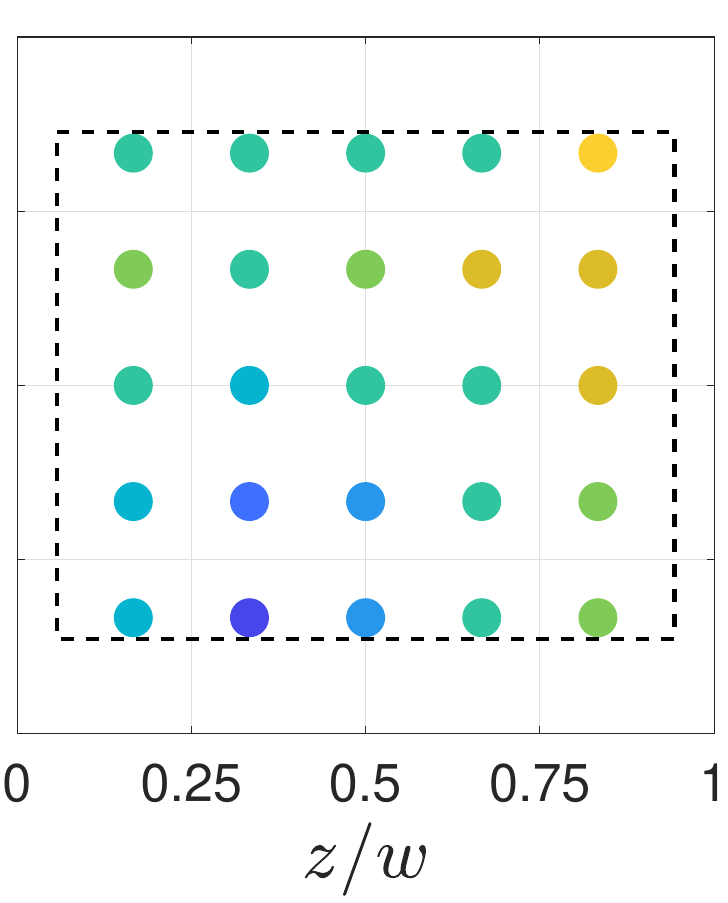}
      \caption{}
        \label{UMEAN_middle}
      \end{subfigure}
      \begin{subfigure}{0.36\linewidth}
        \centering
         \includegraphics[width=\linewidth]{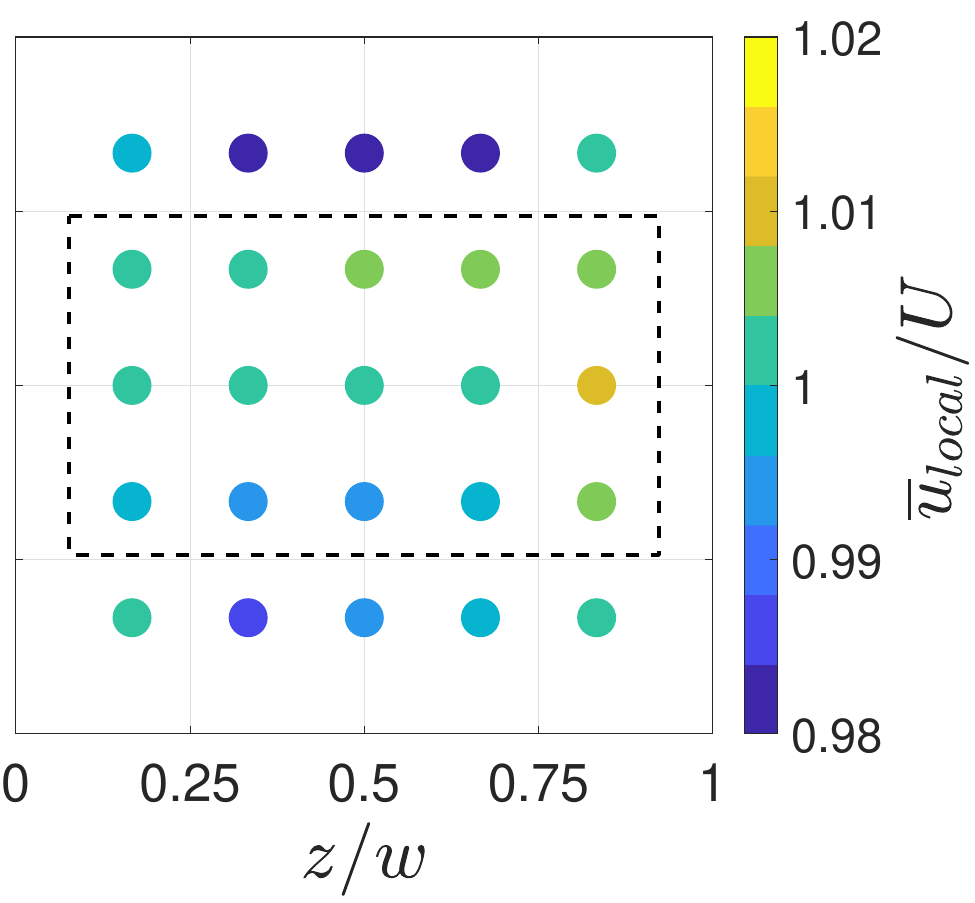}
                \caption{}
                  \label{UMEAN_down}
      \end{subfigure}
    \vfill
\begin{subfigure}{0.34\linewidth}
  \centering
  \includegraphics[width=\linewidth]{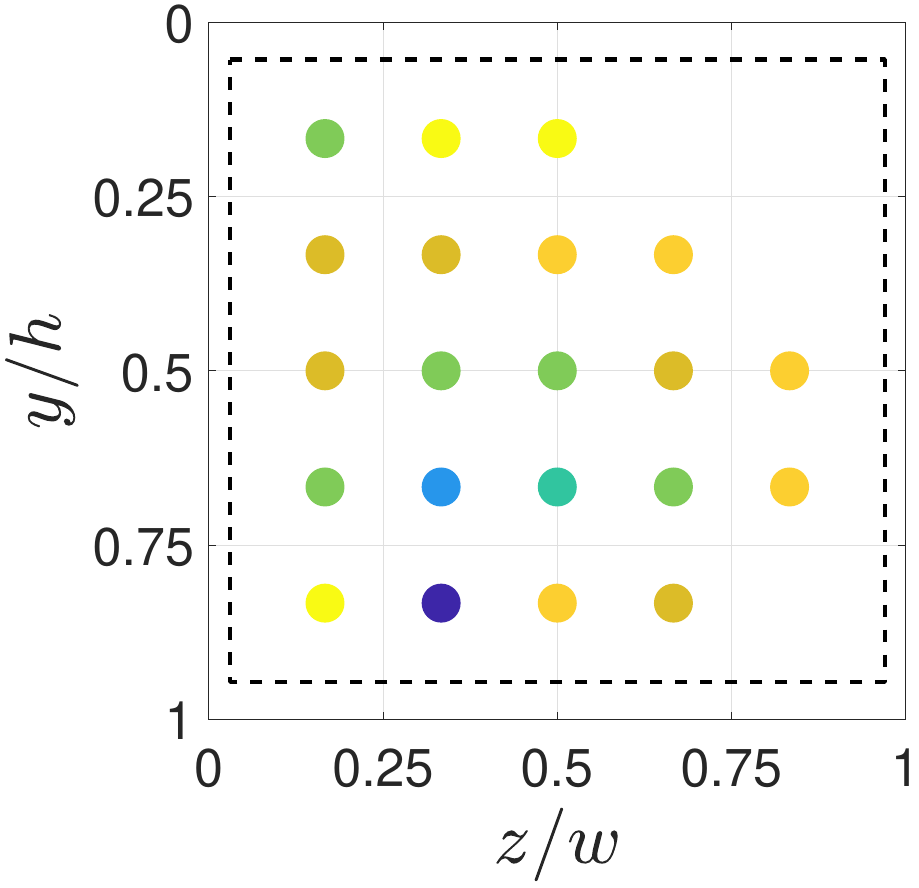}
  \caption{}
    \label{VMEAN_up}
  \end{subfigure}
  \begin{subfigure}{0.27\linewidth}
  \centering
    \includegraphics[width=\linewidth]{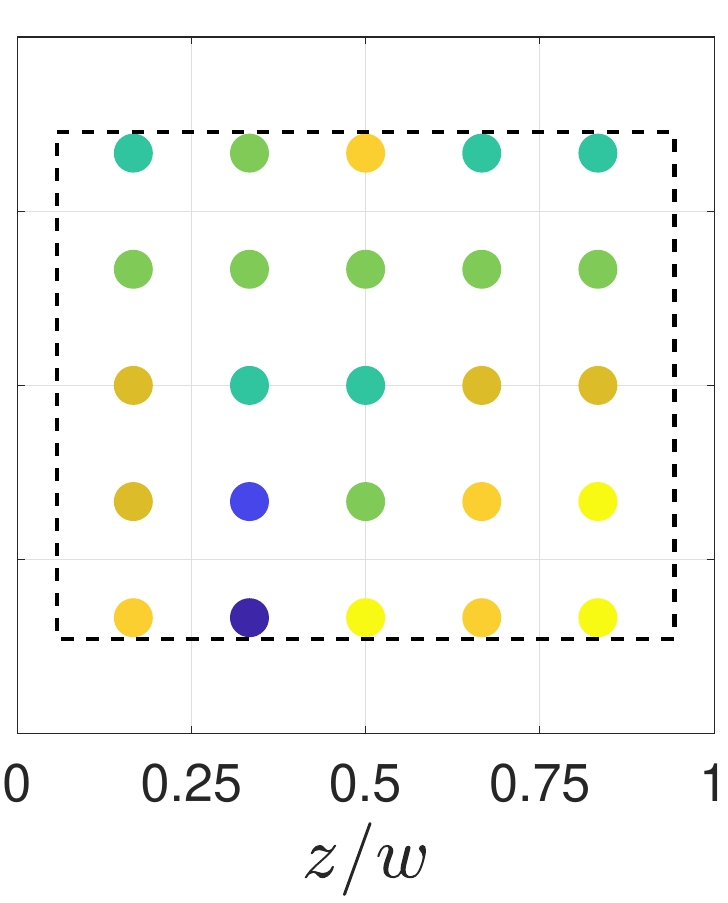}
      \caption{}
        \label{VMEAN_middle}
      \end{subfigure}
        \begin{subfigure}{0.37\linewidth}
        \centering
         \includegraphics[width=\linewidth]{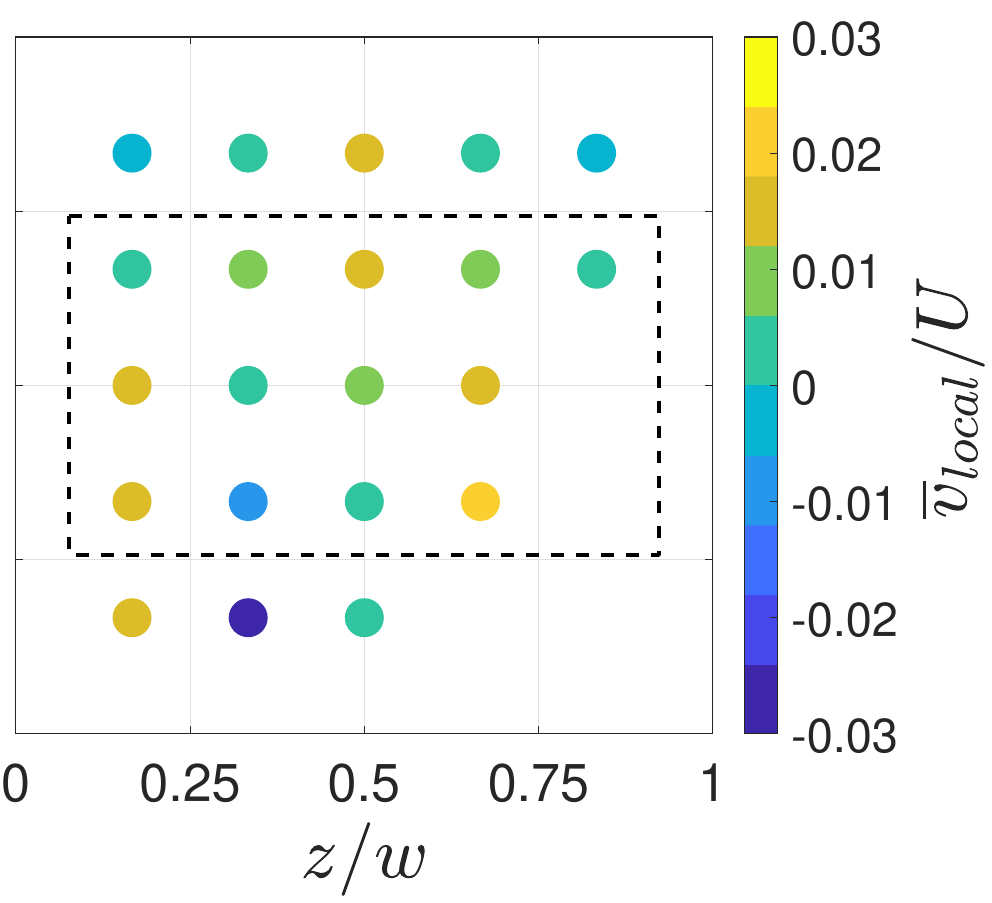}
                \caption{}
                \label{VMean_down}
      \end{subfigure}
      \end{minipage}%
      }
      \caption{Local measurements of the mean streamwise velocity $\overline{u}$ (a, b, c) and the mean vertical velocity $\overline{v}$ (d, e, f) in three streamwise locations. Local velocities are normalized by the local mean freestream velocity $U$. Measurements are shown for 285 RPM or $U_{\infty}=5$ m/s. Dashed line indicates the TBL location.}
    \label{285RPM_meanvelo}
  \end{figure}

Local velocity measurements in $y-z$ plane were performed in the three streamwise locations as described in \autoref{measurement plan}. In each measurement point, the local streamwise velocity $u_{local}$ and the vertical velocity $v_{local}$ are time-averaged and normalized with the bulk velocity $U$ of each cross section. The latter is calculated from the central $3 \times 3$ grid points in each cross section. The measurements of the streamwise and vertical velocity for $U_{\infty}=5$ m/s are presented in \autoref{285RPM_meanvelo}. The measured boundary layer limits are indicated with dashed lines  assuming a symmetrical TBL growth: identical on the top and bottom walls, and identical on the side walls. For $x=0.05L$, all measurement points lie outside the TBL, for $x=L/2$, the points closer to the top and bottom wall lie at the edge of the TBL, while for $x=0.95L$, the points closer to the top and bottom wall lie inside the TBL. In the case of measurements further away from the laser probe (larger laser paths, $z/w>0.5$), low data rates were recorded, so that the convergence of statistics was not satisfactory. These points are omitted from the results.

\begin{figure}[!ht]
\centering
\resizebox{0.75\textwidth}{!}{%
\begin{minipage}{\textwidth}
 %\textbf{$x=0.05L$ \hspace{2cm} $x=L/2$ \hspace{2cm} $x=0.95L$}\par\medskip
\noindent
\begin{tabular}{p{0.36\linewidth} p{0.22\linewidth} p{0.34\linewidth}}
\centering \textbf{$x=0.05L$} &
\centering \textbf{$x=L/2$} &
\centering \textbf{$x=0.95L$}
\end{tabular}
\par\medskip
\begin{subfigure}{0.36\linewidth}
  \centering
  \includegraphics[width=\linewidth]{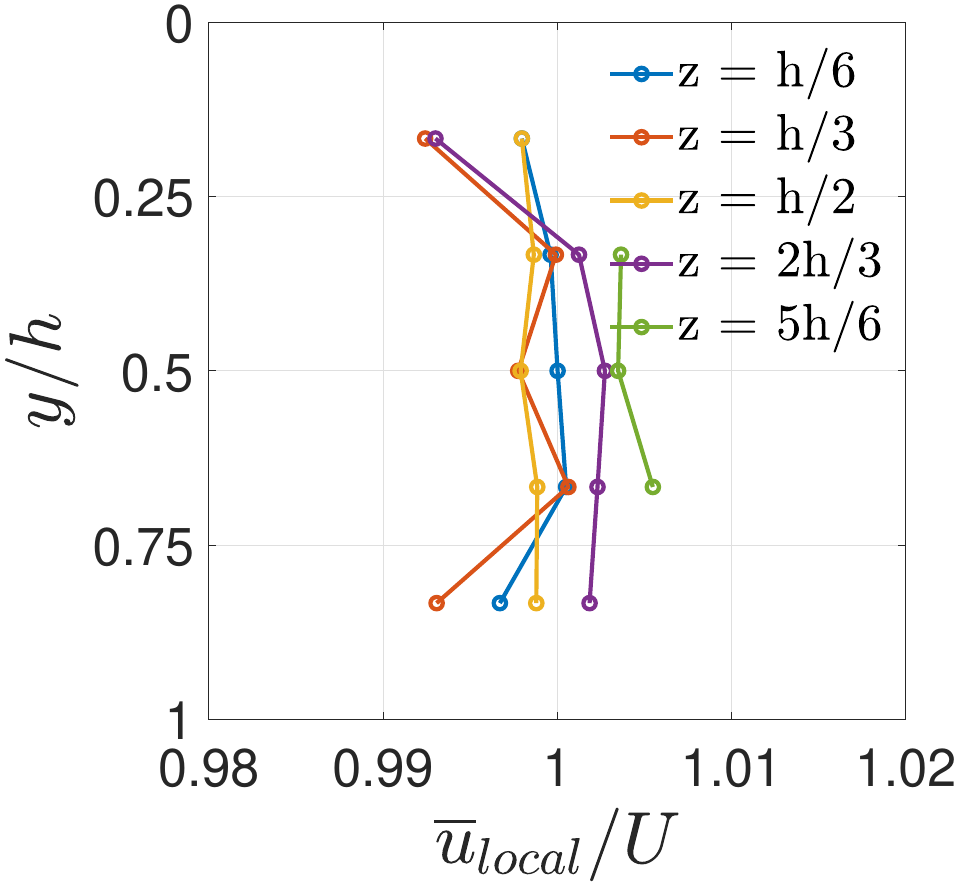}
  \caption{}
  \label{UMean_profile_up}
  \end{subfigure}
  %\hspace{0.01\textwidth}
  \begin{subfigure}{0.30\linewidth}
  \centering
    \includegraphics[width=\linewidth]{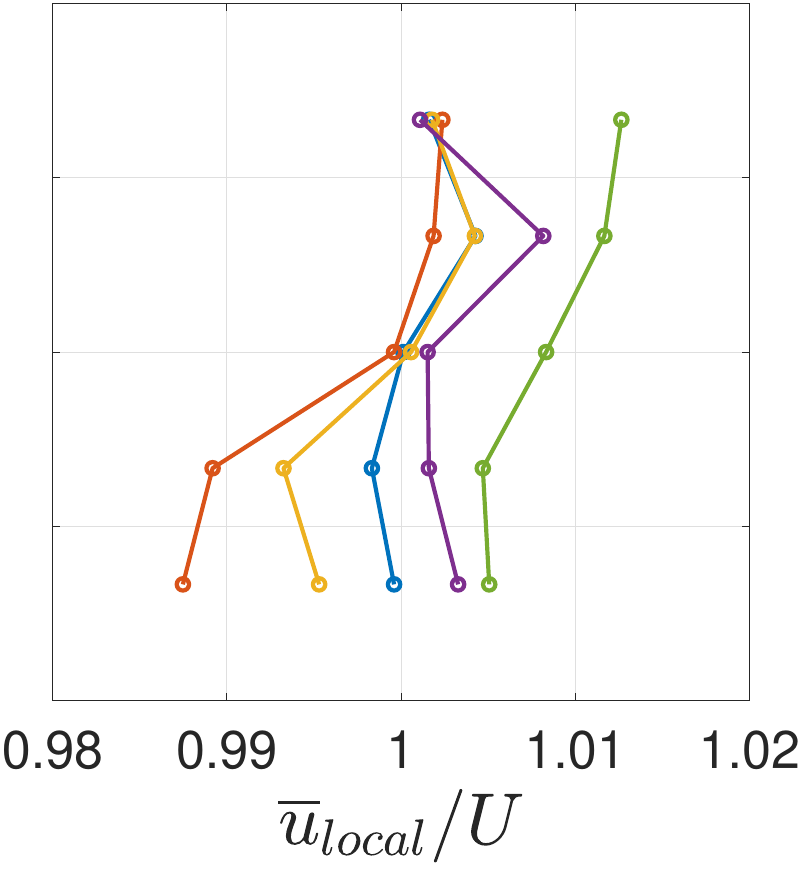}
      \caption{}
        \label{UMean_profile_middle}
      \end{subfigure}
      \begin{subfigure}{0.30\linewidth}
        \centering
         \includegraphics[width=\linewidth]{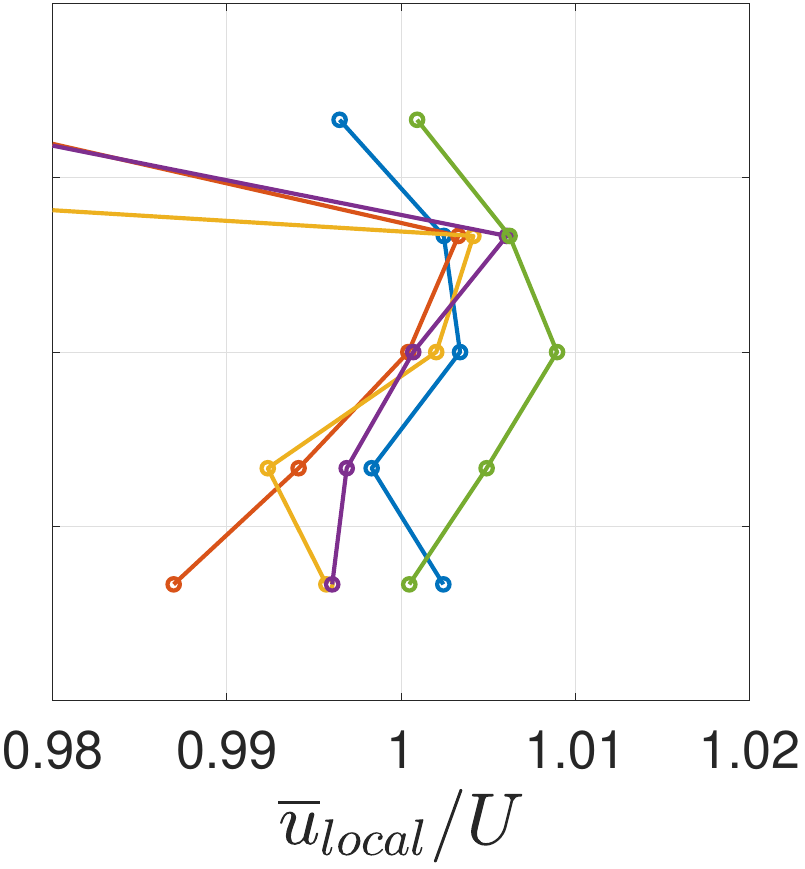}
                \caption{}
                  \label{UMean_profile_down}
      \end{subfigure}
    \vfill
\begin{subfigure}{0.36\linewidth}
  \centering
  \includegraphics[width=\linewidth]{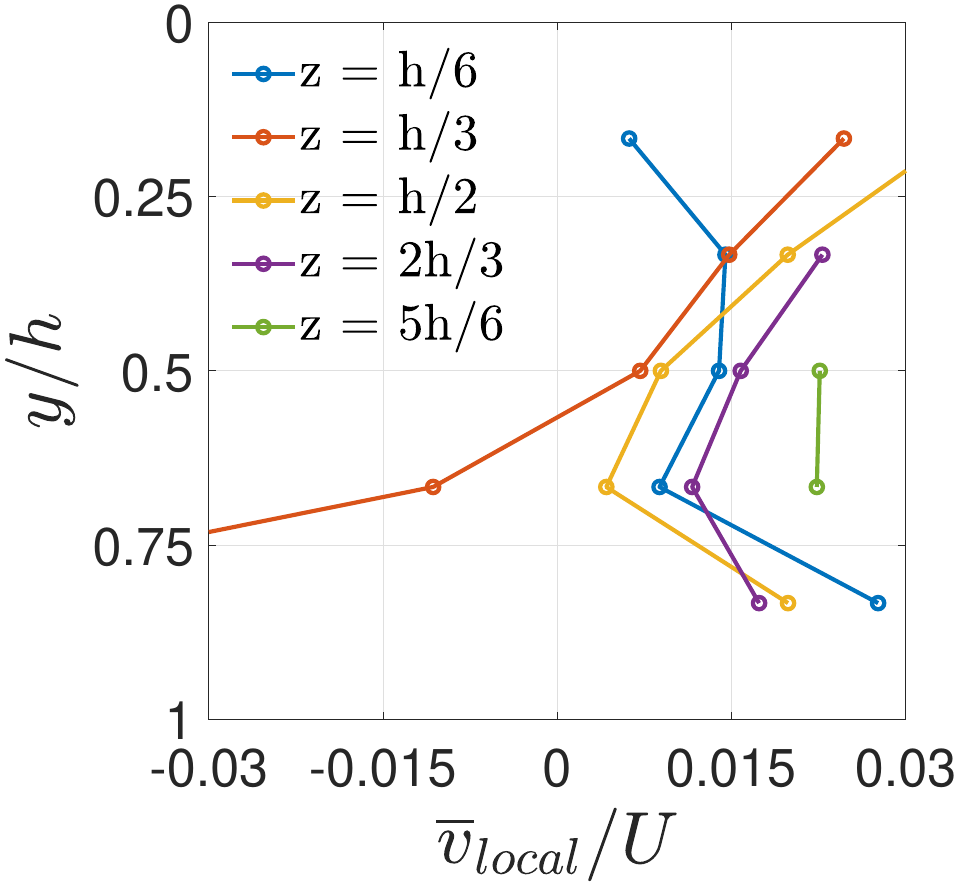}
  \caption{}
                    \label{VMean_profile_up}
  \end{subfigure}
  %\hspace{0.01\textwidth}
  \begin{subfigure}{0.30\linewidth}
  \centering
    \includegraphics[width=\linewidth]{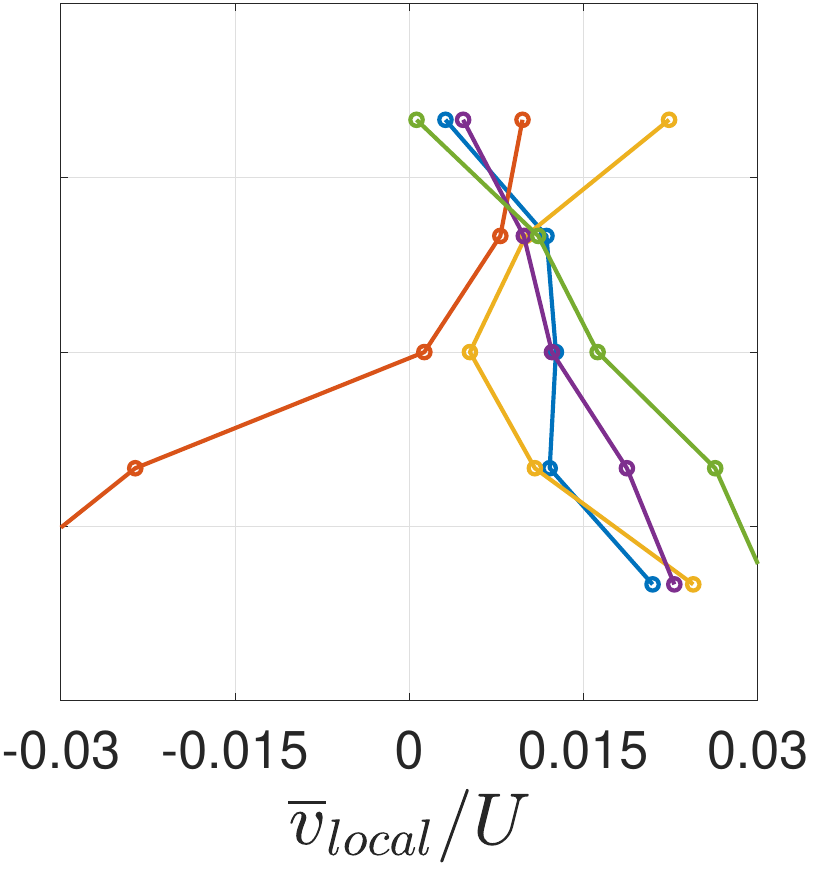}
      \caption{}
                          \label{VMean_profile_middle}
      \end{subfigure}
      \begin{subfigure}{0.30\linewidth}
        \centering
         \includegraphics[width=\linewidth]{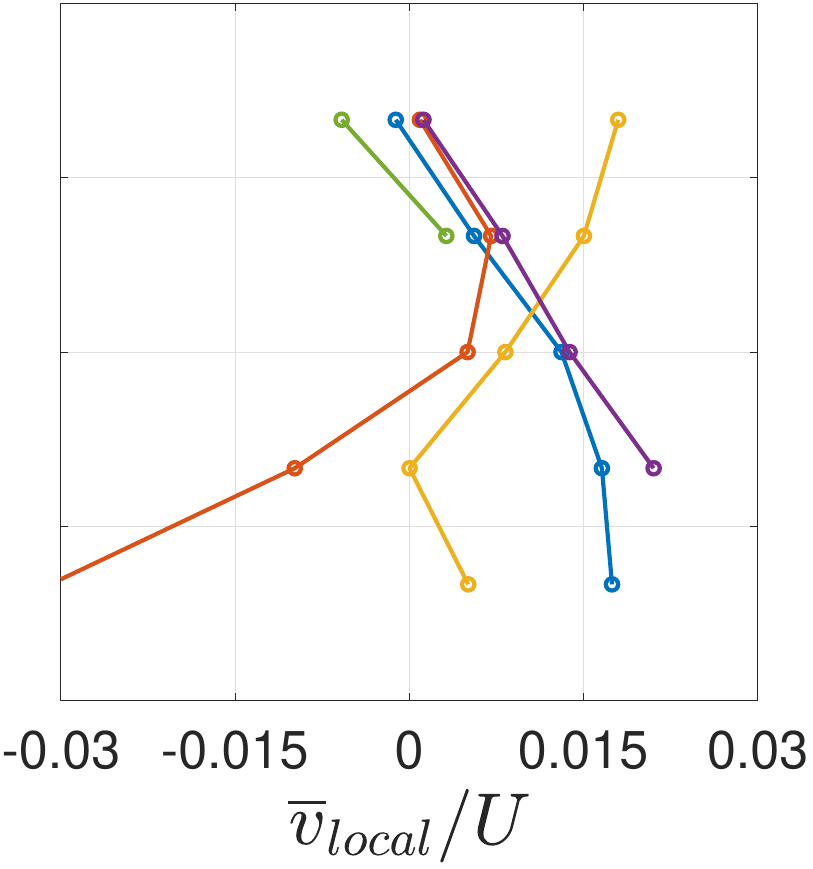}
                \caption{}
                \label{VMean_profile_down}
      \end{subfigure}
      \end{minipage}
      }
      \caption{Local mean streamwise $\overline{u}$ (a, b, c) and vertical $\overline{v}$ (d, e, f) velocity profiles in three streamwise locations for 285 RPM or $U_{\infty}=5$ m/s. See also \autoref{285RPM_meanvelo}.}
    \label{285RPM_profiles}
  \end{figure}

Overall, the local streamwise velocity $u_{local}$ shows minimal deviation from the corresponding bulk velocity $U$ across all positions, indicating good flow uniformity in the core region (\autoref{UMEAN_up}--\autoref{UMEAN_down}). Larger deviations (with $u_{local}<U$) are observed in the most downstream measurement plane ($x=0.95L$) which can be attributed to the influence of the growing TBL. The corresponding velocity profiles across the tunnel width reflect the aforementioned observations (\autoref{UMean_profile_up}--\autoref{UMean_profile_down}). No distinct flow patterns are evident.

The vertical velocity $v_{local}$ remains small throughout the measurement domain ($<1\%$ of $U_{\infty}$), typically of the order of a few cm/s (\autoref{VMEAN_up}-\autoref{VMean_down}). Higher values are observed in points located inside or close to the TBL, which can be attributed to an imperfect alignment of the laser probe, measurement noise and/or a small pressure gradient imposed from the sloping bottom of the test section. More specifically, at $z=w/3$ the two points closer to the bottom wall showcase elevated vertical velocities directed away from the bottom across all streamwise locations (\autoref{VMEAN_up}-\autoref{VMean_down}). This localized vertical motion, which was also observed in subsequent PIV measurements under similar conditions, could suggest a persistent local inflow non-uniformity (inspection of the honeycomb upstream of the test section did not reveal any defects). Regardless, measurements confined at the top half of the test section, will not be affected by this.

\begin{figure}[!h]
\centering
\resizebox{0.75\textwidth}{!}{%
\begin{minipage}{\textwidth}
 %\textbf{$x=0.05L$ \hspace{2cm} $x=L/2$ \hspace{2cm} $x=0.95L$}\par\medskip
\noindent
\begin{tabular}{p{0.36\linewidth} p{0.16\linewidth} p{0.32\linewidth}}
\centering \textbf{$x=0.05L$} &
\centering \textbf{$x=L/2$} &
\centering \textbf{$x=0.95L$}
\end{tabular}
\par\medskip
\begin{subfigure}{0.34\linewidth}
  \centering
  \includegraphics[width=\linewidth]{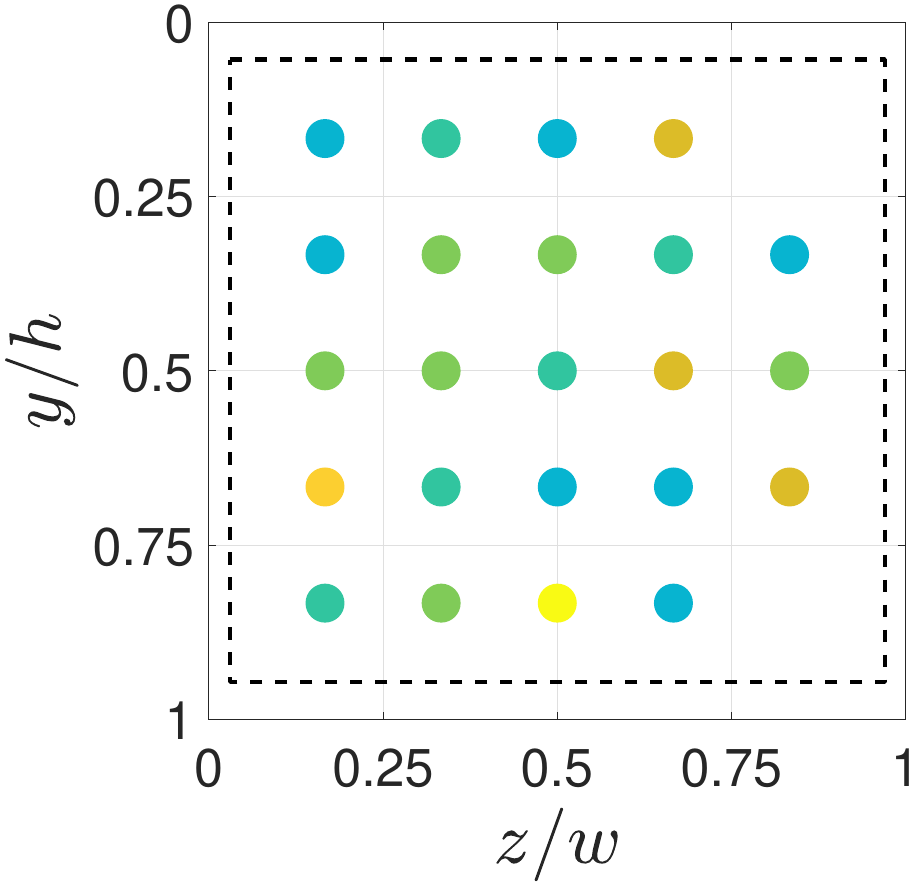}
  \caption{}
  \end{subfigure}
  \begin{subfigure}{0.27\textwidth}
  \centering
    \includegraphics[width=\linewidth]{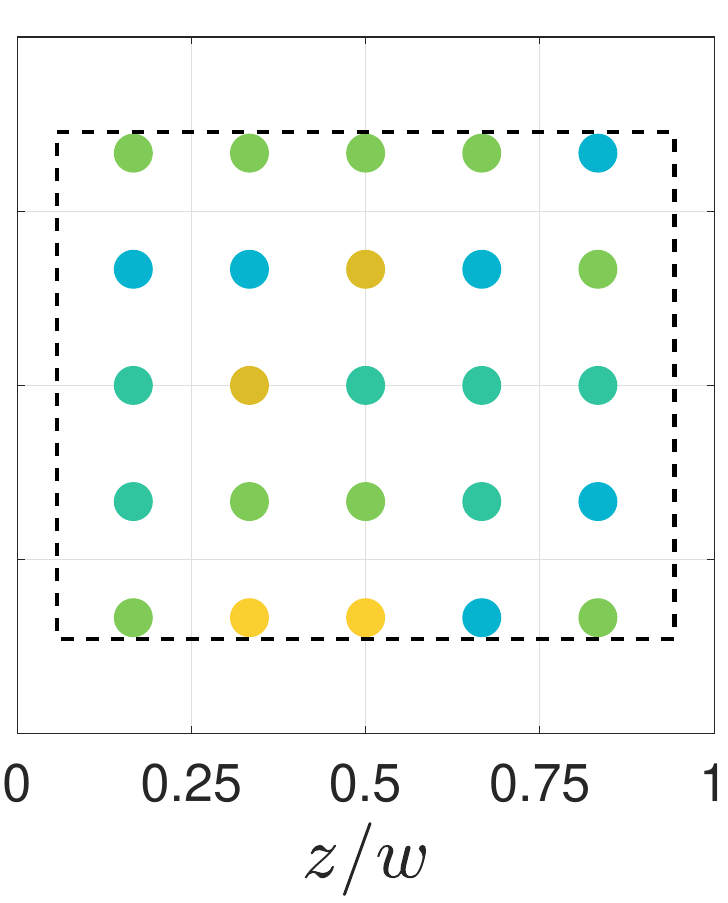}
      \caption{}
      \end{subfigure}
      \begin{subfigure}{0.36\linewidth}
        \centering
         \includegraphics[width=\textwidth]{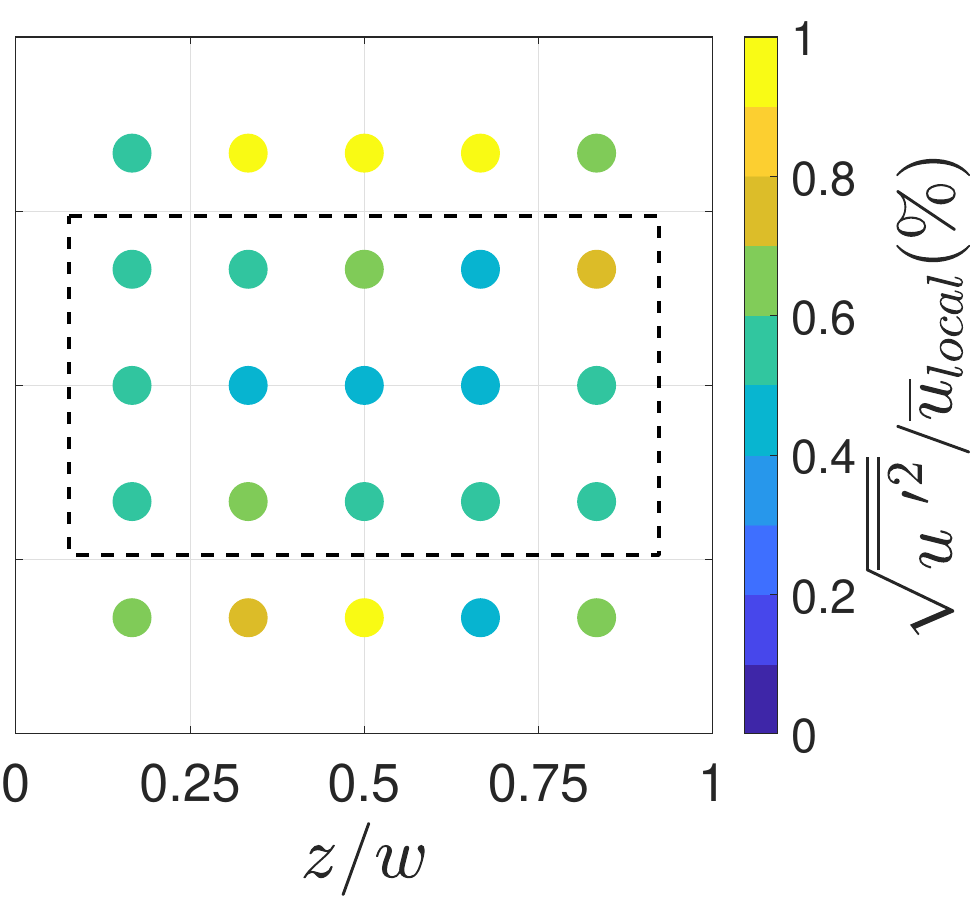}
                \caption{}
      \end{subfigure}
    \vfill
\begin{subfigure}{0.34\linewidth}
  \centering
  \includegraphics[width=\linewidth]{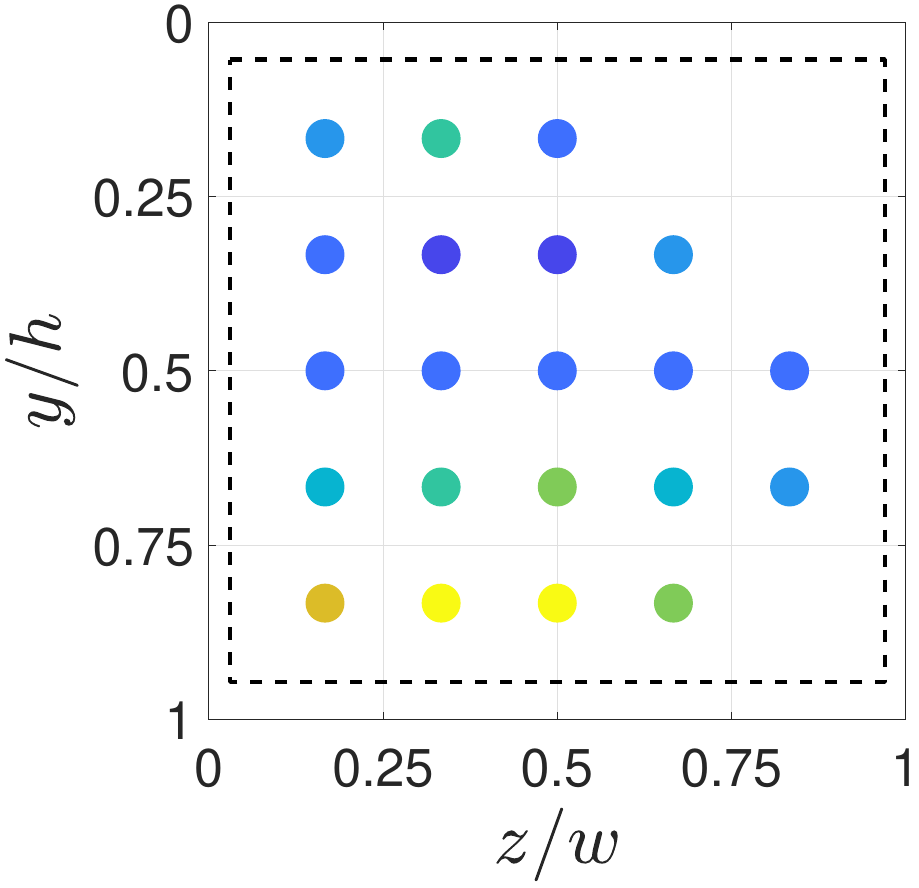}
  \caption{}
  \end{subfigure}
  \begin{subfigure}{0.27\textwidth}
  \centering
    \includegraphics[width=\linewidth]{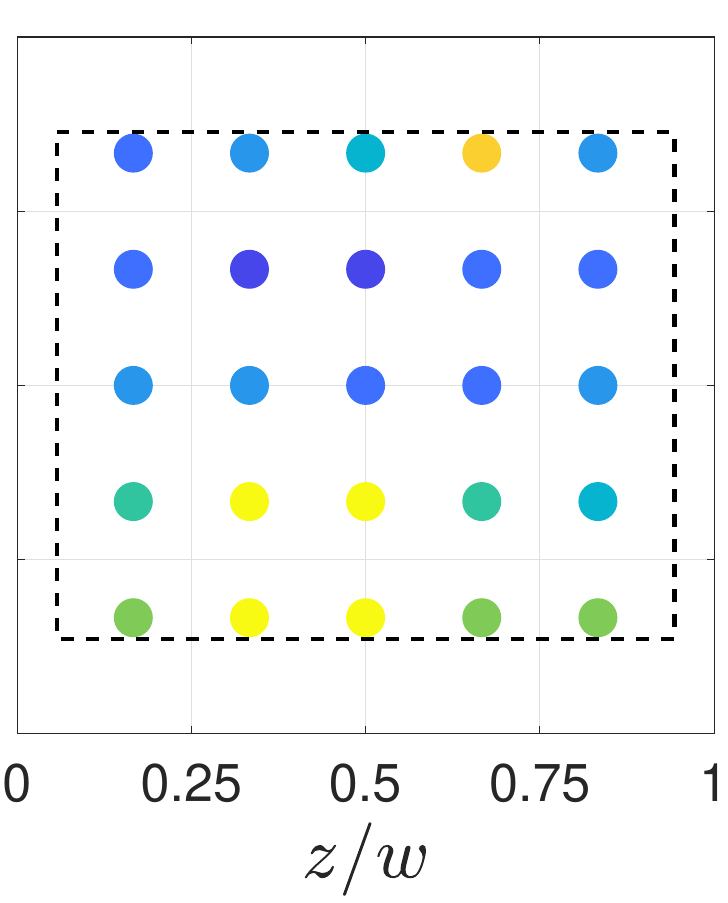}
      \caption{}
      \end{subfigure}
      \begin{subfigure}{0.36\linewidth}
        \centering
         \includegraphics[width=\textwidth]{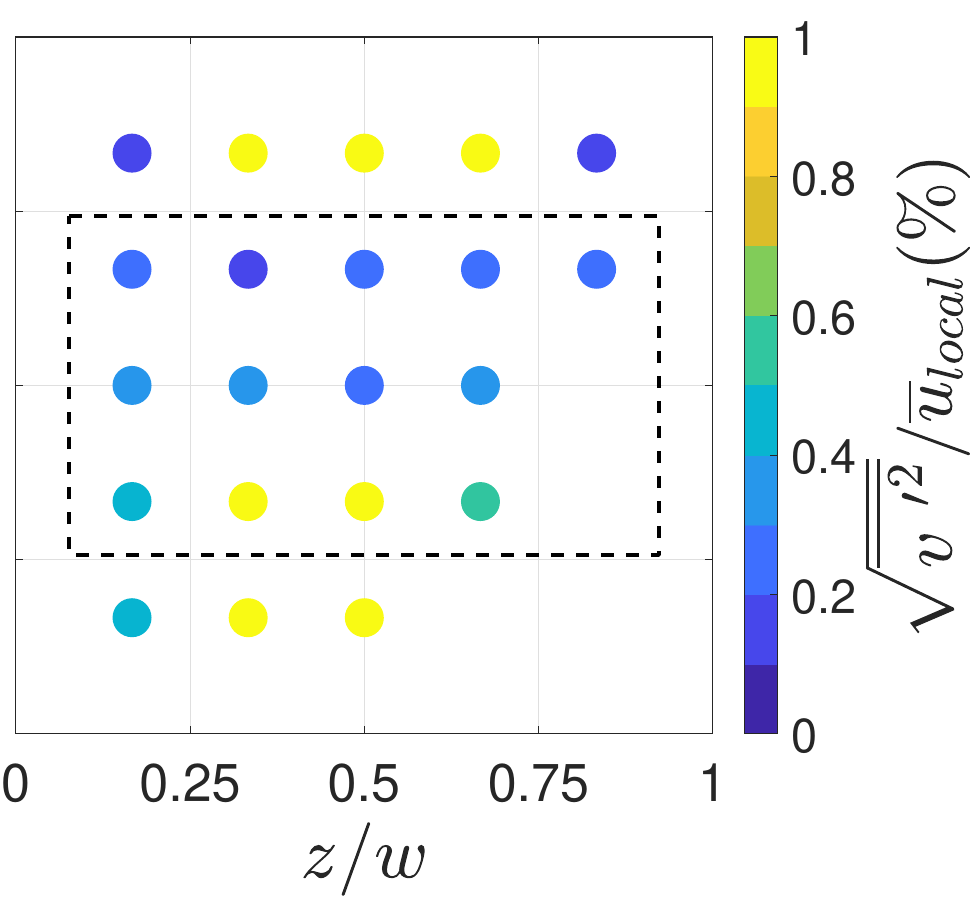}
                \caption{}
      \end{subfigure}
      \end{minipage}
      }
      \caption{Local turbulence intensity of the  streamwise (a, b, c) and vertical (d, e, f) velocity  in three streamwise locations for 285 RPM or $U_{\infty}=5$ m/s. Dashed lines indicate the TBL location.}
    \label{285RPM_TI}
  \end{figure}

Subsequently, the turbulence intensity of the streamwise and vertical velocity components is presented in \autoref{285RPM_TI}. The velocity data were analyzed using the slotting technique \citep{mayo1974digital,tummers2001spectral}. This method is capable of estimating the power spectra
of non-equidistant time series often obtained from LDA. Once the autocovariance function is estimated using this technique, the uncorrelated noise can be removed by replacing the value of autocovariace function at $\tau =0$ (where $\tau$ is the time lag) by a value extrapolated from a line fitted to the autocovariance function. Following this procedure, a turbulence intensity of $0.5\%-1\%$ was found for the points outside the TBL.

  \begin{figure}[!t]
\centering
\resizebox{0.8\textwidth}{!}{%
\begin{minipage}{\textwidth}
 %\textbf{$x=0.05L$ \hspace{2cm} $x=L/2$ \hspace{2cm} $x=0.95L$}\par\medskip
\noindent
\begin{tabular}{p{0.42\linewidth} p{0.12\linewidth} p{0.40\linewidth}}
\centering \textbf{$x=0.05L$} &
\centering \textbf{$x=L/2$} &
\centering \textbf{$x=0.95L$}
\end{tabular}
\par\medskip
\begin{subfigure}{0.37\linewidth}
  \centering
  \includegraphics[width=\linewidth]{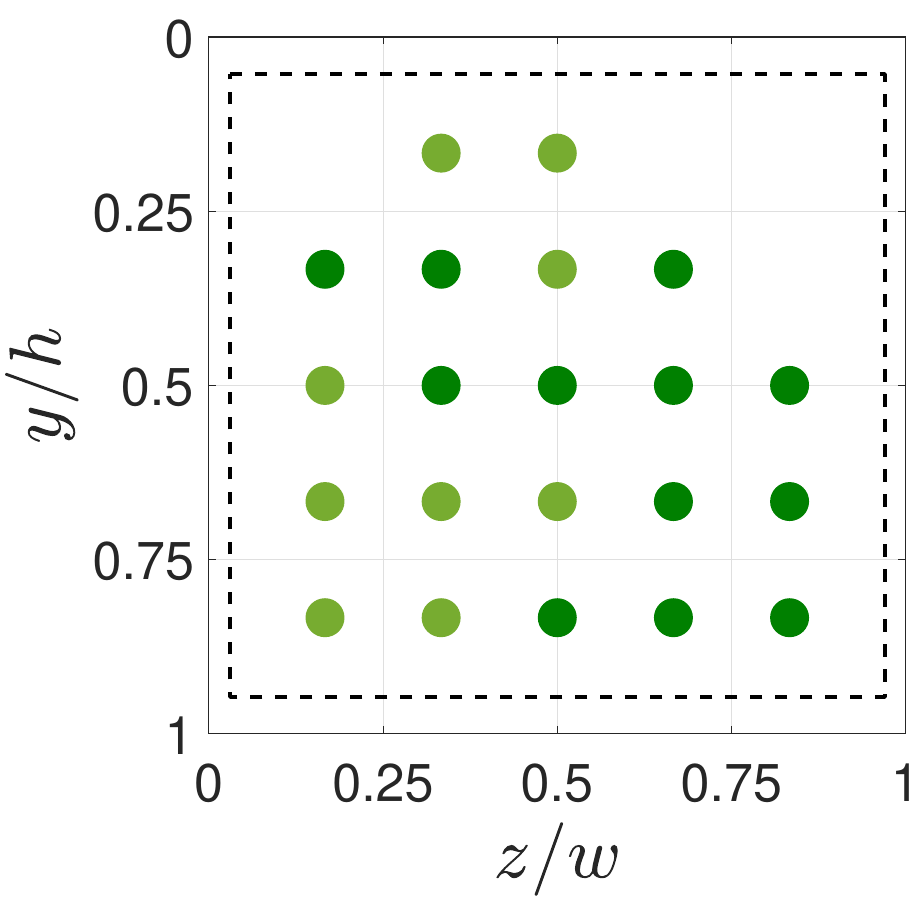}
  \caption{}
  \end{subfigure}
  \begin{subfigure}{0.29\linewidth}
  \centering
    \includegraphics[width=\linewidth]{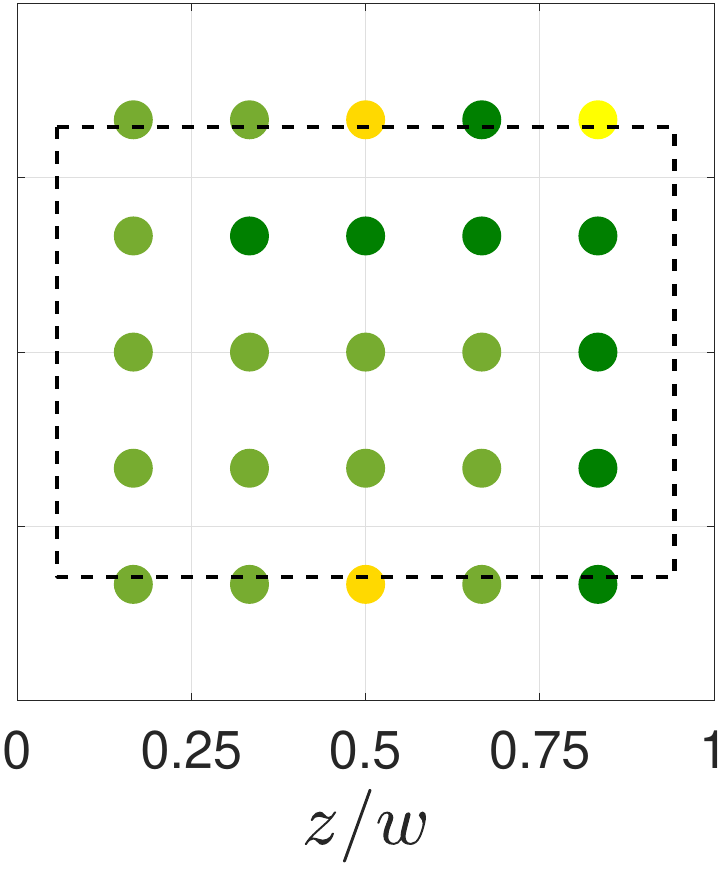}
      \caption{}
      \end{subfigure}
      \begin{subfigure}{0.29\linewidth}
        \centering
         \includegraphics[width=\linewidth]{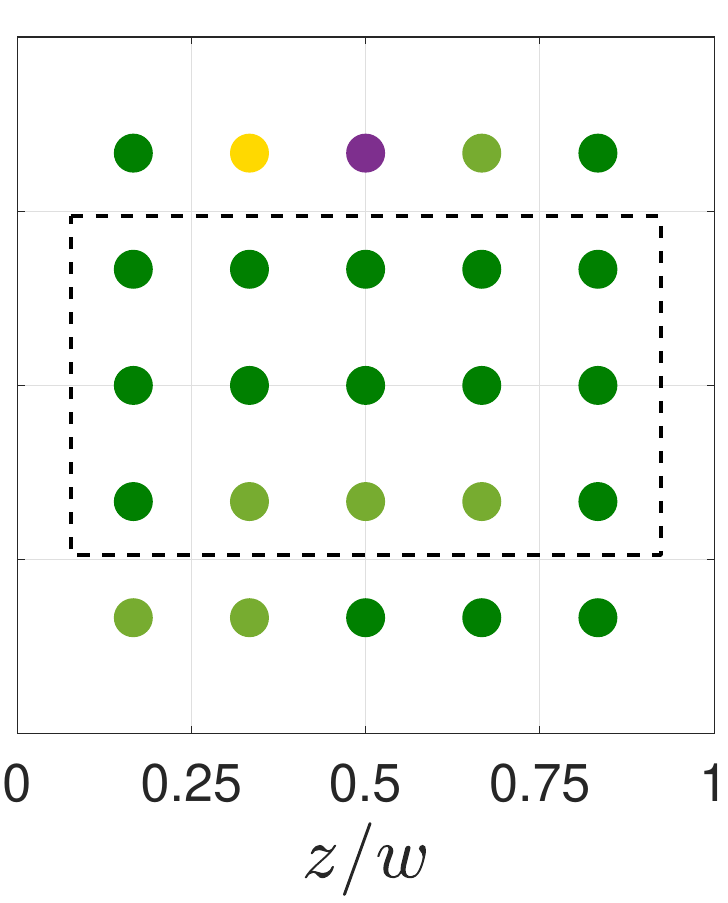}
                \caption{}
      \end{subfigure}
    \vfill
\begin{subfigure}{0.37\linewidth}
  \centering
  \includegraphics[width=\linewidth]{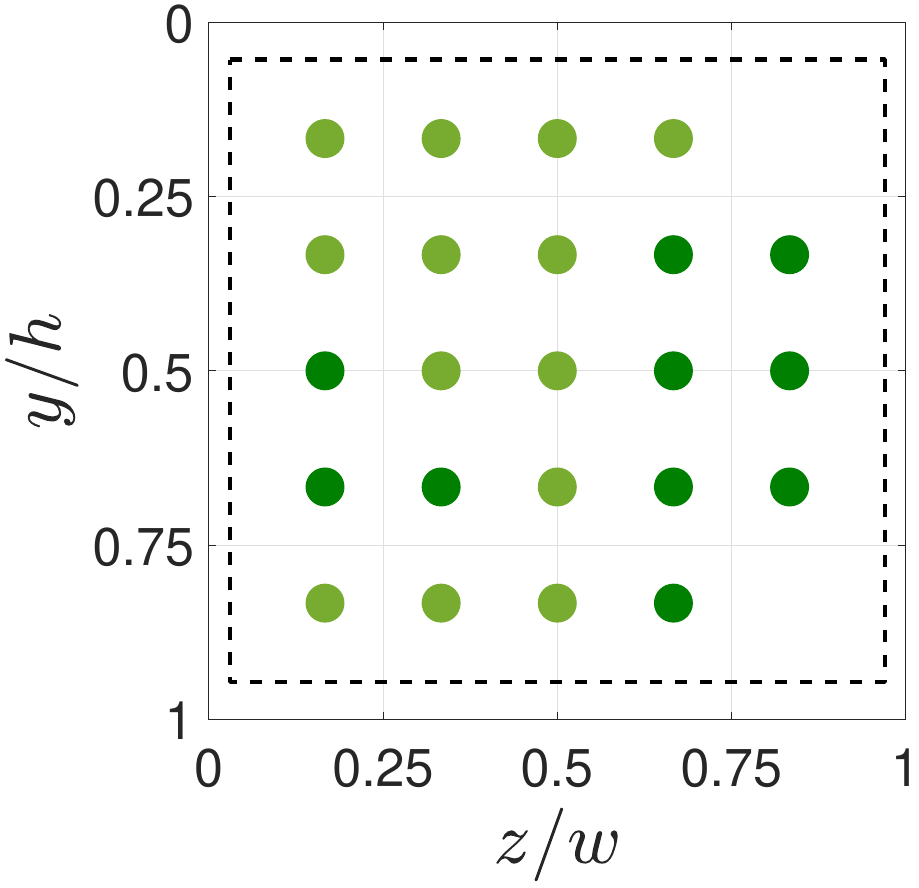}
  \caption{}
  \end{subfigure}
  \begin{subfigure}{0.29\linewidth}
  \centering
    \includegraphics[width=\linewidth]{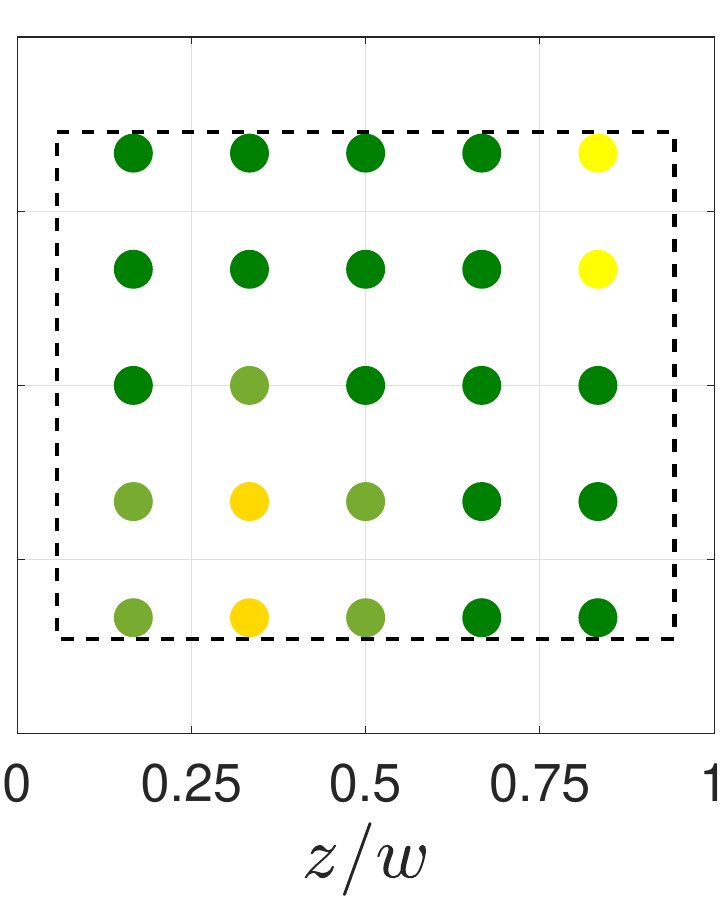}
      \caption{}
      \end{subfigure}
      \begin{subfigure}{0.29\linewidth}
        \centering
         \includegraphics[width=\linewidth]{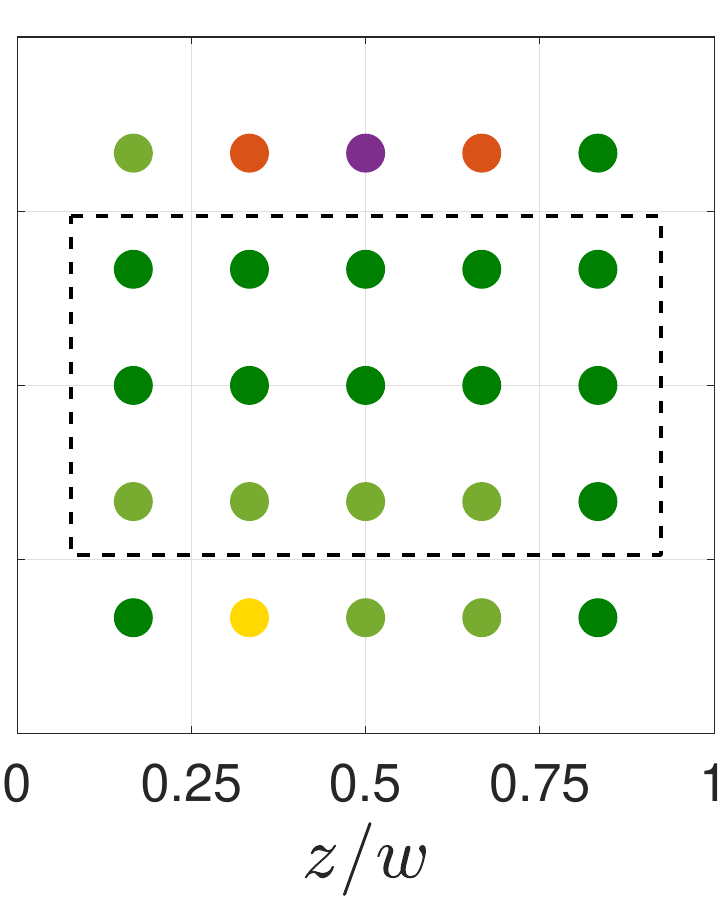}
                \caption{}
      \end{subfigure}
       \vfill
\begin{subfigure}{0.32\linewidth}
  \centering
  \includegraphics[width=\linewidth]{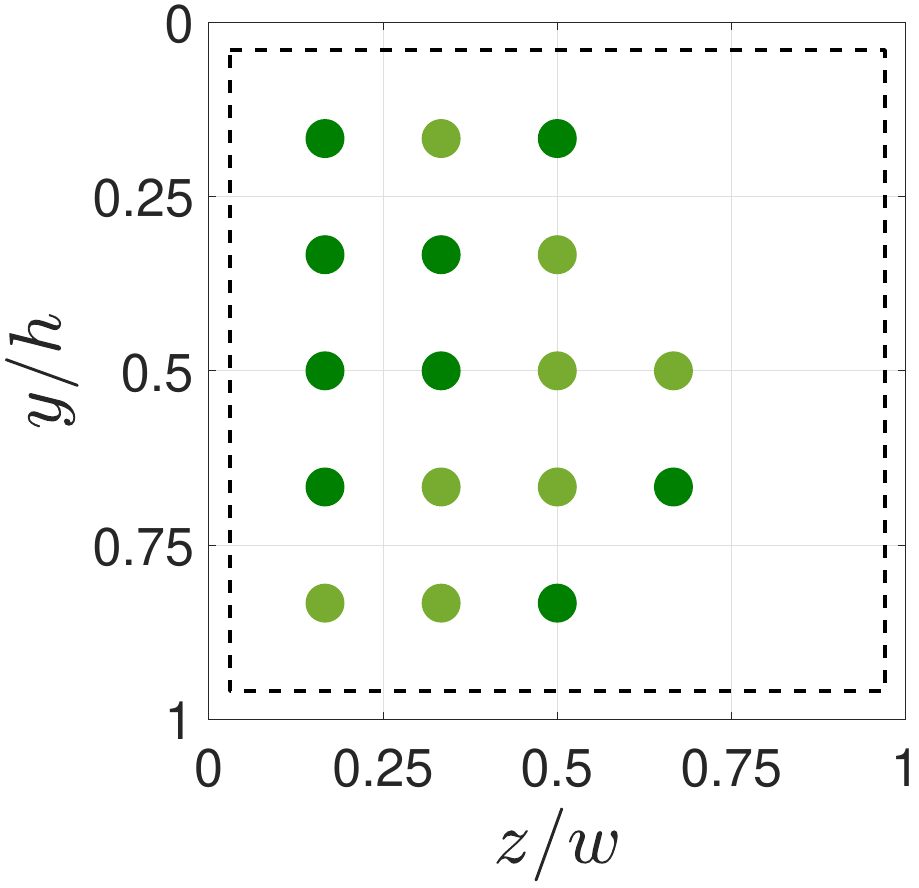}
  \caption{}
  \end{subfigure}
  \begin{subfigure}{0.25\linewidth}
  \centering
    \includegraphics[width=\linewidth]{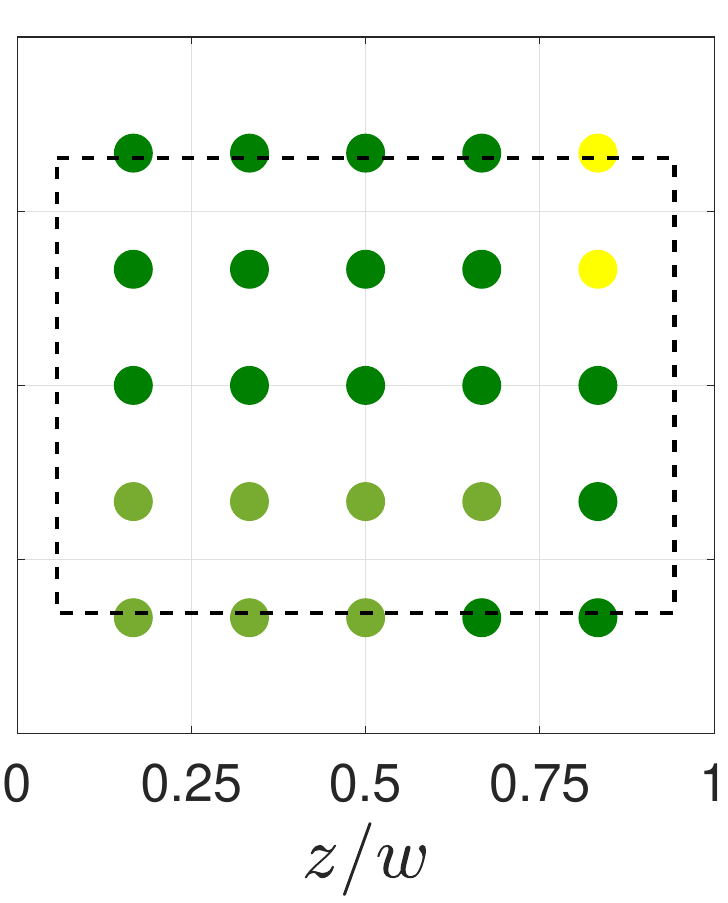}
      \caption{}
      \end{subfigure}
      \begin{subfigure}{0.36\linewidth}
        \centering
         \includegraphics[width=\linewidth]{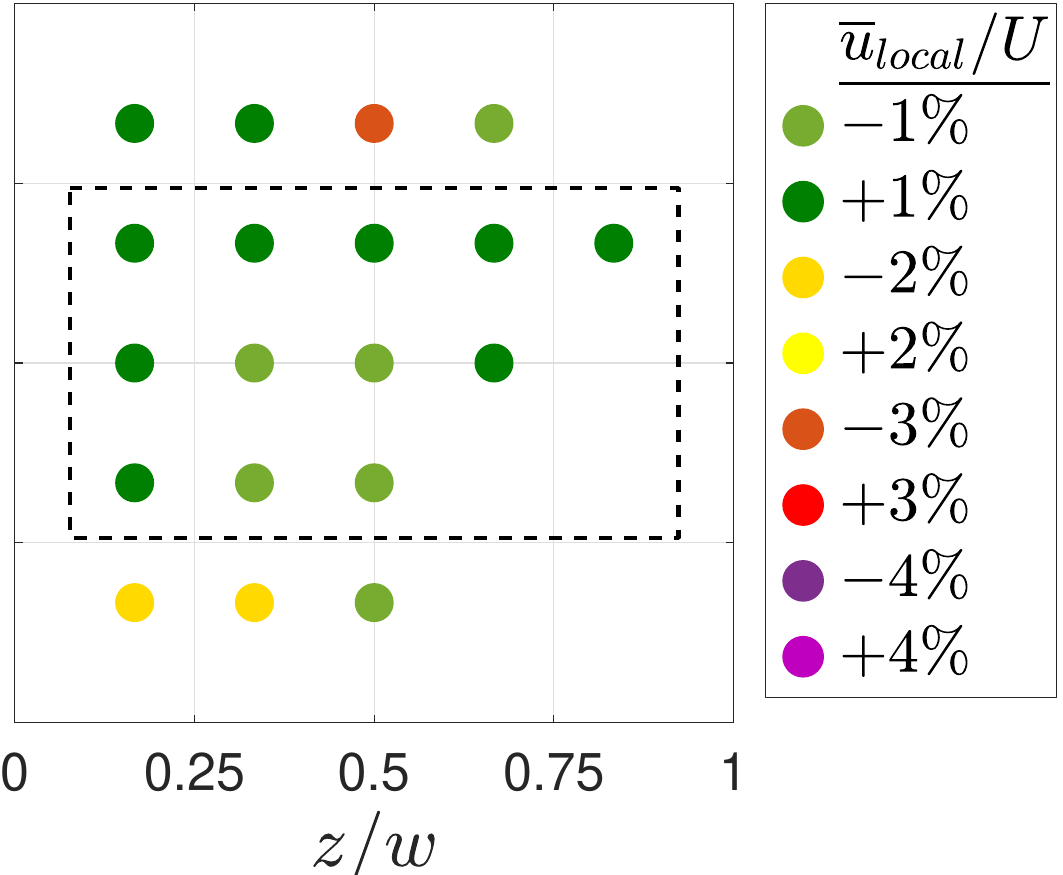}
                \caption{}
      \end{subfigure}
      \end{minipage}
      }
      \caption{Ratio  of the local streamwise velocity and freestream velocity for $U_{\infty}=3.5$ m/s or 200 RPM (a, b, c), $U_{\infty}=5$ m/s or 285 RPM (d, e, f) and $U_{\infty}=7$ m/s or 400 RPM (g, h, i) in three streamwise locations. Dashed lines indicate the TBL location. The ratio is color-coded indicating the flow uniformity.}
        \label{color_coding}
  \end{figure}

So far, results of the local measurements of the mean velocity and local turbulent intensity are presented for $U_{\infty}=5$ m/s. Similar trends as the ones discussed were found for $U_{\infty}=3.5$ m/s and $U_{\infty}=7$ m/s (\autoref{extraLDA}) as well. For reference, results of $u_{local}$ for all three $U_{\infty}$ are presented in a color-coded way in \autoref{color_coding}. It is seen that approximately for all points outside the TBL, $u_{local}$ does that differ more than $1\%$ from $U$. Outliers are limited to the most downstream position ($x=0.95L$) for points inside the TBL and therefore affected by its growth.

 \begin{figure}[!t]
\centering
\resizebox{0.9\textwidth}{!}{%
\begin{minipage}{\textwidth}
 %\textbf{$x=0.05L$ \hspace{2cm} $x=L/2$ \hspace{2cm} $x=0.95L$}\par\medskip
\noindent
\begin{tabular}{p{0.36\linewidth} p{0.23\linewidth} p{0.34\linewidth}}
\centering \textbf{$x=0.05L$} &
\centering \textbf{$x=L/2$} &
\centering \textbf{$x=0.95L$}
\end{tabular}
\par\medskip
\begin{subfigure}{0.35\linewidth}
  \centering
  \includegraphics[width=\linewidth]{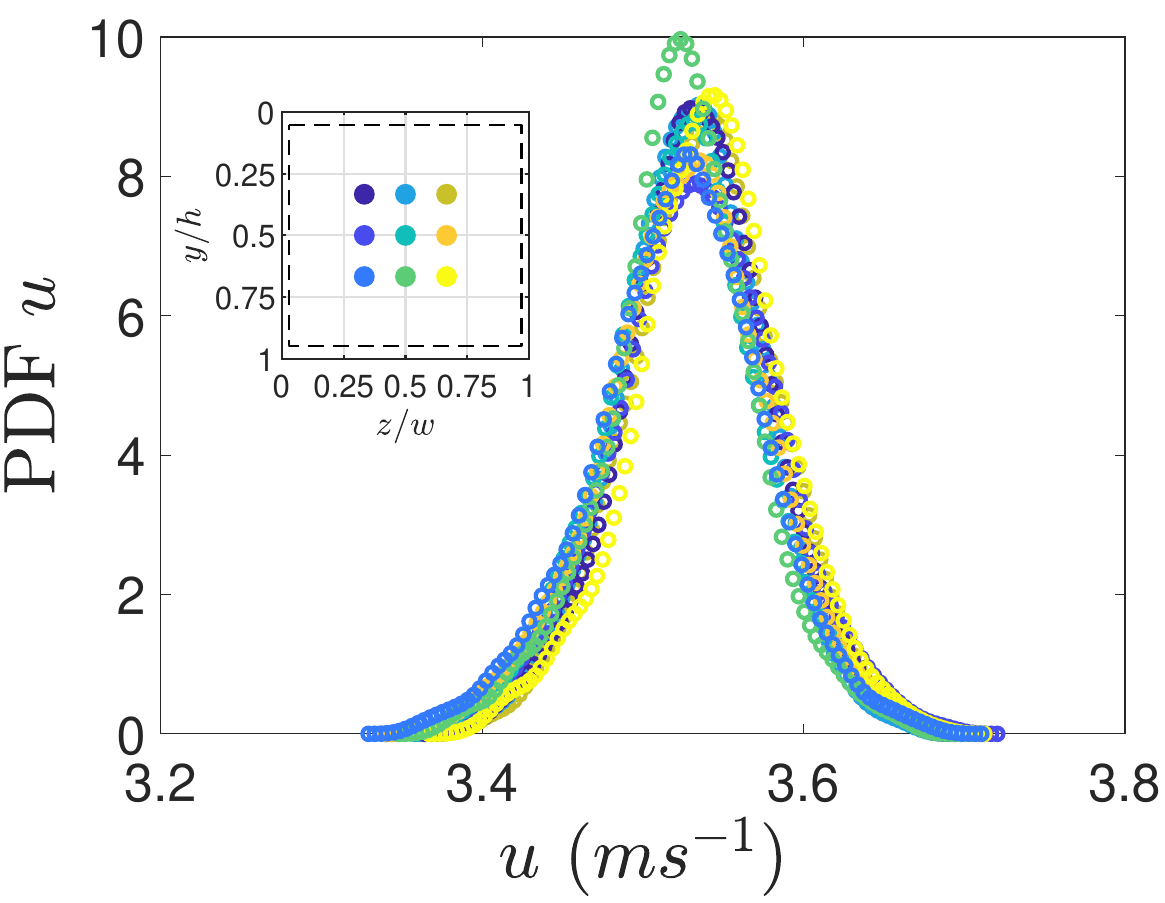}
  \caption{}
  \end{subfigure}
  \begin{subfigure}{0.31\linewidth}
  \centering
    \includegraphics[width=\linewidth]{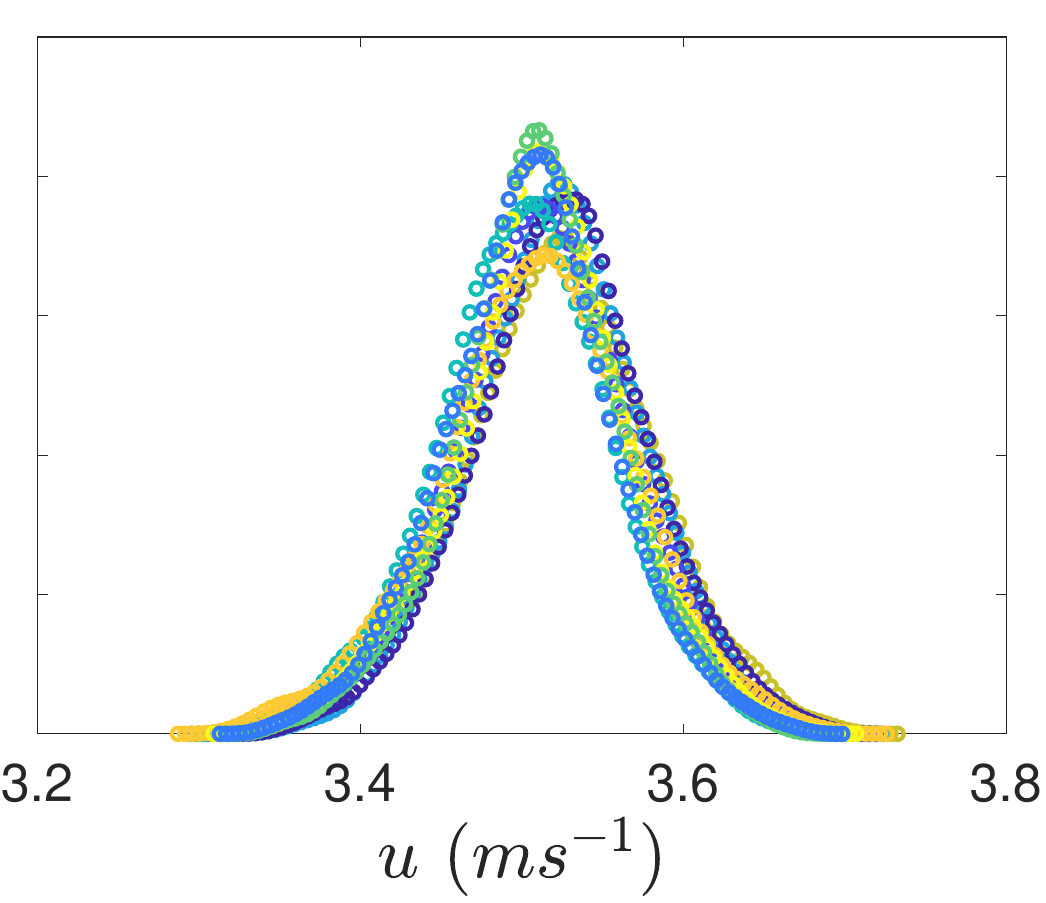}
      \caption{}
      \end{subfigure}
      \begin{subfigure}{0.31\linewidth}
        \centering  
         \includegraphics[width=\textwidth]{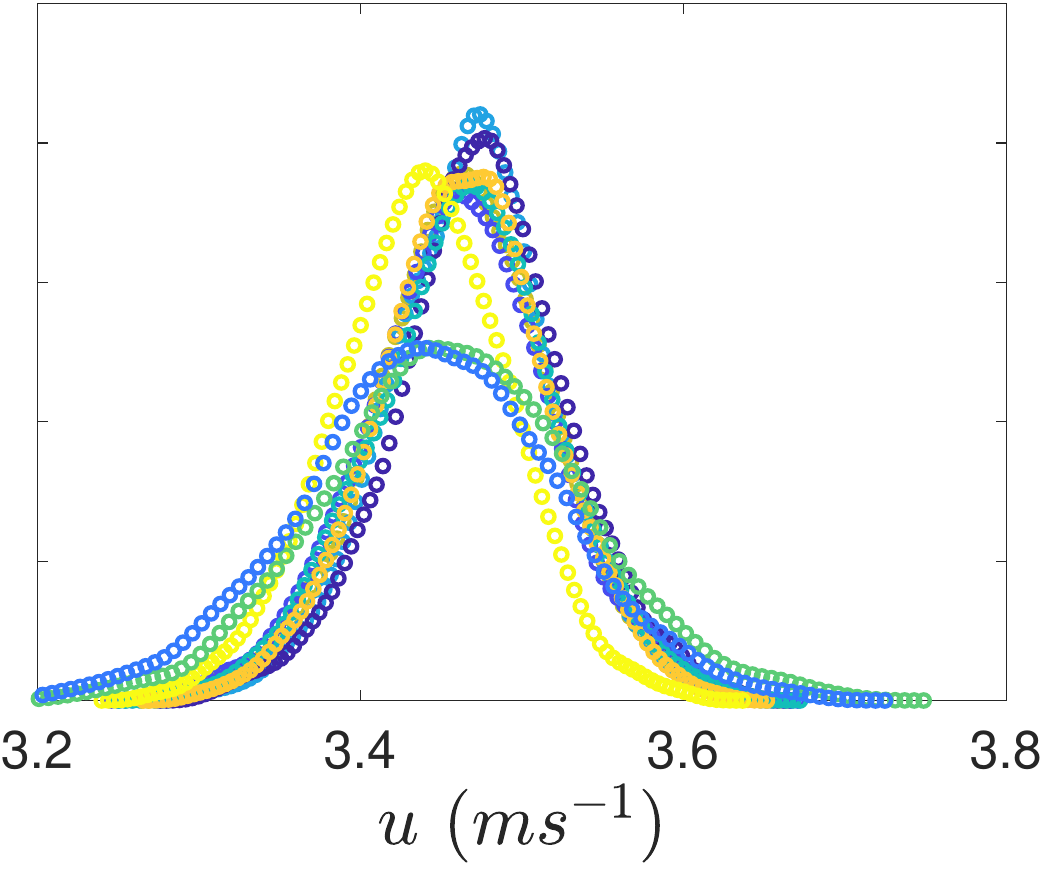}
                \caption{}
      \end{subfigure}
    \vfill
\begin{subfigure}{0.35\linewidth}
  \centering
    \includegraphics[width=\linewidth]{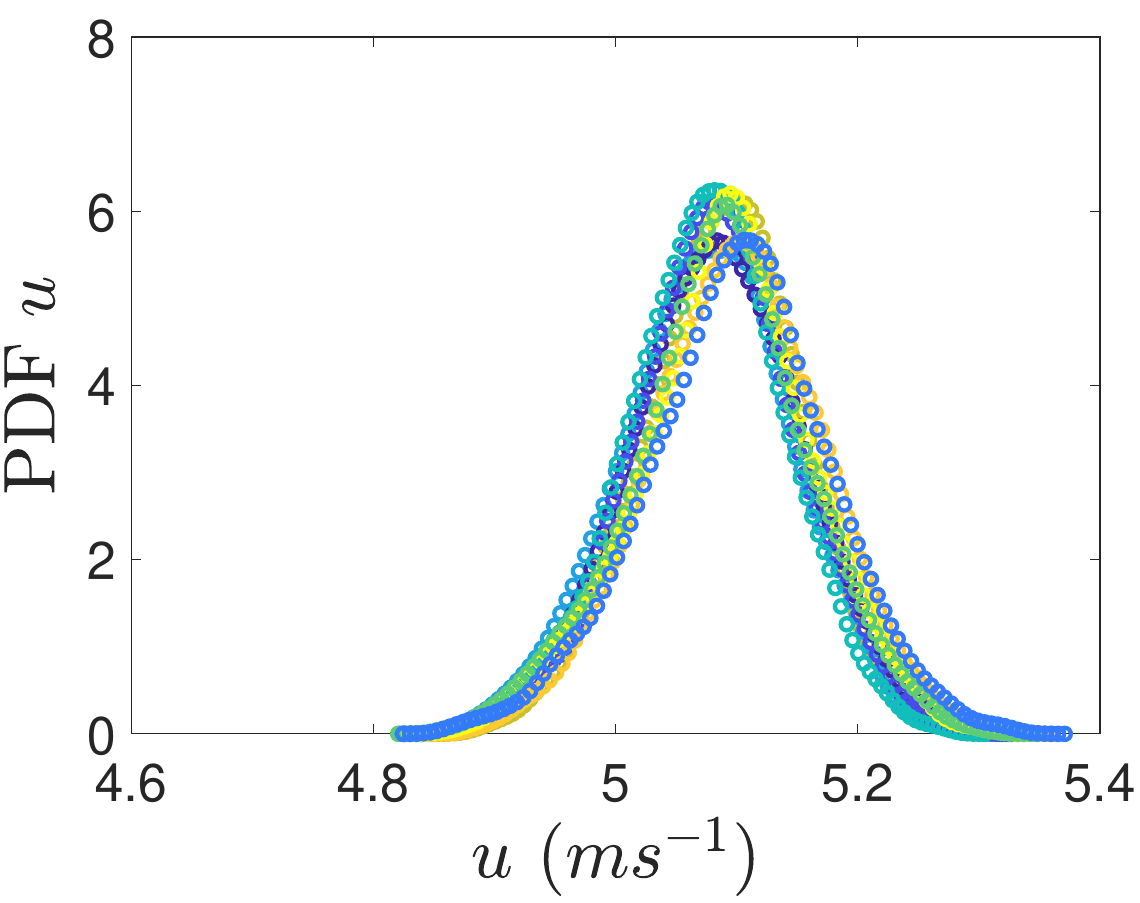}
  \caption{}
  \end{subfigure}
  \begin{subfigure}{0.31\linewidth}
  \centering
    \includegraphics[width=\linewidth]{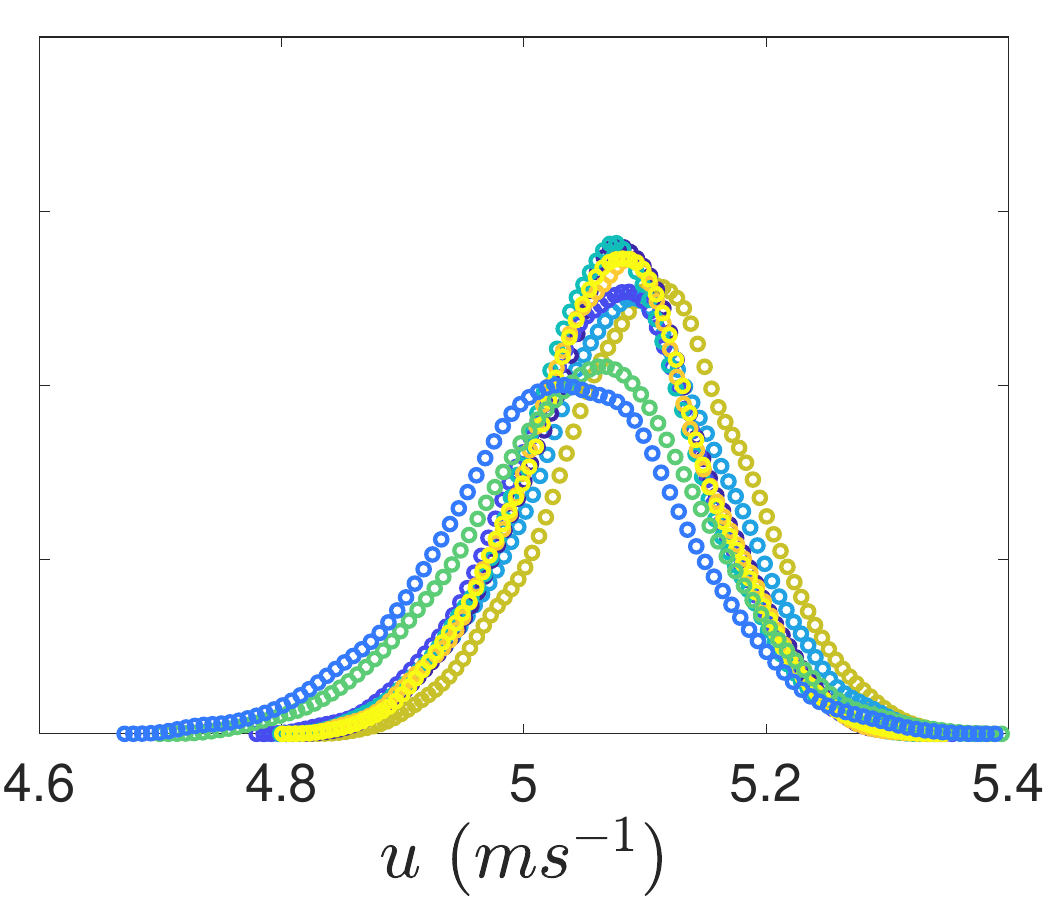}
      \caption{}
      \end{subfigure}
      \begin{subfigure}{0.31\linewidth}
        \centering
         \includegraphics[width=\linewidth]{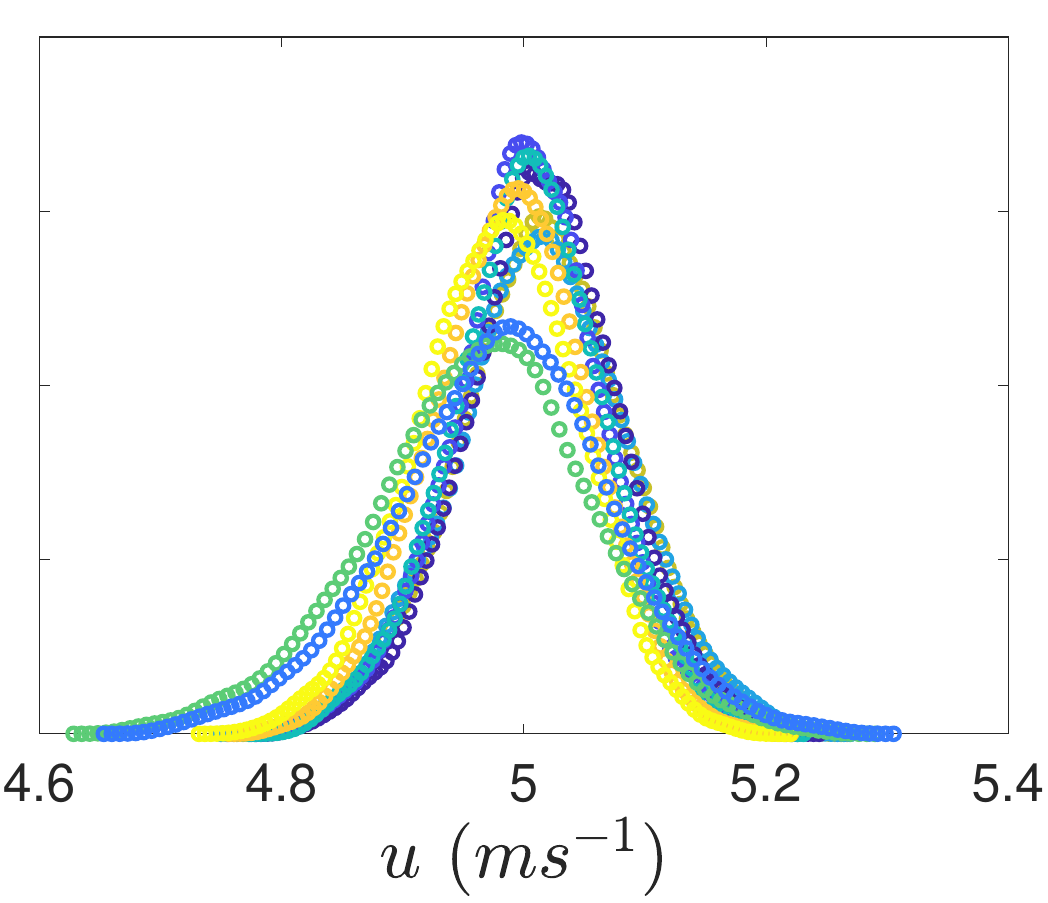}
                \caption{}
      \end{subfigure}
       \vfill
\begin{subfigure}{0.35\linewidth}
  \centering
  \includegraphics[width=\linewidth]{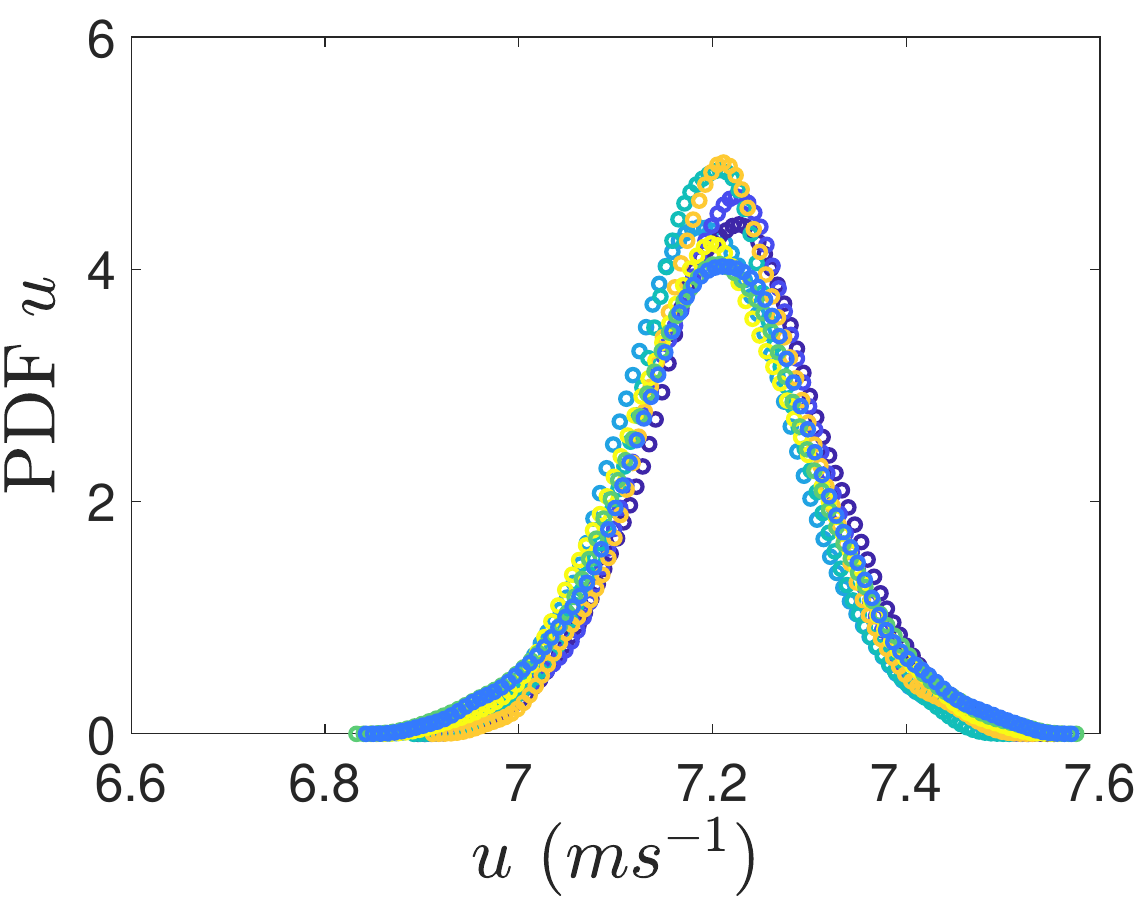}
  \caption{}
  \end{subfigure}
  \begin{subfigure}{0.31\textwidth}
  \centering
    \includegraphics[width=\linewidth]{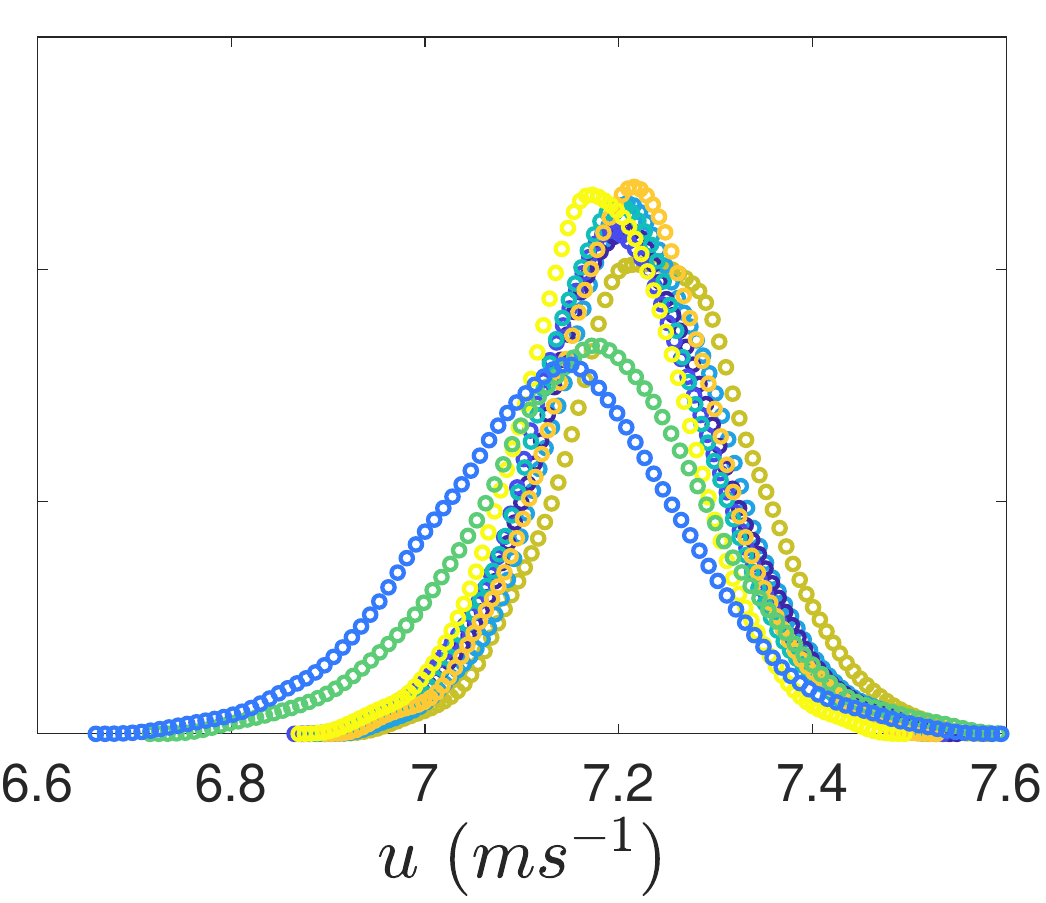}
      \caption{}
      \end{subfigure}
      \begin{subfigure}{0.31\linewidth}
        \centering
         \includegraphics[width=\linewidth]{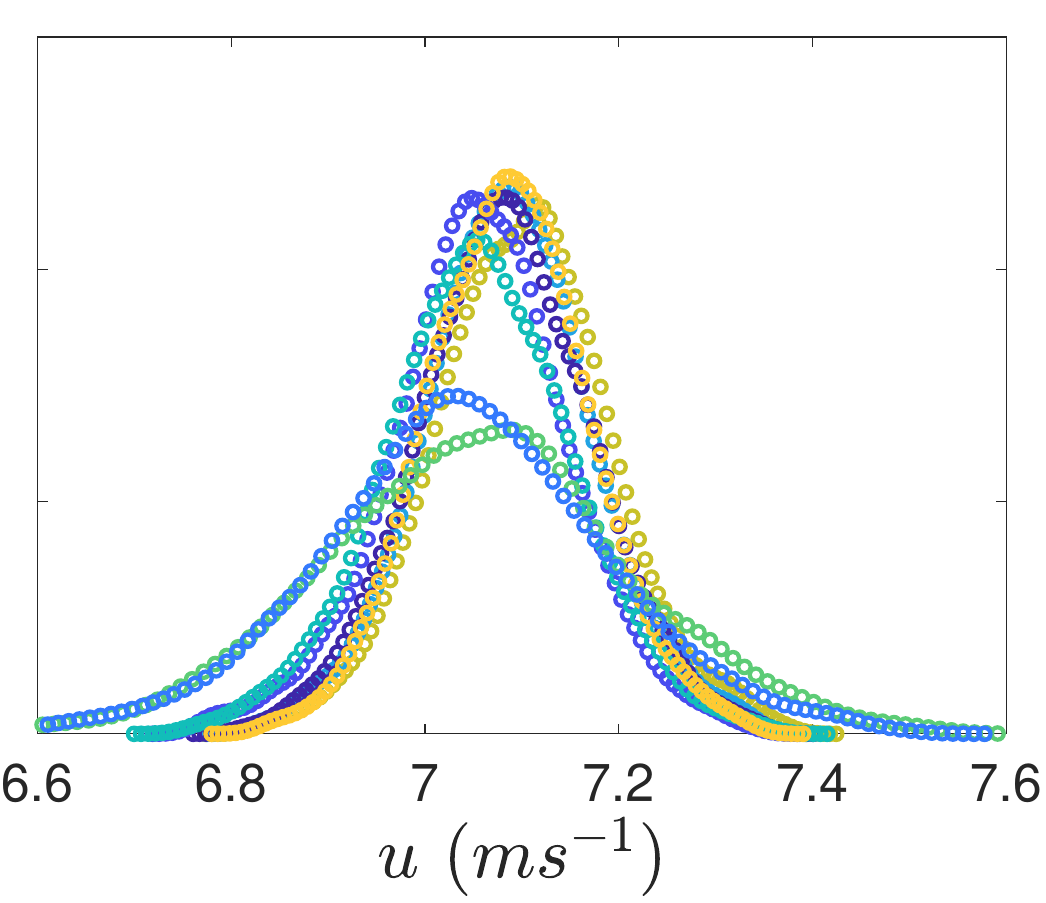}
                \caption{}
      \end{subfigure}
      \end{minipage}
      }
      \caption{PDF of the local streamwise velocity $u$ for $U_{\infty}=3.5$ m/s or 200 RPM (a, b, c), $U_{\infty}=5$ m/s or 285 RPM (d, e, f) and $U_{\infty}=7$ m/s or 400 RPM (g, h, i) in three streamwise locations.}
            \label{pdfs}
  \end{figure}

Finally, the probability density functions (PDFs) of the instantaneous streamwise velocity $u$, measured at the central $3 \times 3$ grid points, are shown in \autoref{pdfs} for all three $U_\infty$. Overall, the distributions overlap, indicating good flow uniformity. A few outliers are visible at $x = L/2$, particularly near the edges of the TBL. However, these deviations do not appear in the corresponding mean velocity or turbulence intensity (TI) profiles. This may be attributed to the fact that the TI data are noise-corrected, while the instantaneous velocities used for the PDFs are not---suggesting that the outliers could be noise-induced. In contrast, for $x = 0.95L$, where both the PDFs and the TI profiles show signs of increased fluctuations near the wall, the outliers in the PDFs are more likely due to a physical local increase in turbulence intensity.

\subsection{Long term variations}
\label{long measurements}

\begin{figure}[!t]
\begin{center}
\begin{subfigure}{0.48\linewidth}
  \centering
  \includegraphics[width=\linewidth]{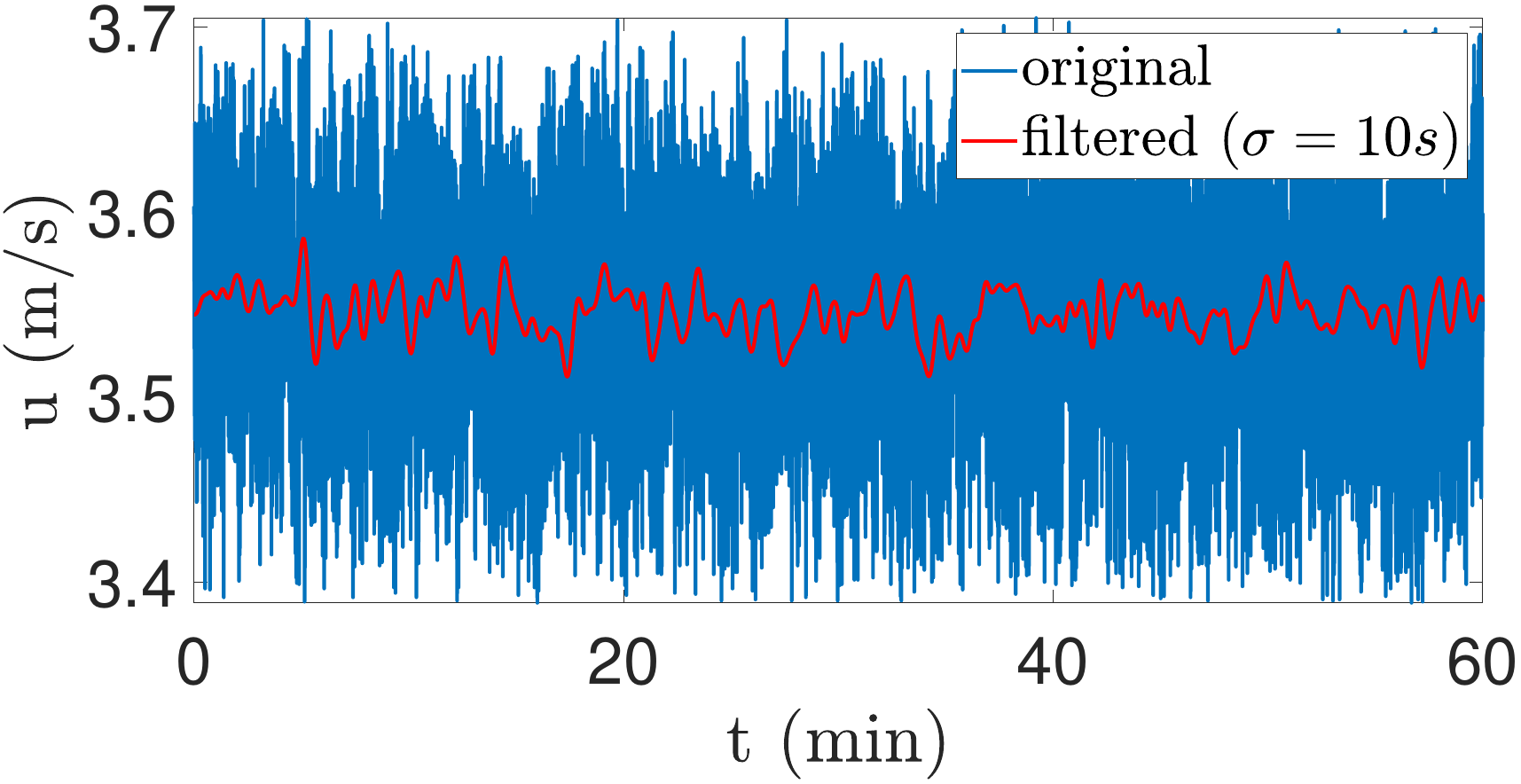}
  \caption{$U_{\infty}=3.5$ m/s}
  \label{}
  \end{subfigure}
  \begin{subfigure}{0.48\linewidth}
  \centering
  \includegraphics[width=\linewidth]{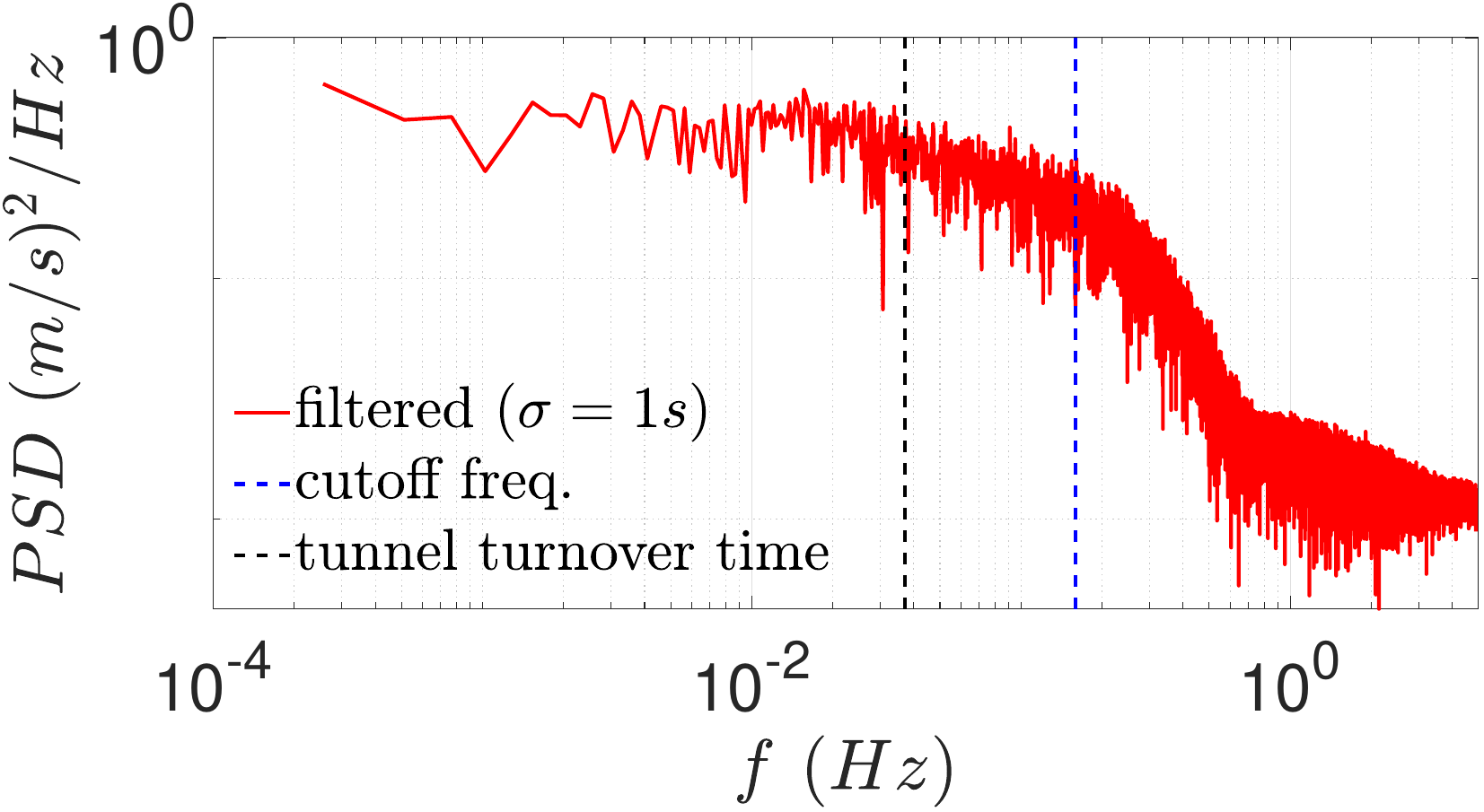}
  \caption{$U_{\infty}=3.5$ m/s}
  \label{}
  \end{subfigure}
  \vspace{0.05\textwidth}
  \begin{subfigure}{0.48\linewidth}
  \centering
  \includegraphics[width=\linewidth]{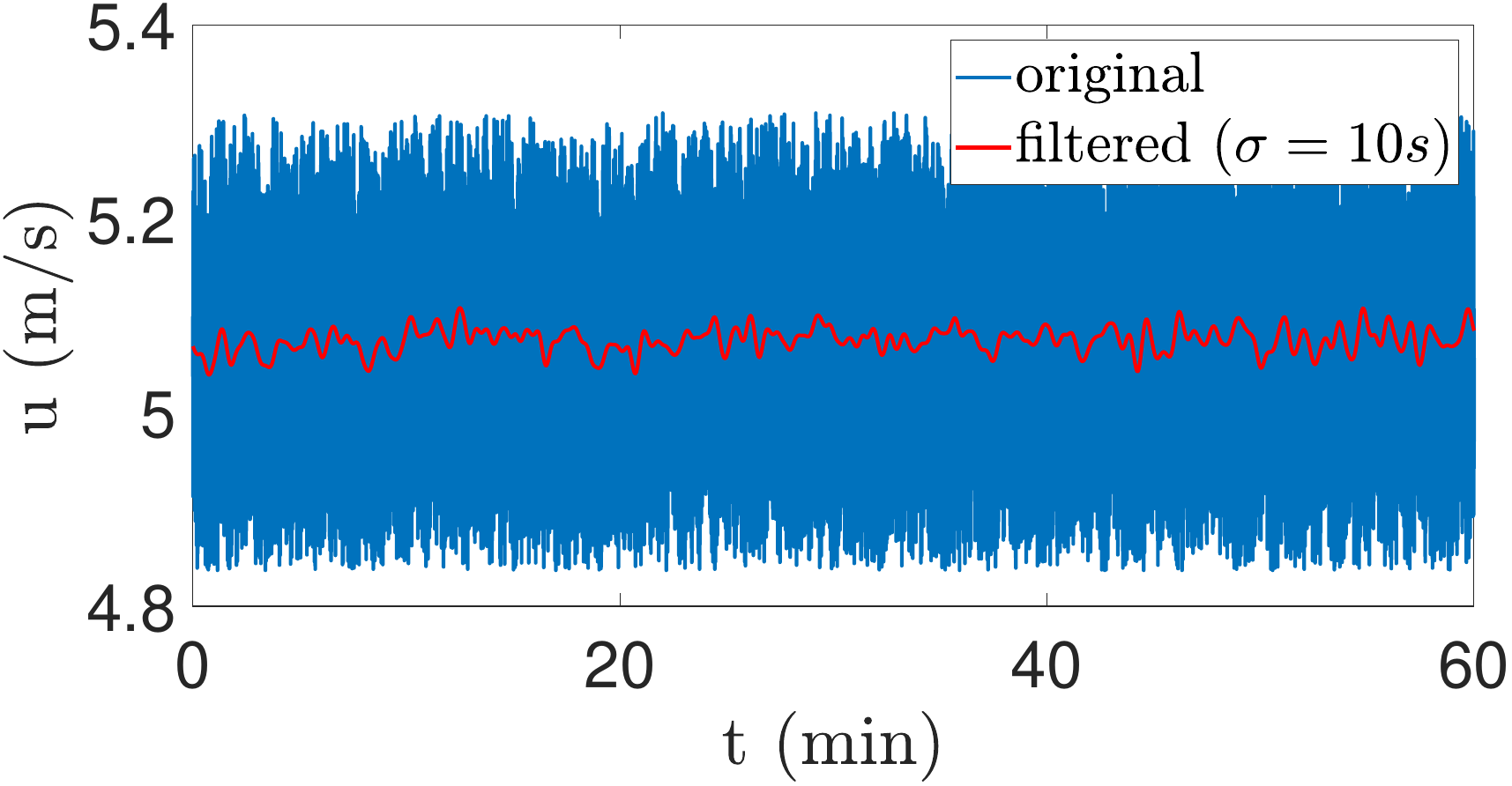}
  \caption{$U_{\infty}=5$ m/s}
  \label{}
  \end{subfigure}
  \begin{subfigure}{0.48\linewidth}
  \centering
  \includegraphics[width=\linewidth]{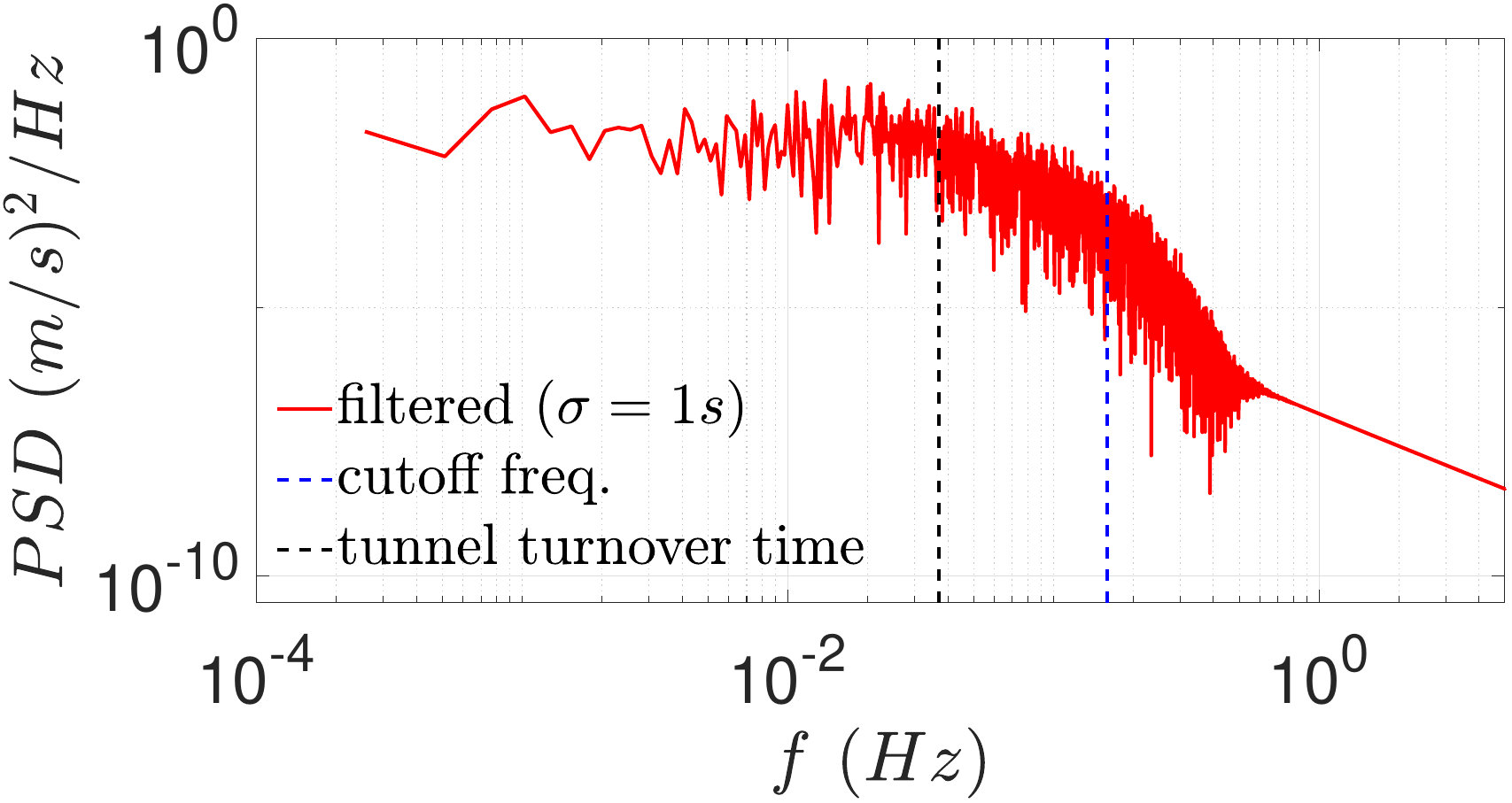}
  \caption{$U_{\infty}=5$ m/s}
  \label{}
  \end{subfigure}
    \vspace{0.05\textwidth}
  \begin{subfigure}{0.48\linewidth}
  \centering
  \includegraphics[width=\linewidth]{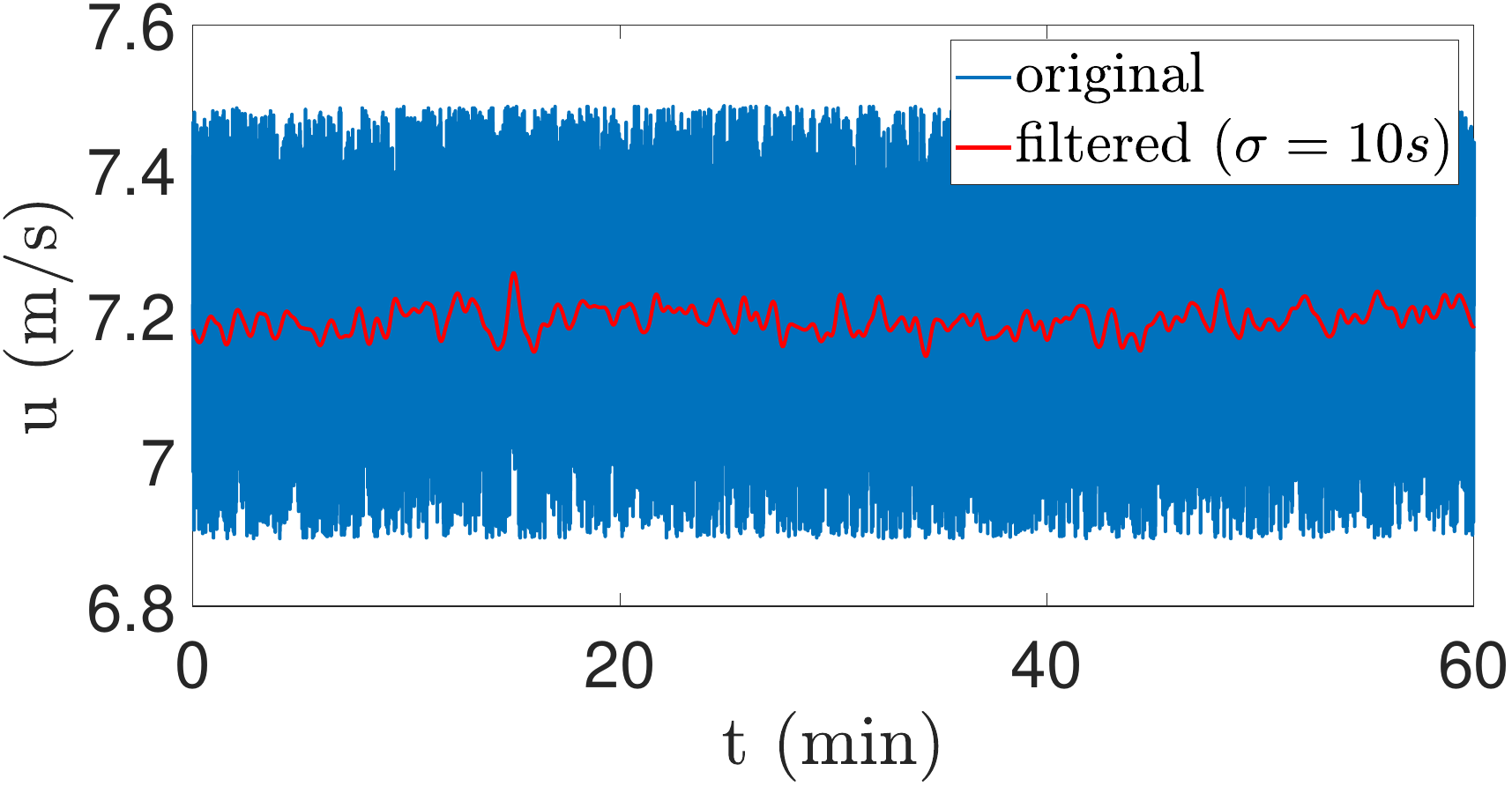}
  \caption{$U_{\infty}=7$ m/s}
  \label{}
  \end{subfigure}
  \begin{subfigure}{0.48\linewidth}
  \centering
  \includegraphics[width=\linewidth]{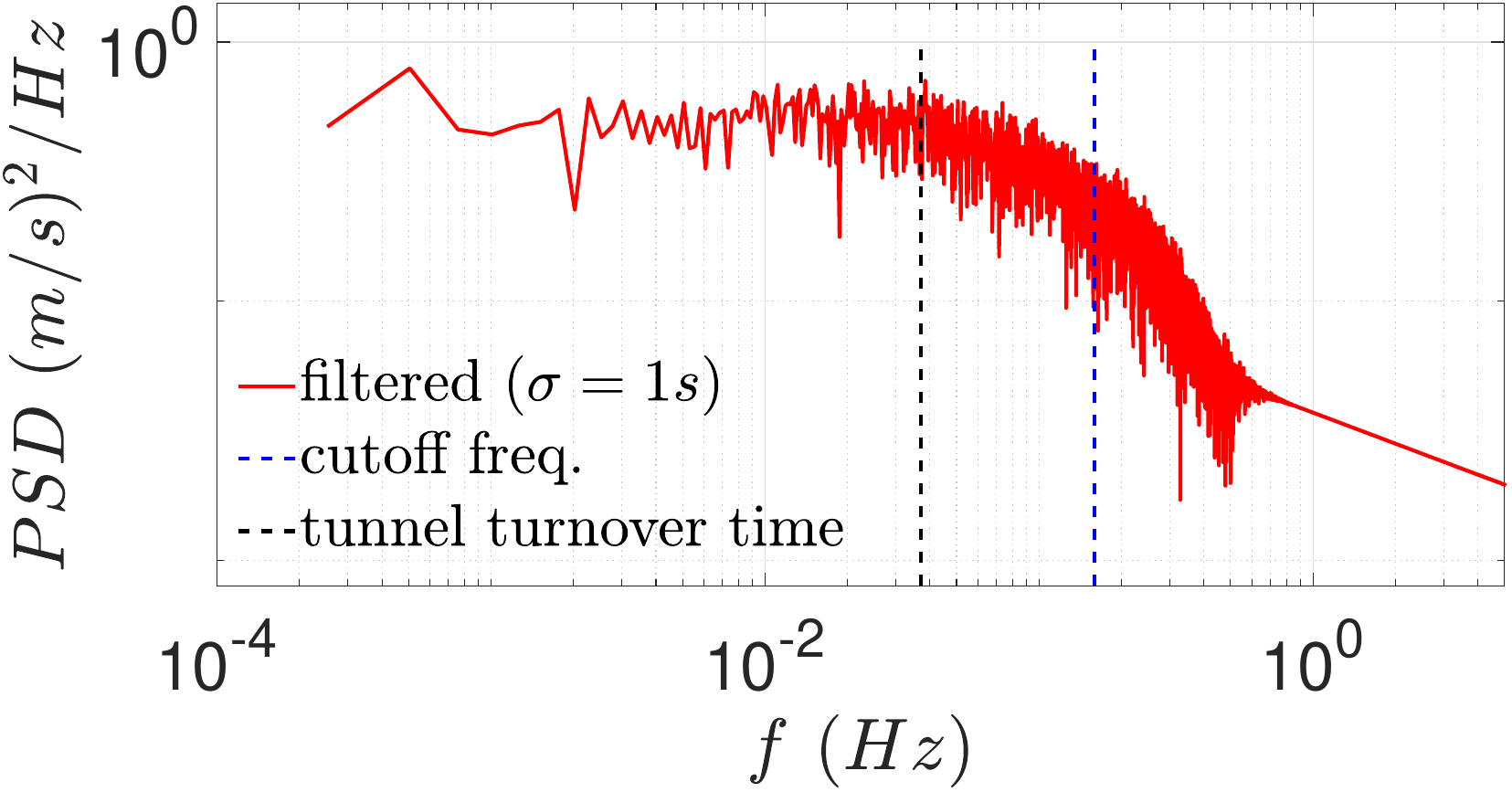}
  \caption{$U_{\infty}=7$ m/s}
  \label{}
  \end{subfigure}
      \caption{Original and filtered velocity signal for three $U_{\infty}$ (a, c, e) and the corresponding power spectral densities (b, d, f). Measurement point is at $x=L/2$ in the midpoint of the $y-z$ plane.}
      \label{timesignalandPSD}
      \end{center}
  \end{figure}

Longer measurements ($\approx$ 60 minutes) were performed to investigate potential large scale variations in the mean velocity. These measurements were taken along the centerline of the test section ($x=L/2$, $y=h/2$, $z=w/2$) for three different $U_{\infty}$ (3,5 and 7 m/s or 200, 285 and 400 RPM). To help visualize low frequency trends, the original signal was first smoothed using a Gaussian filter with a standard deviation of $\sigma_t = 10$ s (\autoref{timesignalandPSD}), as the focus here is not on high-frequency and short-term fluctuations. The instantaneous velocity $u$ deviated by maximum 0.74\% at 3 m/s, 0.28\% at 5 m/s, and 0.35\% at 7 m/s from $U_{\infty}$ across the full signal duration. Although there is no evidence of mean drift, some minor periodicity on the order of one minute can be observed.

To further investigate any dominant frequencies, the power spectral density (PSD) was computed (\autoref{timesignalandPSD}). To that end, a Gaussian filter with a smaller standard deviation ($\sigma_t = 1$ s) was applied to the signal. Then, the data was resampled to a uniform time grid using piecewise cubic Hermite polynomial interpolation (PCHIP). The effective cutoff frequency introduced by the Gaussian filter can be approximated by $f_c = 1/(2\pi\sigma_t)$ and it is indicated in \autoref{timesignalandPSD}. For reference, the frequency associated with the tunnel turnover time, computed as $\tau_{tunnel} = V_{tunnel}/(U_{\infty} h w)$, is also shown, as this could potentially introduce periodic effects.

In general, the PSDs for all three $U_{\infty}$ do not reveal any dominant peaks. For the lowest $U_{\infty}$, a local minimum appears around 0.02 Hz corresponding roughly to the one-minute fluctuation seen in the time signal. Additionally, a local minimum is observed near the tunnel turnover frequency in the intermediate $U_{\infty}$ case. However, these features are subtle and may result from signal processing effects such as filtering or spectral resolution.

\FloatBarrier

\subsection{Inflow measurements at the midpoint}

\subsubsection{Freestream turbulence intensity}

\begin{figure}[!t]
\centering
\includegraphics[width= 0.7\linewidth]{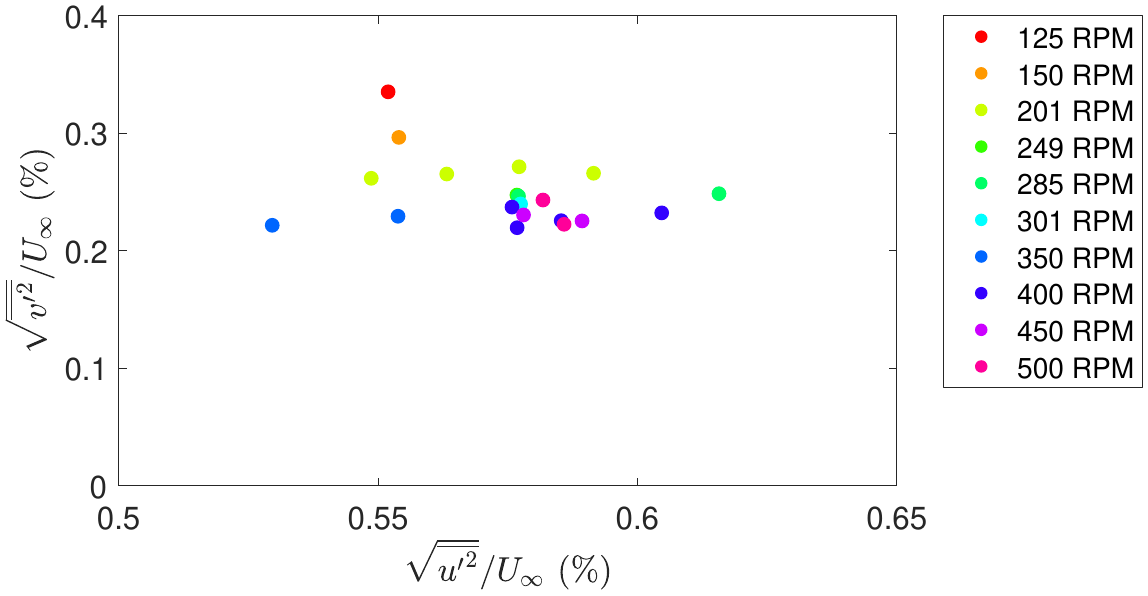}
\caption{Turbulent intensity in the freestream ($x=0.05L$) for a wide range of RPM of the pump (125-500 RPM), corresponding to $U_{\infty}$ ranging from 2-9 m/s.}
\label{RPM_Uti_vti}
\end{figure}

In order to estimate the freestream turbulent intensity, 30 minute long measurements were performed close to the inlet of the test section ($x=0.05L$)  at the midpoint of the $y-z$ plane ($y=h/2$ and $z=w/2$). These were performed for a wider range of RPM of the pump (125-500 RPM) than the flow uniformity, TBL and long term measurements, corresponding to $U_{\infty}$ ranging from 2-9 m/s. For each $U_{\infty}$, the velocity time series ($u$ and $v$), were filtered to remove outliers (based on $5\sigma$) and then the mean velocity was computed. Then, the mean velocity was subtracted from the instantaneous one. Consequently, the velocity fluctuations were analyzed using the slotting technique  (\cite{mayo1974digital,tummers2001spectral}) as for the flow uniformity measurements. Following this procedure, a turbulence intensity of
$0.5\%$ - $0.6\%$ was found in the freestream for all cases (\autoref{RPM_Uti_vti}).

\FloatBarrier
\subsubsection{RPM - $U_{\infty}$ correlation}

\begin{figure}[!htp]
\centering
\includegraphics[width= 0.7\linewidth]{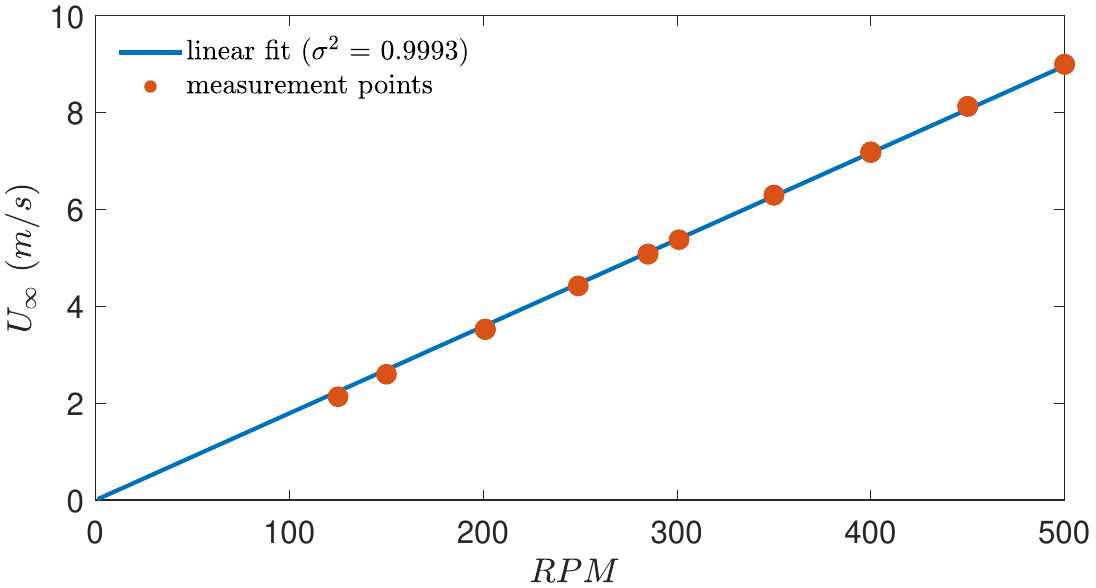}
\caption{Correlation of the freestream velocity and the rotation frequency of the VIT.}
\label{Uinf_RPM}
\end{figure}

The same measurements were used in order to establish the relation between the RPM and the freestream velocity at the tunnel (\autoref{Uinf_RPM}). A linear relation between the RPM and $U_{\infty}$ was found, which can be approximated by: 

\begin{equation}
    U_{\infty}=0.0179\times RPM
\end{equation}

Based on this correlation, the maximum freestream velocity expected in the test section is 13.5 m/s for the maximum rotational frequency of the VIT (755 RPM).

\FloatBarrier
\subsubsection{Calibration of the differential pressure sensor}
 
Since the differential pressure over the large contraction (see \autoref{mpft_archit}) is expected to be used as a bulk indication of the flow velocity at the test section, a comparison with the velocity acquired from LDA at the inlet of the test section ($x=L/2$, $y=h/2$ and $z=w/2$) is made. 

\begin{figure}[htp!]
\centering
\includegraphics[width= 0.7\linewidth]{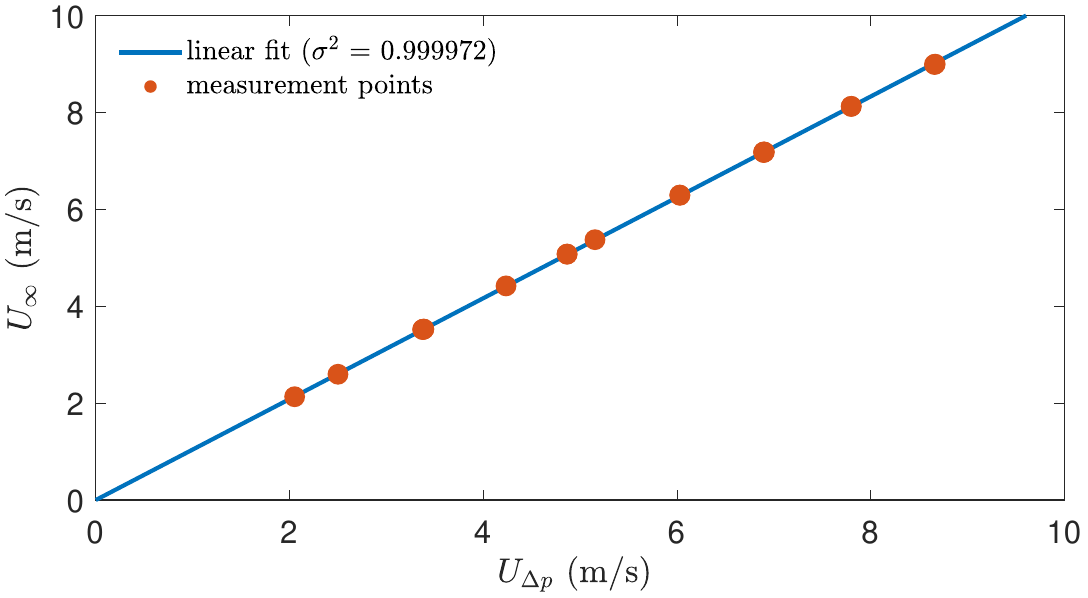}
\caption{Correlation of the freestream velocity at $x=0.05L$ and at the midpoint of the $y-z$ plane and with the differential pressure over the contraction (see \autoref{mpft_archit}).}
\label{Uinf_dp}
\end{figure}

Based on the differential pressure over the contraction, $U_{\Delta p}$ was calculated via Bernoulli (see \autoref{udp}). A linear correction is found between $U_{\infty}$ and $U_{\Delta p}$ over a wide range of conditions. More specifically:

\begin{equation}
    U_{\infty}=1.0415\times U_{\Delta p}
\end{equation}

The $4.15\%$ difference, found between $U_{\Delta p}$ and $U_{\infty}$ can be partially explained considering that using Bernoulli, a uniform velocity over the depth is assumed, whereas in reality there is a logarithmic profile in the tunnel due to the growing boundary layers of the top and side walls. Since we have already established an estimate of the TBL thickness of the top and side walls (see \autoref{BLmeasurements}), assuming symmetry we can reconstruct the velocity profile over the depth of the tunnel. Subsequently, by integrating the velocity profile over the depth we arrive at $U_{\infty}/U_{\Delta p}\approx 2.8\%$.
\FloatBarrier

\subsection{Repeatability of the measurements}

\begin{table*}[!ht]
\centering
\begin{tabular}{c c c c c}
& & \\ 
 \hline
 \hline
 \multicolumn{5}{c}{$U_{\infty}=3.5$ m/s} \\
 \hline
 \hline
x  & $Re_{x}\times10^6$ & $U_{\infty}$ from BL meas. & $U_{\infty}$ from grid meas. & $U_{\infty}$ from long meas.\\
\hline
0.05L & 1.7 & - & 3.51 & 3.53\\
L/2 & 5.4 & 3.51 & 3.50 & 3.53\\
0.95L & 9.1 & - & 3.44 & - \\
\hline
\hline
\multicolumn{5}{c}{$U_{\infty}=5$ m/s} \\
 \hline
 \hline
x  & $Re_{x}\times10^6$ & $U_{\infty}$  from BL meas. & $U_{\infty}$ from grid meas. & $U_{\infty}$ from long meas.\\
\hline
0.05L & 2.4 & 5.04 & 5.07 & 5.08\\
L/2 & 7.6 & 5.07 & 5.06 & 5.07\\
0.95L & 12.8 & 5.00 & 4.98 & - \\
\hline
\hline
\multicolumn{5}{c}{$U_{\infty}=7$ m/s} \\
 \hline
 \hline
x  & $Re_{x}\times10^6$ & $U_{\infty}$ from BL meas. & $U_{\infty}$ from grid meas. & $U_{\infty}$ from long meas.\\
\hline
0.05L & 0.8 & 7.14 & 7.16 & 7.18\\
L/2 & 8.2 & 7.19 & 7.15 & 7.15\\
0.95L & 15.6 & - & 7.01 & - \\
\hline
\hline
\end{tabular}
\caption{Repeatability of $U_{\infty}$.}
\label{repeatability}
\end{table*}

As a result of the multiple measurements (TBL measurements, flow uniformity and long term measurements), each targeting different flow qualities, $U_{\infty}$ was recorded multiple times. A summary is given in \autoref{repeatability}, where $U_{\infty}$ is measured at the midpoint of the test section at all cases. Across all three measured $U_{\infty}$, the difference is lower than $1\%$, equal to the uncertainty level of the LDA.

\section{Summary \& Conclusions}

Velocity measurements were performed using Laser Doppler Anemometry to assess the flow quality in the test section of the new Multiphase Flow Tunnel at Delft University of Technology.  Preliminary boundary layer measurements revealed that the TBL growth is not strictly canonical and the TBL starts developing upstream of the test section. Smaller TBL thickness was measured along the side wall than the top wall. Local measurements of the streamwise and vertical velocity along multiple streamwise locations in the test section revealed that the flow is uniform (within 1\% of the bulk velocity) with few exceptions near the walls. The corresponding turbulent intensity of these local measurements was found to be 0.5\%-1\%. Longer term measurements (approximately 60 minutes) performed in the middle of the test section, did not reveal any large scale fluctuations of the mean velocity, although some periodicity of the order of 1 minute was detected. Shorter measurements (approximately 30 minutes) at the inlet of the test section and over a large range of conditions (corresponding to 2 - 9 m/s), revealed that the freestream velocity exhibited a linear relation with rotational frequency of the VIT. Based on the same measurements, a 4.2\% difference is found between the freestream velocity in the test section measured with LDA and the velocity derived from the differential pressure over the large contraction. That can largely be explained by the underlying assumptions of converting the differential pressure to velocity at the test section. Finally, the freestream turbulent intensity was found from the same measurement location to be 0.5\% - 0.6\% for all $U_{\infty}$. 

\section*{\small Acknowledgments}

The authors would like to thank Prof. M.J. Tummers for valuable instruction in Laser Doppler Anemometry measurements and the technical staff of the laboratory of Ship Hydromechanics for their support and assistance.

\section*{\small Funding}
This work is part of the public–private research program “Water Quality in Maritime Hydrodynamics” (AQUA) project P17-07. The support by the Netherlands Organisation for Scientific Research (NWO) Domain Applied and Engineering Sciences, and project partners is gratefully acknowledged.

\appendix

\section{Additional local velocity measurements}
\label{extraLDA}

\begin{figure}[!h]
\centering
\resizebox{0.75\textwidth}{!}{%
\begin{minipage}{\textwidth}
\noindent
\begin{tabular}{p{0.36\linewidth} p{0.16\linewidth} p{0.32\linewidth}}
\centering \textbf{$x=0.05L$} &
\centering \textbf{$x=L/2$} &
\centering \textbf{$x=0.95L$}
\end{tabular}
\par\medskip
\begin{subfigure}{0.34\linewidth}
  \centering
  \includegraphics[width=\linewidth]{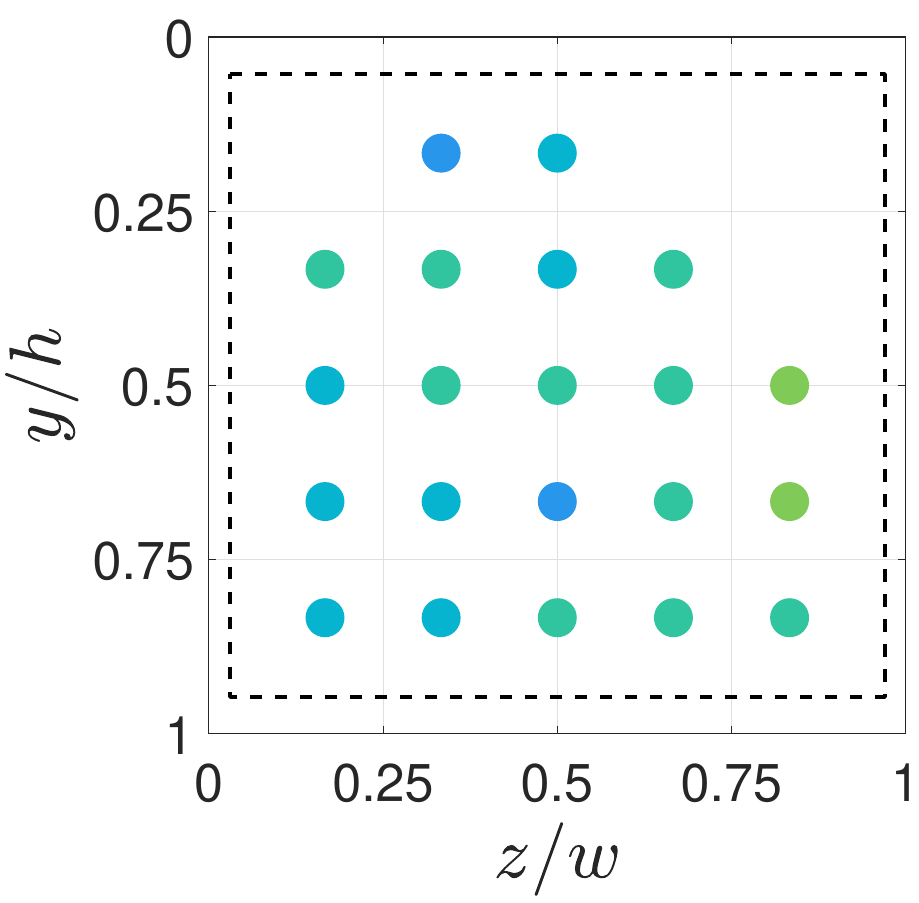}
  \caption{}
  \end{subfigure}
  \begin{subfigure}{0.27\textwidth}
  \centering
    \includegraphics[width=\linewidth]{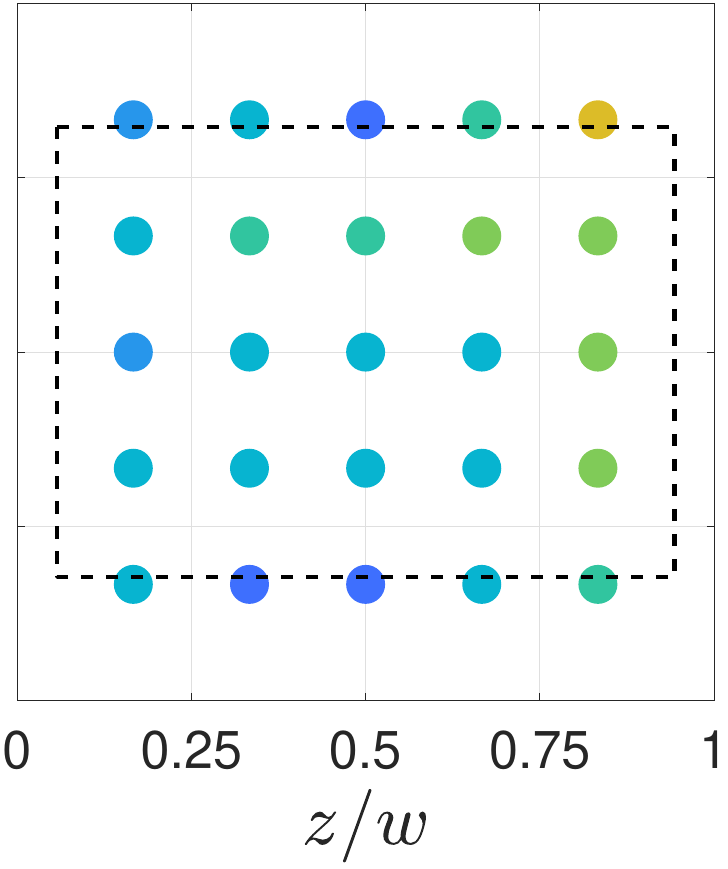}
      \caption{}
      \end{subfigure}
      \begin{subfigure}{0.36\linewidth}
        \centering
         \includegraphics[width=\textwidth]{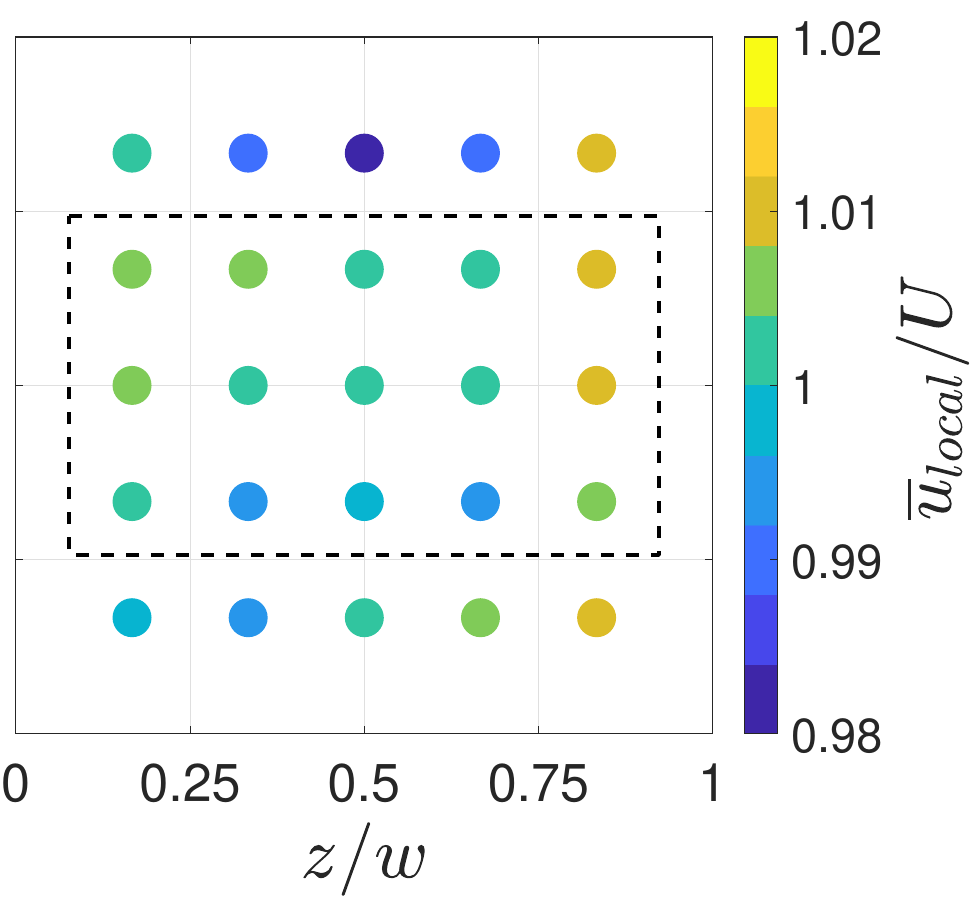}
                \caption{}
      \end{subfigure}
    \vfill
\begin{subfigure}{0.34\linewidth}
  \centering
  \includegraphics[width=\linewidth]{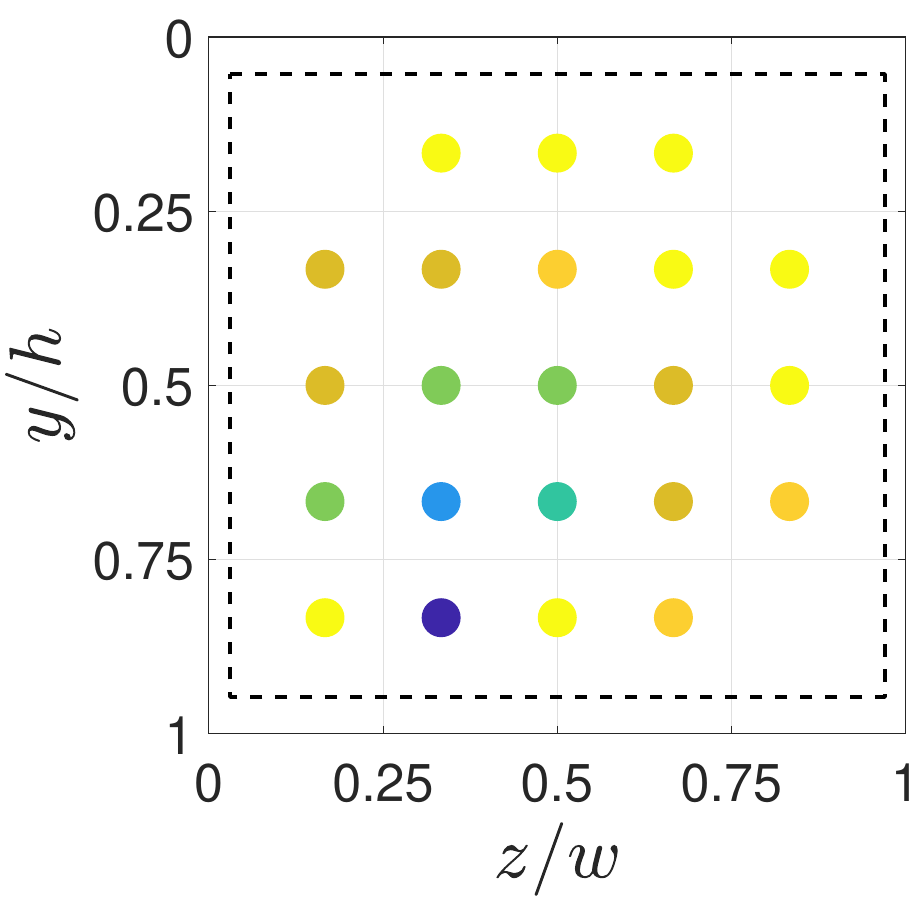}
  \caption{}
   \label{Vmean_up_200}
  \end{subfigure}
  \begin{subfigure}{0.27\textwidth}
  \centering
    \includegraphics[width=\linewidth]{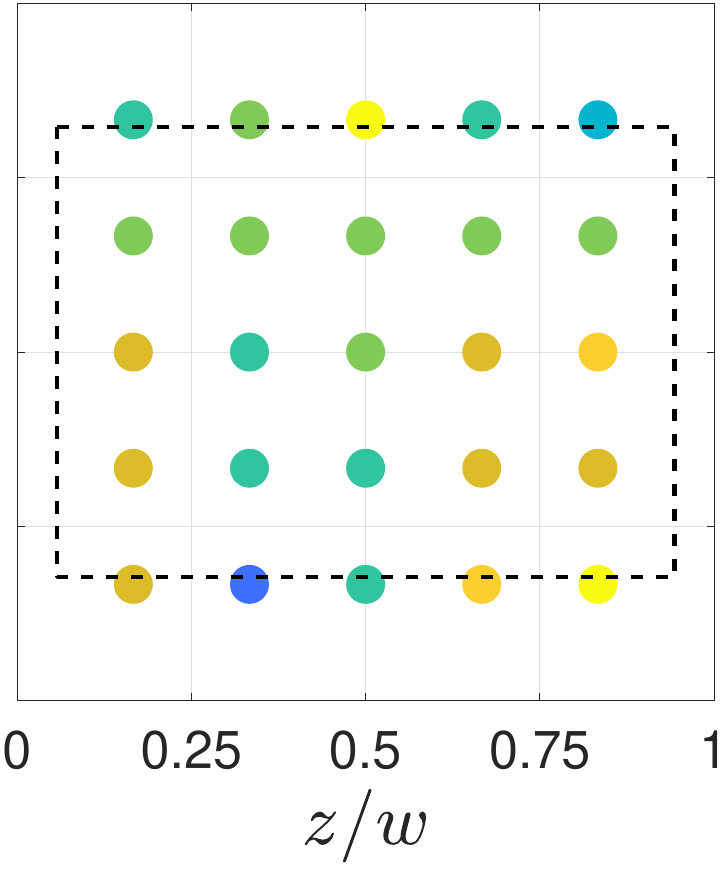}
      \caption{}
       \label{Vmean_middle_200}
      \end{subfigure}
      \begin{subfigure}{0.36\linewidth}
        \centering
         \includegraphics[width=\textwidth]{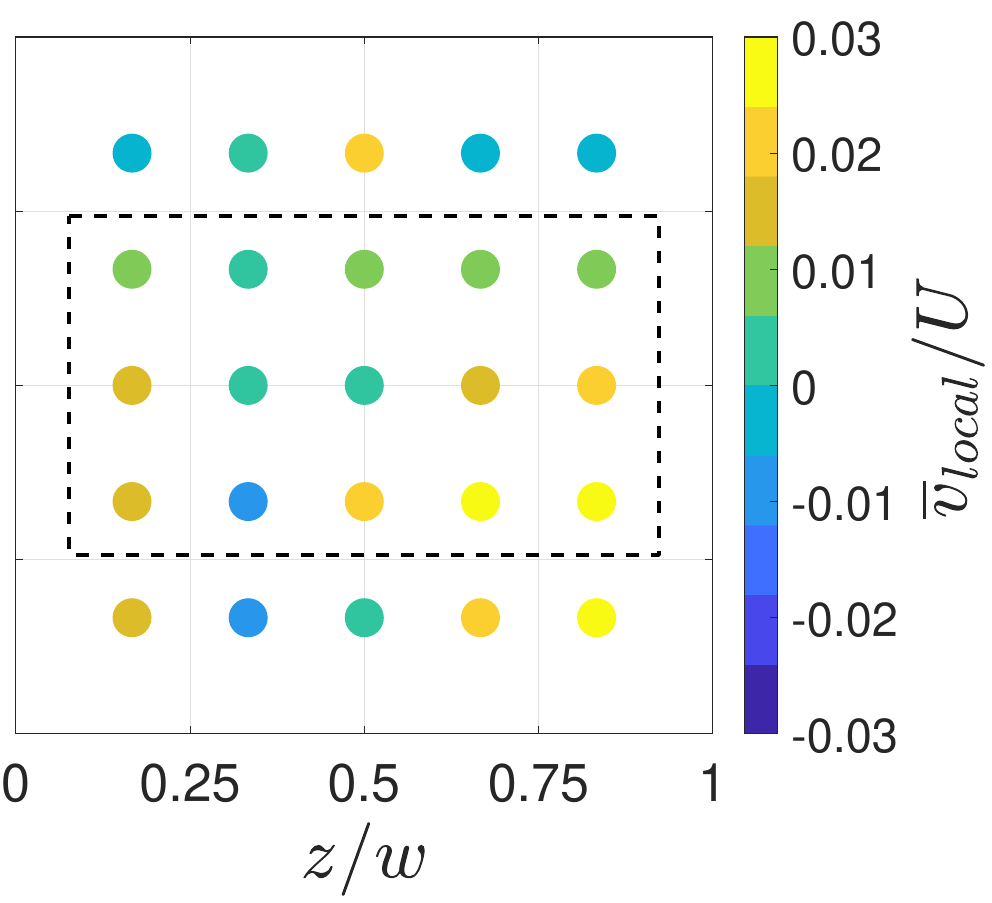}
                \caption{}
                \label{Vmean_down_200}
      \end{subfigure}
      \end{minipage}
}
            \caption{Local measurements of the mean streamwise velocity $\overline{u}$ (a, b, c) and the mean vertical velocity $\overline{v}$ (d, e, f) in three streamwise locations. Local velocities are normalized by the local mean freestream velocity $U$. Measurements are shown for 200 RPM or $U_{\infty}=3.5$ m/s. Dashed line indicates the TBL location.}
      \label{200RPM_velos}
  \end{figure}

\begin{figure}[!h]
\centering
\resizebox{0.75\textwidth}{!}{%
\begin{minipage}{\textwidth}
 %\textbf{$x=0.05L$ \hspace{2cm} $x=L/2$ \hspace{2cm} $x=0.95L$}\par\medskip
\noindent
\begin{tabular}{p{0.36\linewidth} p{0.16\linewidth} p{0.32\linewidth}}
\centering \textbf{$x=0.05L$} &
\centering \textbf{$x=L/2$} &
\centering \textbf{$x=0.95L$}
\end{tabular}
\par\medskip
\begin{subfigure}{0.34\linewidth}
  \centering
  \includegraphics[width=\linewidth]{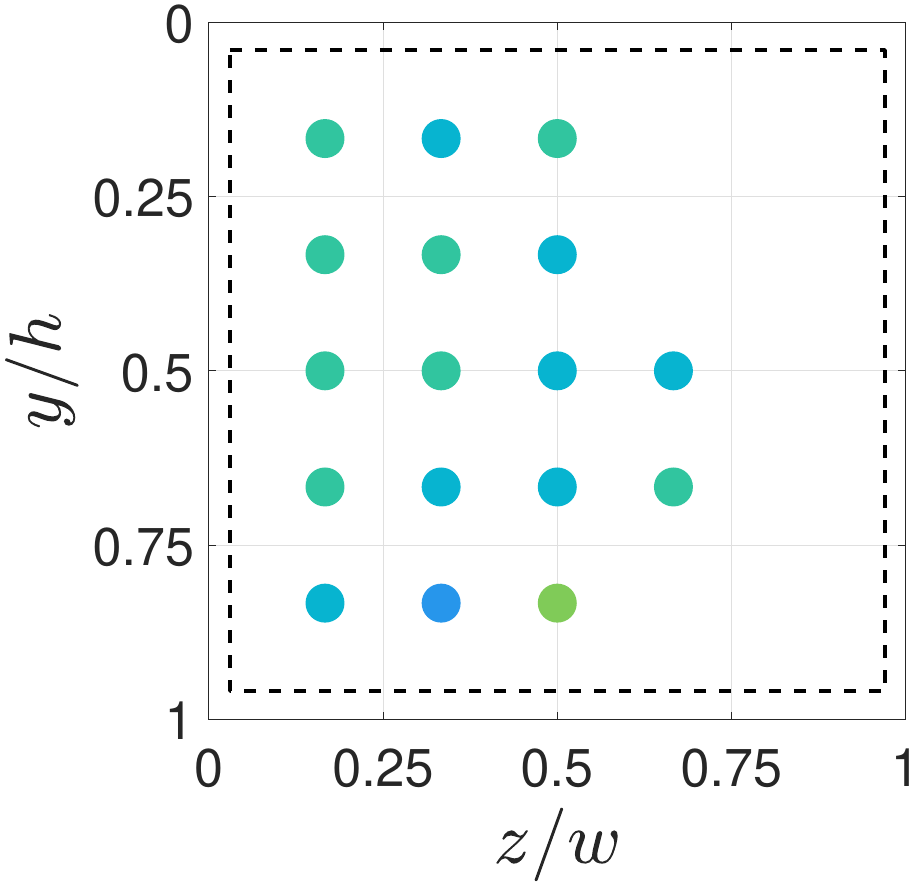}
  \caption{}
  \end{subfigure}
  \begin{subfigure}{0.27\textwidth}
  \centering
    \includegraphics[width=\linewidth]{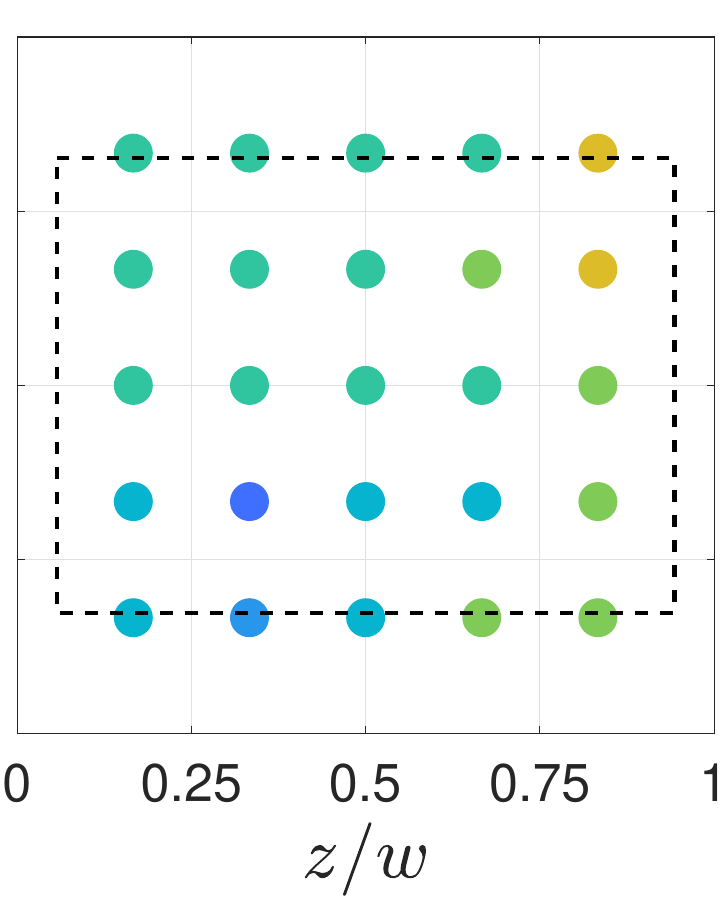}
      \caption{}
      \end{subfigure}
      \begin{subfigure}{0.36\linewidth}
        \centering
         \includegraphics[width=\textwidth]{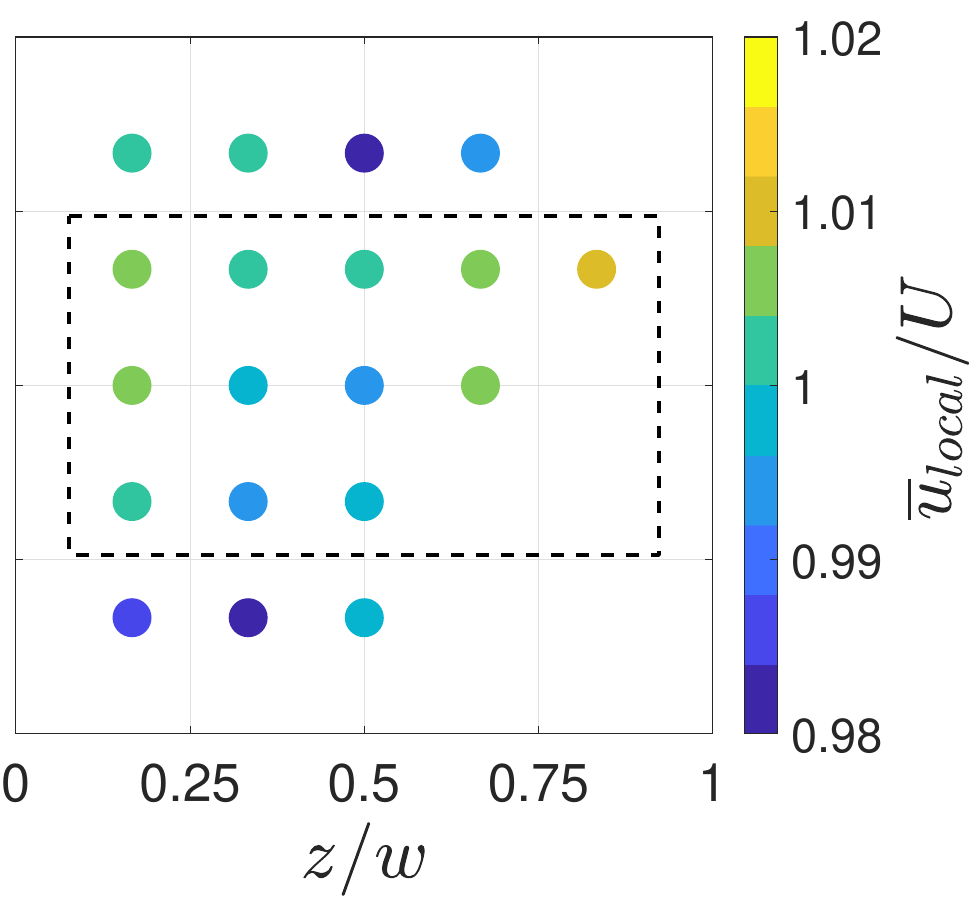}
                \caption{}
      \end{subfigure}
    \vfill
\begin{subfigure}{0.34\linewidth}
  \centering
  \includegraphics[width=\linewidth]{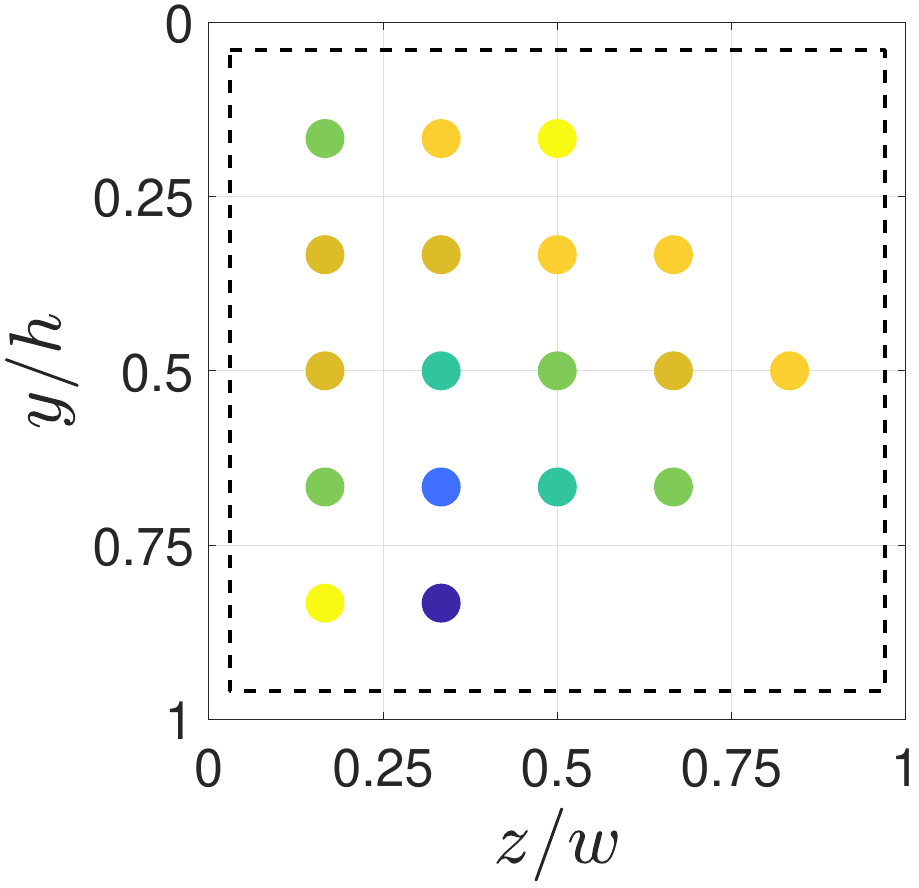}
  \caption{}
   \label{Vmean_up_400}
  \end{subfigure}
  \begin{subfigure}{0.27\textwidth}
  \centering
    \includegraphics[width=\linewidth]{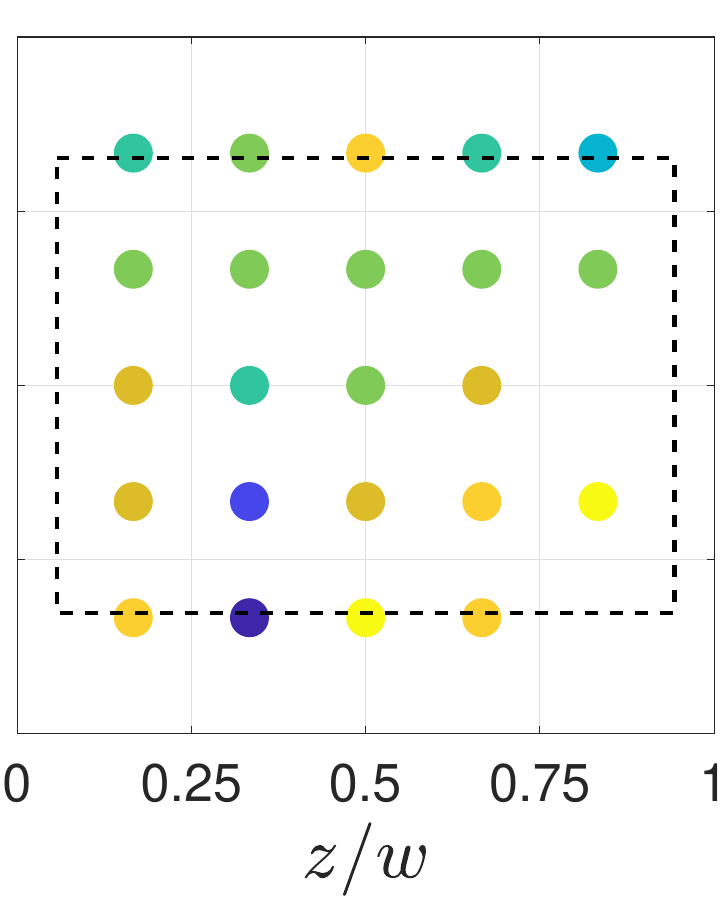}
      \caption{}
       \label{Vmean_middle_400}
      \end{subfigure}
      \begin{subfigure}{0.36\linewidth}
        \centering
         \includegraphics[width=\textwidth]{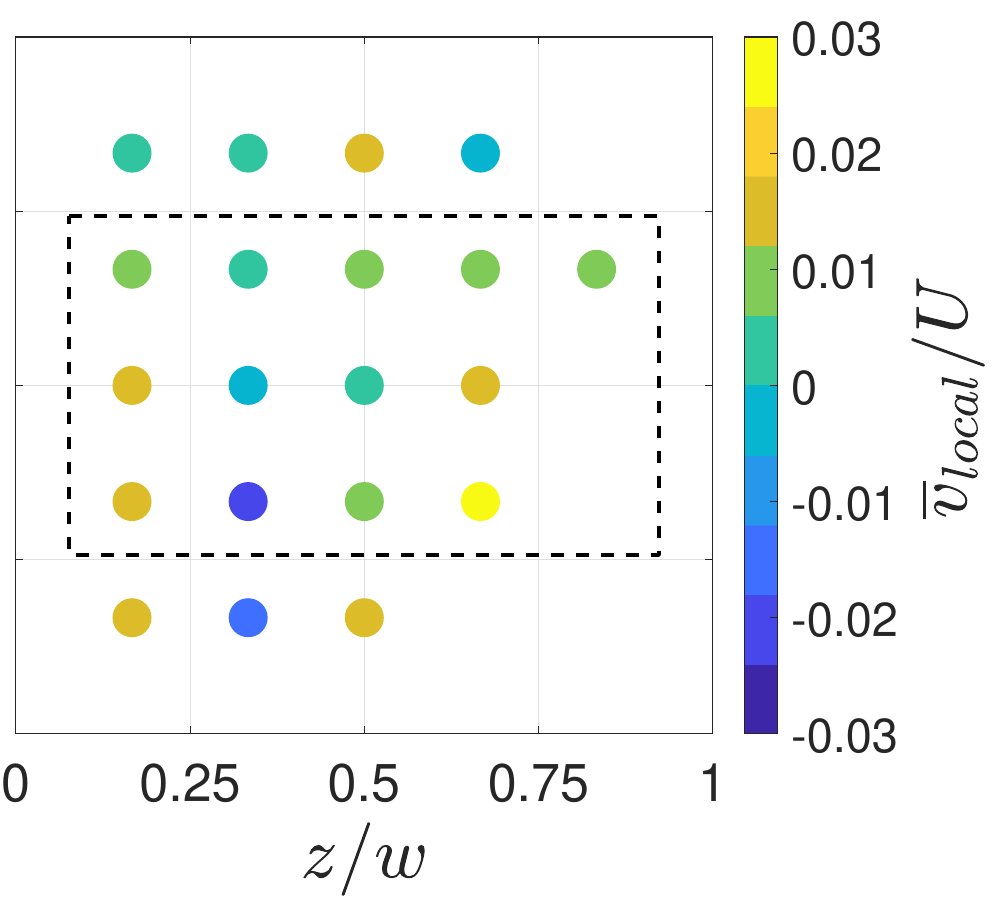}
                \caption{}
                 \label{Vmean_down_400}
      \end{subfigure}
      \end{minipage}%
        }
      \caption{Local measurements of the mean streamwise velocity $\overline{u}$ (a, b, c) and the mean vertical velocity $\overline{v}$ (d, e, f) in three streamwise locations. Local velocities are normalized by the local mean freestream velocity $U$. Measurements are shown for 400 RPM or $U_{\infty}=7$ m/s. Dashed line indicates the TBL location.}
    \label{400RPM_velos}
  \end{figure}

\begin{figure}[!h]
\centering
\resizebox{0.75\textwidth}{!}{%
\begin{minipage}{\textwidth}
 %\textbf{$x=0.05L$ \hspace{2cm} $x=L/2$ \hspace{2cm} $x=0.95L$}\par\medskip
\noindent
\begin{tabular}{p{0.36\linewidth} p{0.16\linewidth} p{0.32\linewidth}}
\centering \textbf{$x=0.05L$} &
\centering \textbf{$x=L/2$} &
\centering \textbf{$x=0.95L$}
\end{tabular}
\par\medskip
 \begin{subfigure}{0.34\linewidth}
  \centering
  \includegraphics[width=\linewidth]{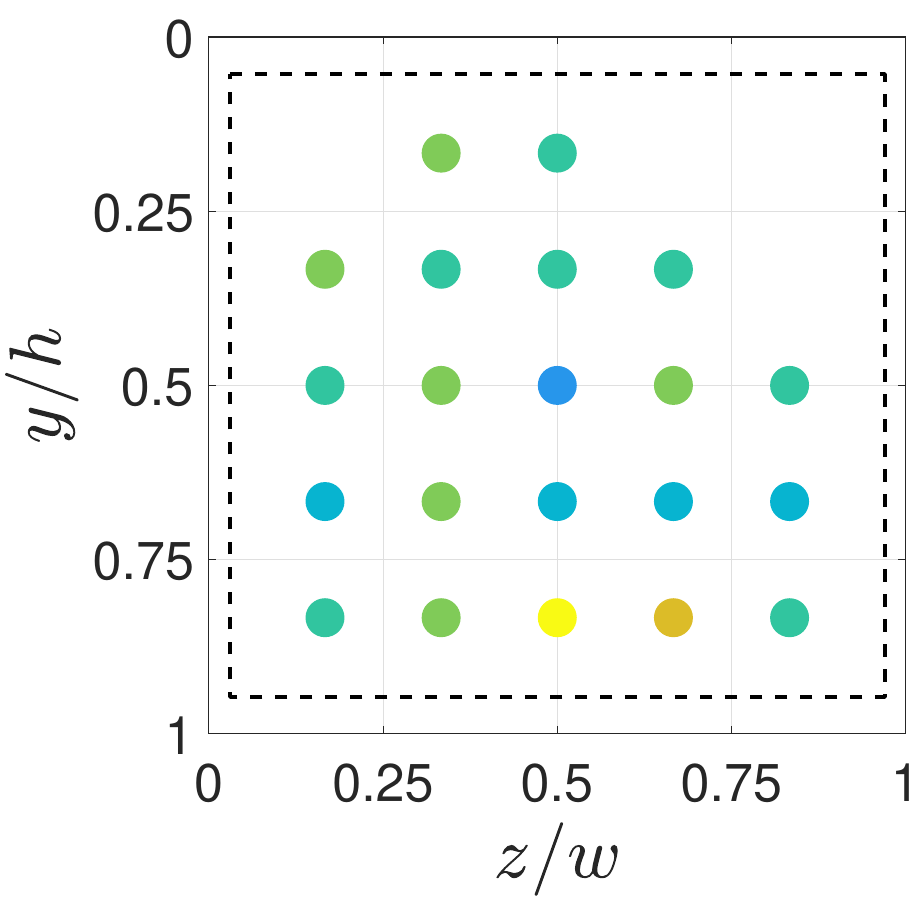}
  \caption{}
  \end{subfigure}
  %\hspace{0.01\textwidth}
  \begin{subfigure}{0.27\textwidth}
  \centering
    \includegraphics[width=\linewidth]{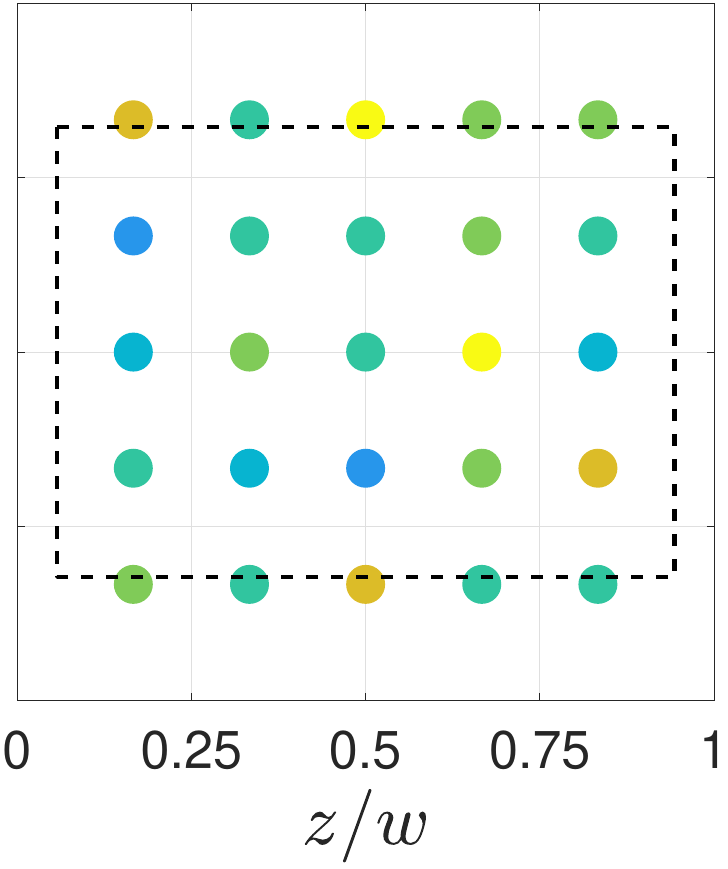}
      \caption{}
      \end{subfigure}
      \begin{subfigure}{0.36\linewidth}
        \centering
         \includegraphics[width=\textwidth]{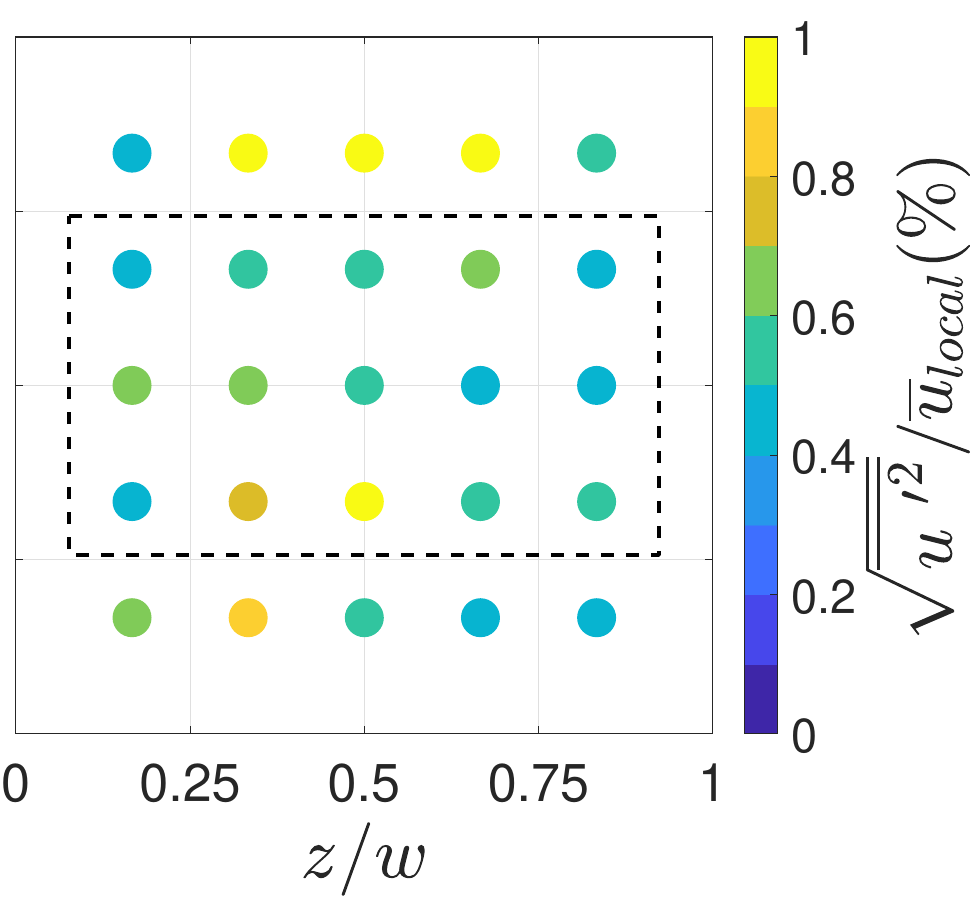}
                \caption{}
      \end{subfigure}
    \vfill
\begin{subfigure}{0.34\linewidth}
  \centering
  \includegraphics[width=\linewidth]{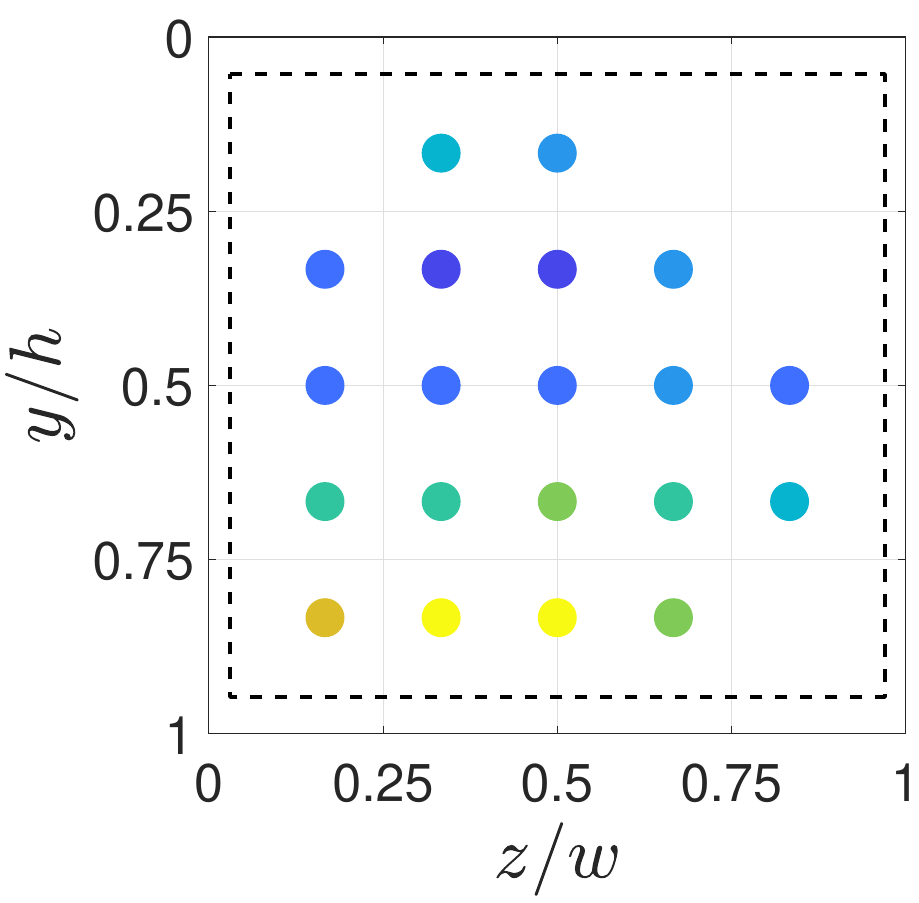}
  \caption{}
  \end{subfigure}
  \begin{subfigure}{0.27\textwidth}
  \centering
    \includegraphics[width=\linewidth]{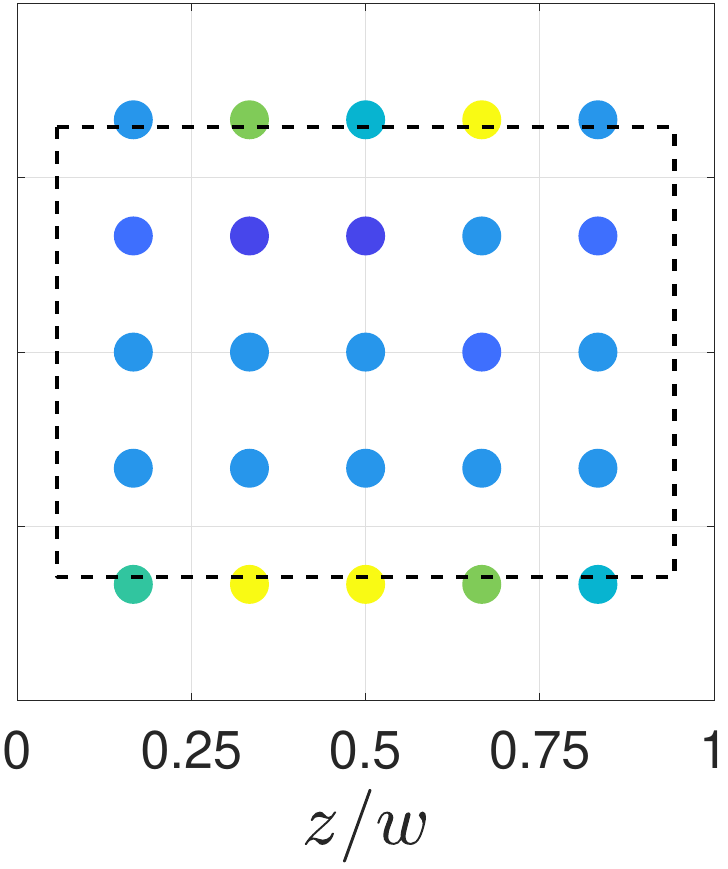}
      \caption{}
      \end{subfigure}
      \begin{subfigure}{0.36\linewidth}
        \centering
         \includegraphics[width=\textwidth]{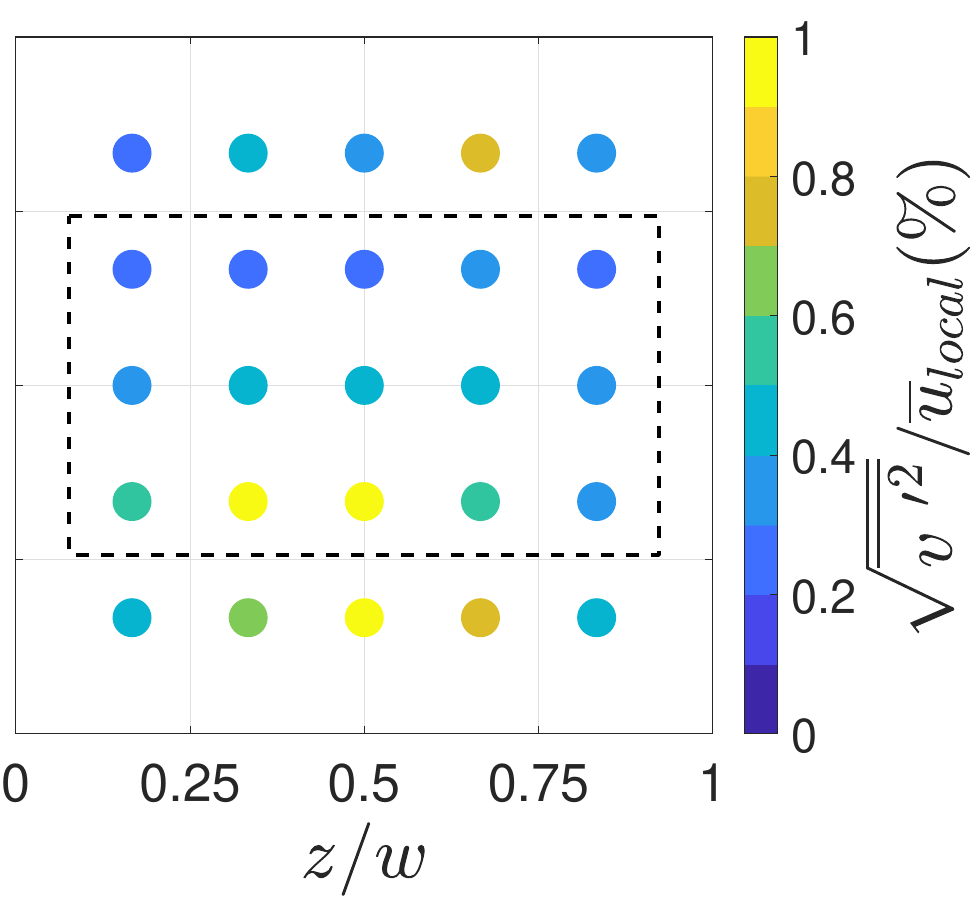}
                \caption{}
      \end{subfigure}
      \end{minipage}
      }
      \caption{Local turbulence intensity of the  streamwise (a, b, c) and vertical (d, e, f) velocity  in three streamwise locations for 200 RPM or $U_{\infty}=3.5$ m/s. Dashed lines indicate the TBL location.}
                        \label{200RPM_TI}
  \end{figure}

\begin{figure}[!h]
\centering
\resizebox{0.75\textwidth}{!}{%
\begin{minipage}{\textwidth}
 %\textbf{$x=0.05L$ \hspace{2cm} $x=L/2$ \hspace{2cm} $x=0.95L$}\par\medskip
\noindent
\begin{tabular}{p{0.36\linewidth} p{0.16\linewidth} p{0.32\linewidth}}
\centering \textbf{$x=0.05L$} &
\centering \textbf{$x=L/2$} &
\centering \textbf{$x=0.95L$}
\end{tabular}
\par\medskip
\begin{subfigure}{0.34\linewidth}
  \centering
  \includegraphics[width=\linewidth]{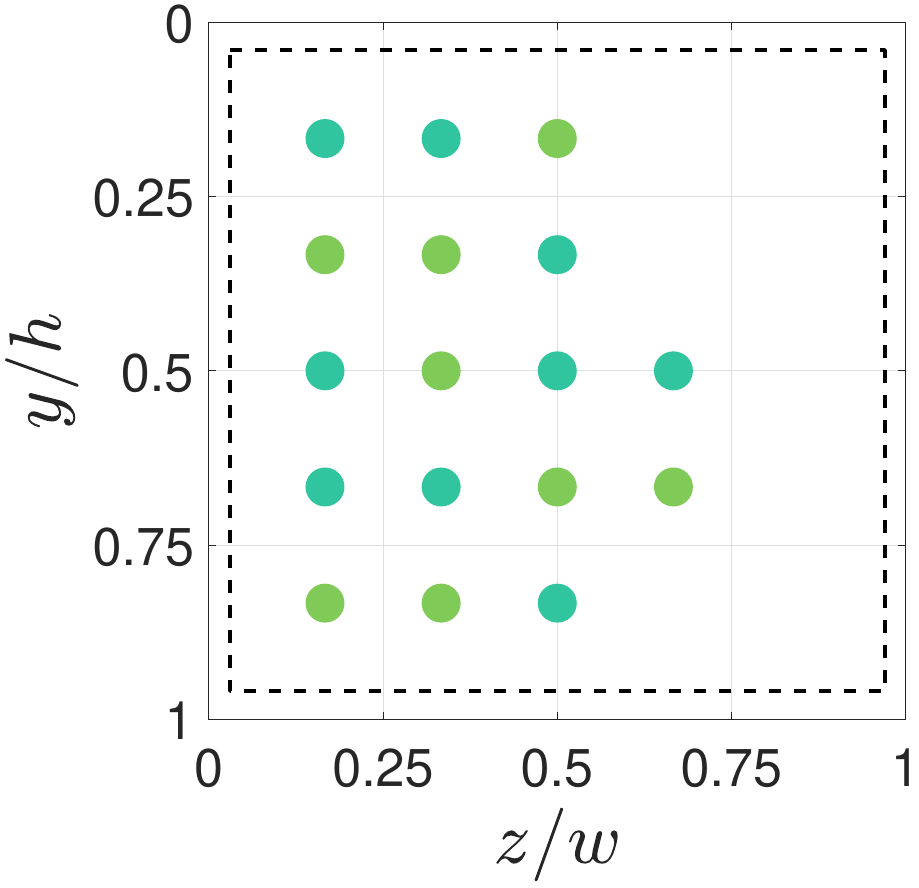}
  \caption{}
  \end{subfigure}
  \begin{subfigure}{0.27\textwidth}
  \centering
    \includegraphics[width=\linewidth]{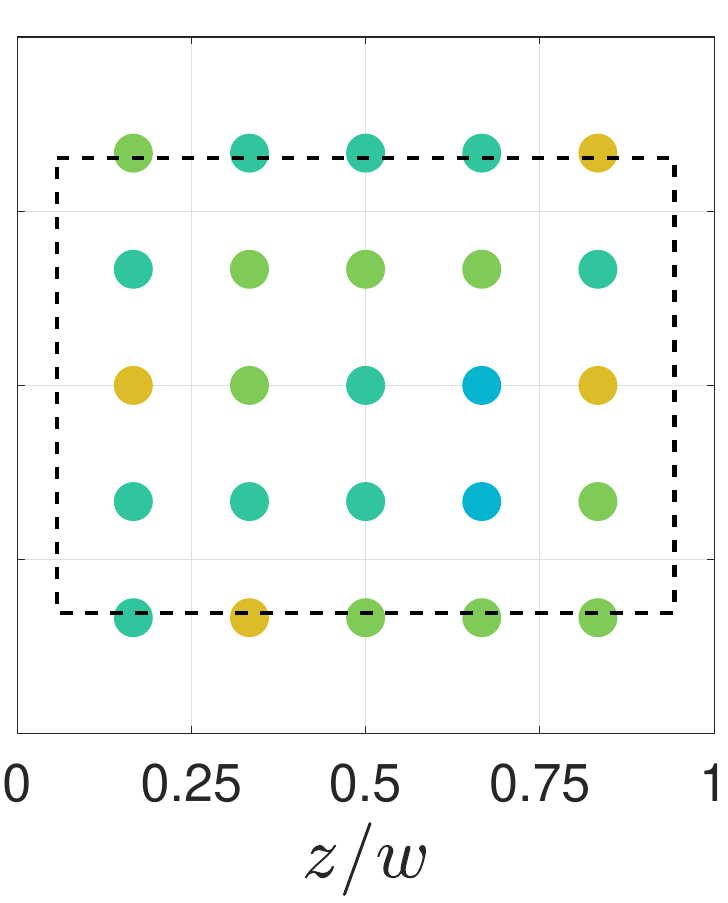}
      \caption{}
      \end{subfigure}
      \begin{subfigure}{0.36\linewidth}
        \centering
         \includegraphics[width=\textwidth]{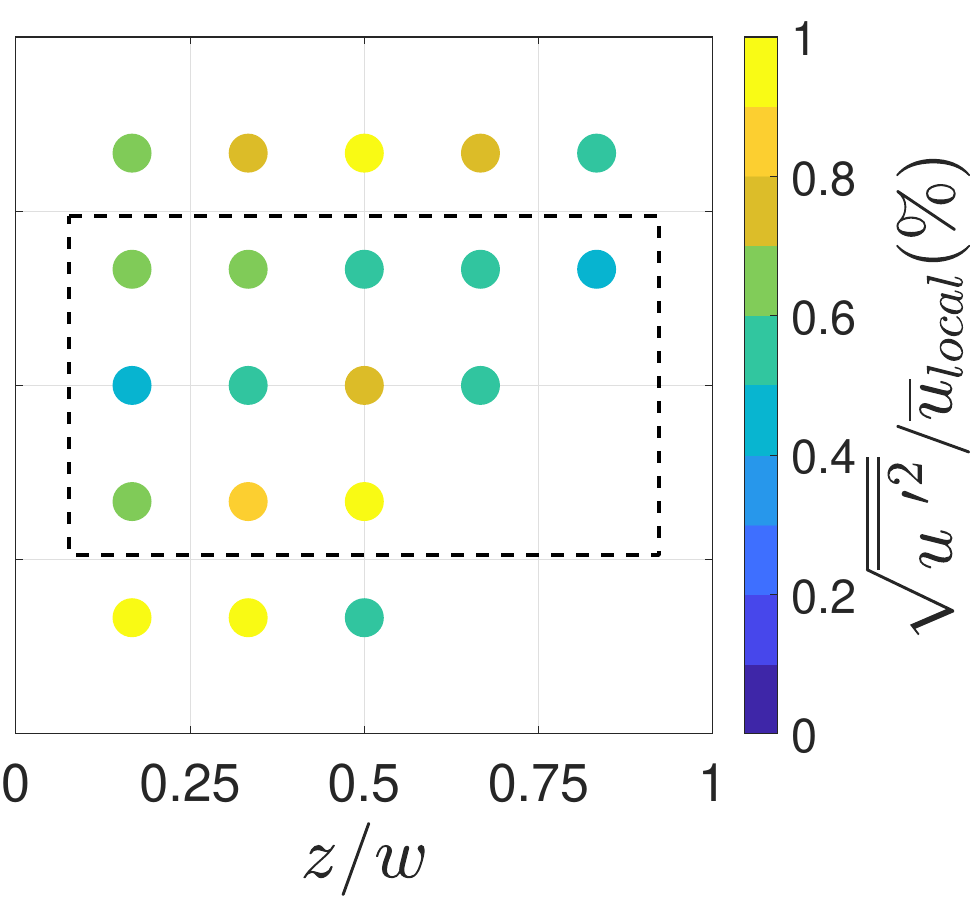}
                \caption{}
      \end{subfigure}
    \vfill
\begin{subfigure}{0.34\linewidth}
  \centering
  \includegraphics[width=\linewidth]{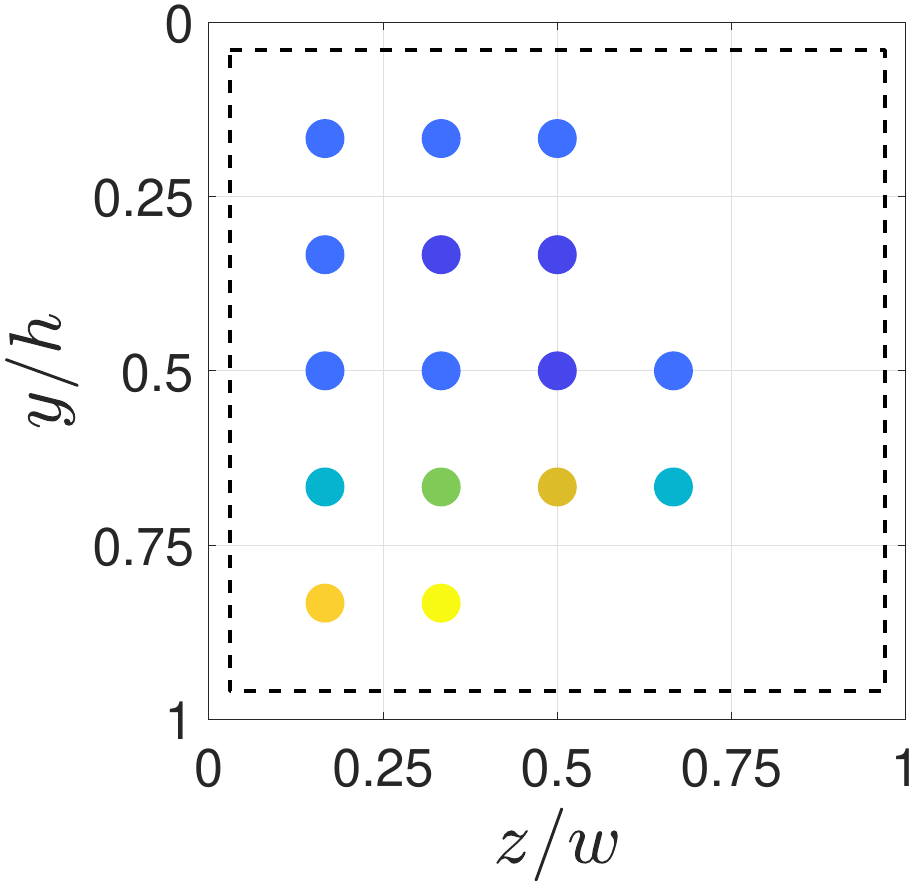}
  \caption{}
  \end{subfigure}
  \begin{subfigure}{0.27\textwidth}
  \centering
    \includegraphics[width=\linewidth]{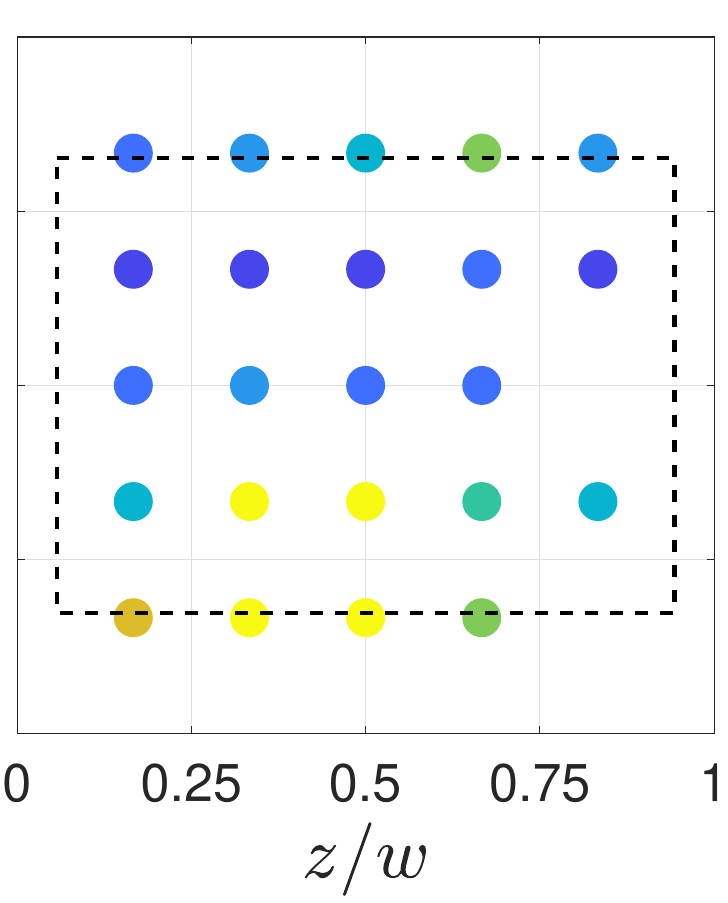}
      \caption{}
      \end{subfigure}
      \begin{subfigure}{0.36\linewidth}
        \centering
         \includegraphics[width=\textwidth]{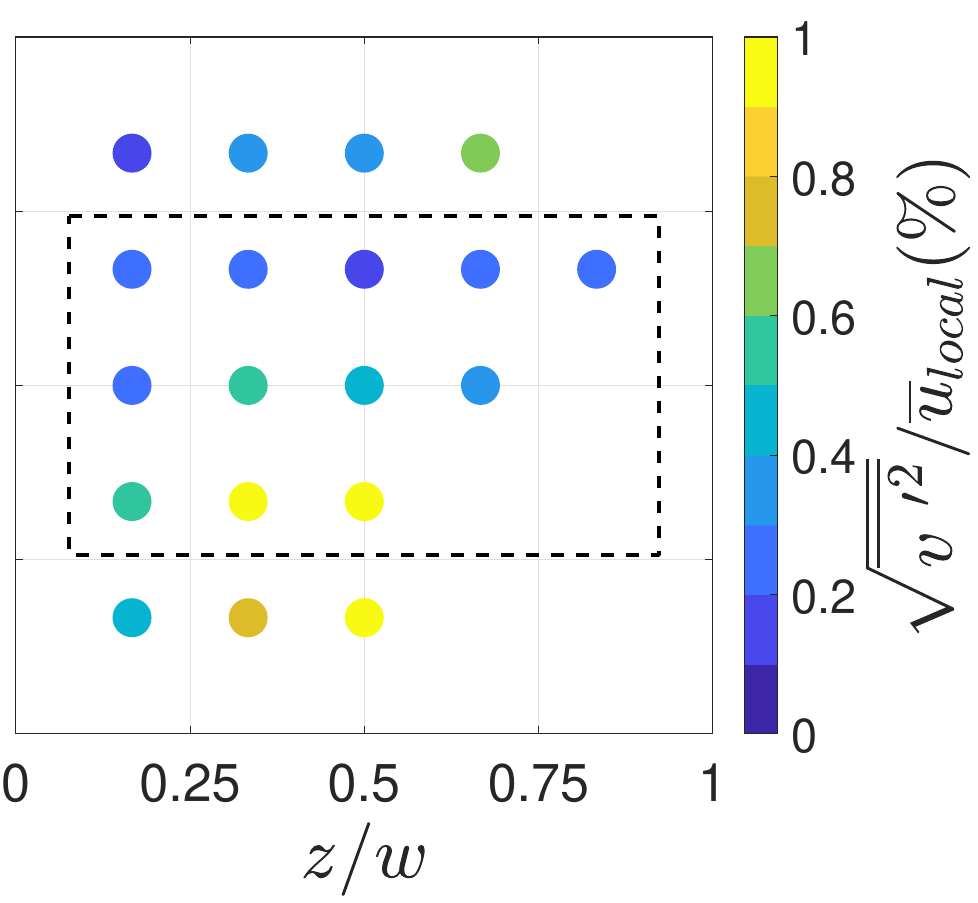}
                \caption{}
      \end{subfigure}
      \end{minipage}
      }
      \caption{Local turbulence intensity of the  streamwise (a, b, c) and vertical (d, e, f) velocity  in three streamwise locations for 400 RPM or $U_{\infty}=7$ m/s. Dashed lines indicate the TBL location.}
                              \label{400RPM_TI}

  \end{figure}

Local velocity measurements of mean streamwise and vertical velocity for two additional rotational frequencies of the VIT are presented in \autoref{200RPM_velos}  and  \autoref{400RPM_velos}. No big deviations are found from the bulk velocity in the core region of the $y-z$ plane for the streamwise velocity. In the case of the vertical velocity,  at $z=h/3$ the two points closer to the bottom wall showcase elevated vertical velocities directed away from the bottom across all streamwise locations (\autoref{Vmean_up_200}-\autoref{Vmean_down_200} and \autoref{Vmean_up_400}-\autoref{Vmean_down_400}). This localized vertical motion,  was also observed in the case of 285 RPM or 5 m/s (see \autoref{VMEAN_up}-\autoref{VMean_down}). The local turbulence intensity was found to be below 1\% for both 200 RPM  (\autoref{200RPM_TI}) and 400 RPM (\autoref{400RPM_TI}).

\clearpage

\FloatBarrier

\bibliographystyle{apalike}
\bibliography{sample}

@manual{TSI_manual,
    title = {Operations Manual Phase Doppler Particle Analyzer (PDPA)/ Laser Doppler Velocimeter (LDV). REVISION D.},
 author={{TSI Incorporated}},
year = {2005}
}

@book{buchhave1978bias,
  title={Bias corrections in turbulence measurements by the laser Doppler anemometer},
  author={Buchhave, PREBEN and George, William K},
  year={1978},
  publisher={State University of New York at Buffalo, Faculty of Engineering and Applied Sciences, Turbulence Research Laboratory}
}

@article{tummers2001spectral,
  title={Spectral analysis of biased LDA data},
  author={Tummers, MJ and Passchier, DM},
  journal={Measurement Science and Technology},
  volume={12},
  number={10},
  pages={1641},
  year={2001},
  publisher={IOP Publishing}
}

@book{ittc2006,
  author       = {{ITTC}},
  title        = {Testing and Extrapolation Methods, General Density and Viscosity of
Water. ITTC - Recommended Procedures and Guidelines},
  year         = {2006},
  publisher    = {International Towing Tank Conference},
}

@inproceedings{mayo1974digital,
  title={Digital estimation of turbulence power spectra from burst counter LDV data},
  author={Mayo Jr, WT and Shay, MT and Riter, S},
  booktitle={2nd International Workshop on Laser Velocimetry, Volume 1},
  volume={1},
  pages={16--24},
  year={1974}
}

@article{clauser1956turbulent,
  title={The turbulent boundary layer},
  author={Clauser, Francis H},
  journal={Advances in applied mechanics},
  volume={4},
  pages={1--51},
  year={1956},
  publisher={Elsevier}
}

@book{white2006viscous,
  title={Viscous fluid flow},
  author={White, Frank M and Majdalani, Joseph},
  volume={3},
  year={2006},
  publisher={McGraw-Hill New York}
}
\end{document}